\documentclass[12pt]{article}
\usepackage{cite}
\usepackage{wasysym}
\usepackage{stmaryrd}
\usepackage{savesym}
\usepackage{graphicx}
\usepackage{pifont}
\usepackage{float}
\usepackage{subfig}
\usepackage[all]{xy}
\usepackage{multicol}
\usepackage{amsfonts}
\usepackage{picture}
\usepackage{amssymb}
\savesymbol{iint}
\savesymbol{iiint}
\usepackage{amsmath}
\restoresymbol{TXF}{iint}
\restoresymbol{TXF}{iiint}
\usepackage{color}
\usepackage{heck}
\usepackage{hyperref}
\hypersetup{colorlinks=true}
\hypersetup{linkcolor=black}
\hypersetup{citecolor=black}
\hypersetup{urlcolor=black}
\usepackage{setspace}
\usepackage{wrapfig}
\usepackage{rotating}
\usepackage{bbm}

\newcommand{\qed}{\nobreak \ifvmode \relax \else
      \ifdim\lastskip<1.5em \hskip-\lastskip
      \hskip1.5em plus0em minus0.5em \fi \nobreak
      \vrule height0.75em width0.5em depth0.25em\fi}
\usepackage{verbatim}
\numberwithin{equation}{section}
\begin{document}

\date{December, 2011}

\institution{HarvardU}{\centerline{${}^{1}$Jefferson Physical Laboratory, Harvard University, Cambridge, MA 02138, USA}}

\institution{SISSA}{\centerline{${}^{2}$Scuola Internazionale Superiore di Studi Avanzati, Via Bonomea 265, 34100 Trieste, ITALY}}

\title{$\mathcal{N}=2$ Quantum Field Theories and Their BPS Quivers}

\authors{Murad Alim,\worksat{\HarvardU}\footnote{e-mail: {\tt alim@physics.harvard.edu}} Sergio Cecotti,\worksat{\SISSA}\footnote{e-mail: {\tt cecotti@sissa.it}} Clay C\'{o}rdova,\worksat{\HarvardU}\footnote{e-mail: {\tt clay.cordova@gmail.com}} Sam Espahbodi,\worksat{\HarvardU}\footnote{e-mail: {\tt espahbodi@physics.harvard.edu}} Ashwin Rastogi,\worksat{\HarvardU}\footnote{e-mail: {\tt rastogi@physics.harvard.edu}} and Cumrun Vafa\worksat{\HarvardU
}\footnote{e-mail: {\tt vafa@physics.harvard.edu}}}

\abstract{We explore the relationship between four-dimensional $\mathcal{N}=2$ quantum field theories and their associated BPS quivers.  For a wide class of theories including super-Yang-Mills theories, Argyres-Douglas models, and theories defined by M5-branes on punctured Riemann surfaces, there exists a quiver which implicitly characterizes the field theory.  We study various aspects of this correspondence including the quiver interpretation of flavor symmetries, gauging, decoupling limits, and field theory dualities.  In general a given quiver describes only a patch of the moduli space of the field theory, and a key role is played by quantum mechanical dualities, encoded by quiver mutations, which relate distinct quivers valid in different patches.  Analyzing the consistency conditions imposed on the spectrum by these dualities results in a powerful and novel \emph{mutation method} for determining the BPS states.  We apply our method to determine the BPS spectrum in a wide class of examples, including the strong coupling spectrum of super-Yang-Mills with an ADE gauge group and fundamental matter, and trinion theories defined by M5-branes on spheres with three punctures.  }

\maketitle

\enlargethispage{\baselineskip}

\setcounter{tocdepth}{3}
\begin{spacing}{1}
\tableofcontents
\end{spacing}

\section{Introduction}
In the study of four-dimensional quantum field theories with extended supersymmetry, one of the most fruitful and enduring ideas has been the analysis of the spectrum of BPS particles.  An understanding of this protected sector of the Hilbert space is often a key ingredient in testing field theory, and stringy dualities and played an important role in the foundational low-energy solution of $\mathcal{N}=2$ gauge theories \cite{SW1,SW2}.  More recently, significant progress has been made both in mathematics and in physics, in understanding the universal rules that govern potential decay processes of BPS particles \cite{Denef00,DM07,GMN08, JS1,KS}, and continuing progress in this subject \cite{DG,GMN09,CV09,DGS,GMN10,CNV,ADJM1,ADJM2,MPS,GMN11,CV11,CDZ,ACCERV1,DZ11,Sen11} suggests that there are yet undiscovered structures lurking in the BPS spectra of field theories.  However in spite of these dramatic developments, there exists no general method for calculating the BPS spectrum of a given field theory.

In this work we study a wide class of field theories where this difficulty is overcome.  These are theories, whose spectra of BPS states can be calculated from the quantum mechanics of an associated BPS quiver.  Such quivers originally arose in string theory constructions of quantum field theories \cite{DM,Dia99,DFR1,DFR2,FM00,Fiol,Denef}.  In that context, there is a natural class of BPS objects, namely D-branes, and a quantum mechanical description of the BPS spectrum is provided by the worldvolume theory of the relevant branes.  This string theory setup provides a simple way of organizing the spectrum into elementary BPS branes and their bound states and explains the non-abelian degrees of freedom needed in the quiver description.  

While the geometric engineering perspective provides a useful source of examples, our focus in this paper is on analyzing the theory of BPS quiver directly from the point of view of quantum field theory.  The class of BPS quiver theories is broad, and includes gauge theories coupled to massive hypermultiplets,  Argyres-Douglas type field theories \cite{AD}, and all theories defined by M5 branes on punctured Riemann surfaces \cite{Witten1997,Gaiotto,Gaiotto:2009gz,BBT,CD1,CV11,ACCERV1,Xie}.\footnote{At least one puncture is required.  The punctureless case provides examples of theories without BPS quivers \cite{CV11}.}  For all of these theories the quiver appears to provide a simple and unique characterization of the theory, and one of the aims of this work is to illustrate in a variety of examples how simple graphical features and operations at the level of quivers translate into physical properties and constructions such as flavor symmetries, gauging, decoupling limits, and dualities.  As a particular highlight, in section 6.2 we study the quiver version of a strong coupling duality \cite{AS} given by the relationship between the theory of $SU(3)$ coupled to six fundamental flavors, and the $E_{6}$ Minahan-Nemeschansky theory \cite{MN} coupled via gauging an $SU(2)\subset E_{6}$ to $SU(2)$ Yang-Mills with an additional fundamental flavor. 

To accomplish our task of exploring BPS quivers, we begin in section \ref{QM} with a detailed description of the way in which quiver quantum mechanics encodes the spectrum of BPS states.   We then develop the theory of quiver representations, the holomorphic description of quiver quantum mechanics and explain how quivers yield a concrete method for studying wall-crossing phenomenon and review basic examples of these techniques.  This material is known and is included for completeness and to provide context for subsequent extensions.

A significant feature of quiver description of the spectrum is that a fixed quiver typically describes the BPS particles only on a small patch of the moduli space of a given theory.  A key role is then played by quantum mechanical dualities, encoded by quiver mutations, which relate distinct quivers valid in different regions of parameter space.  These relationships between a priori distinct quantum mechanics are a one-dimensional version of Seiberg duality \cite{Seiberg}.  Their basic content is that the BPS spectrum can be decomposed into bound states of primitive particles in more than one way by suitable changes of the set of building block BPS states.    In section 3 we discuss these dualities and analyze the constraints that they impose upon the BPS spectrum.  Remarkably we find that these consistency conditions are so powerful that frequently they completely determine the BPS spectrum.  This results in an algorithm, the \emph{mutation method}, for calculating a spectrum that is far simpler than a direct investigation of the quantum mechanics. 

In sections 4-6 we put the general theory to use by computing the BPS spectrum in a broad class of examples.  We focus our attention on two kinds of theories: non-abelian gauge theories with ADE gauge group and fundamental matter, and trinion theories $\mathcal{T}_{n}$ defined by $n$ M5-branes on spheres with three punctures \cite{Gaiotto,GMN09, BBT, ACCERV1}.  For all such theories we determine the quiver and frequently our mutation method is powerful enough to determine the BPS spectrum in a strongly coupled chamber where there are only finitely many BPS states. The spectrum in all chambers can then be deduced by the application of the wall crossing formula of Kontsevich and Soibelman \cite{KS}.  Let $\mathcal{B}$ denote the set of BPS particles at strong coupling, and $|\mathcal{B}|$ the number of such particles.   Then a summary of the gauge theories whose strong coupling spectra we determine is:
\begin{itemize}
\item $SU(N_{c})$ gauge theory coupled to $N_{f}$ fundamentals.
\[|\mathcal{B}|=N_{c}(N_{c}-1)+N_{f}(2N_{c}-1)\]
\item $SO(2N_{c})$ gauge theory coupled to $N_{v}$ vectors. 
\[|\mathcal{B}|=2N_{c}(N_{c}-1)+N_{v}(4N_{c}+1)\]
\item $E_{6}$ gauge theory coupled to $N_{\mathbf{27}}$ fundamental $\mathbf{27}$'s. 
\[|\mathcal{B}|=72+73N_{\mathbf{27}}\]
\item $E_{7}$ gauge theory.
\[|\mathcal{B}|=126\]
\item $E_{8}$ gauge theory.
\[|\mathcal{B}|=240\]
\end{itemize}
 One elegant feature of the above results can be seen in the limit where there is no matter whatsoever so that one is considering the strong coupling BPS spectrum of pure super-Yang-Mills with an arbitrary ADE gauge group.  Then our results can be summarized by noting that the number of BPS particles is given simply by the number of roots in the associated Lie algebra.
 
The second class of examples investigated in section 6 concerns the theories defined by M5 branes on punctured Riemann surfaces.  As studied in detail in \cite{Gaiotto}, such theories can be understood by decomposing the surface into pairs of pants sewn together along tubes.  Physically this decomposition corresponds building up a complicated quantum field theory from a set of building block theories $\mathcal{T}_{n}$ defined by $n$ M5-branes on pairs of pants, by gauging various flavor symmetries.  We compute the BPS quiver associated to these trinion theories $\mathcal{T}_{n}$ in the case of maximal flavor symmetries.  Further, by making use of our understanding of flavor symmetries from the quiver point of view we explain, following \cite{CV11}, in the case of two and three M5 branes how to gauge flavor symmetries in the quivers themselves and therefore how to produce quivers associated to a large class of field theories determined by various Riemann surfaces.  Along the way we determine that the trinion theory $\mathcal{T}_{3}$ given as a mass deformed version of the $E_{6}$ Minahan-Nemeschansky theory has a minimal finite BPS chamber supporting $24$ hypermultiplet states.

\section{BPS Quiver Quantum Mechanics}
\label{QM}
We begin with a four-dimensional $\mathcal{N}=2$ field theory with Coulomb moduli space $\mathcal{U}$.  Here by a point $u\in \mathcal{U}$ we will mean a specification of all supersymmetric parameters in the theory including Coulomb branch moduli, bare masses, and coupling constants.  At a generic value of the moduli $u \in \mathcal{U}$, this field theory has a $U(1)^{r}$ gauge symmetry, and a low energy solution described by:
\begin{itemize}
\item A lattice $\Gamma$ of electric, magnetic, and flavor charges of rank $2r+f$, where $f$ is the rank of the flavor symmetry.
\item A linear function $\mathcal{Z}_{u}: \Gamma \rightarrow \mathbb{C}$, the central charge function of the theory.\footnote{Here we explicitly indicate the $u$ dependence by including a subscript on the central charge function. For notational simplicity, we will eventually drop the subscript and leave the $u$ dependences implicit}  Central charges which couple to the electric and magnetic charges encode the effective coupling and theta angle of the infrared physics, while the central charges that couple to the flavor symmetries sample possible bare masses of matter in the theory.
\end{itemize}

The behavior of the central charge function as one varies the moduli fixes completely the effective action for the neutral massless fields.  However, the description of the massive charged particles is more subtle. According to the $\mathcal{N}=2$ superalgebra, the central charge provides a lower bound on the masses of charged particles.  The mass of a particle with charge $\gamma \in \Gamma$ satisfies
\begin{equation}
M\geq |\mathcal{Z}_{u}(\gamma)|.
\end{equation}
The lightest charged particles are those that saturate the above bound - these are termed BPS. The spectrum of BPS states is a priori undetermined by the low energy solution of the theory alone, and it is precisely this question that we aim to address. We will describe a class of theories where the BPS spectrum can be computed and studied using the technology of quiver quantum mechanics.

\subsection{Quivers and Spectra}
\label{quiverqm}
In this section we lay the foundations for our ideas by describing the connection between quantum mechanical quiver theories and BPS spectra of four-dimensional quantum field theories. In the course of our analysis we will also discover various restrictions on the class of theories to which these quiver techniques apply. We first describe in section \ref{basis} how the BPS spectrum of the 4d theory at a fixed point in moduli space can frequently be used to define an associated quiver, and therefore to pose a supersymmetric quantum mechanics problem. We will then see in section \ref{bpsqm} that the ground states of this supersymmetric quantum mechanics precisely reproduce the BPS spectrum.    From this point of view, the quiver provides merely a clever way of organizing the BPS spectrum.  However, the true power of the technique is that there exist many ways of producing a BPS quiver that \emph{do not} assume a knowledge of the spectrum.   These are briefly surveyed in section \ref{altconst}. It is through these methods that we can hope in turn to discover previously unknown spectra.

\subsubsection{From BPS Spectra to BPS Quivers}
\label{basis}
Let us begin by fixing a point $u \in \mathcal{U}$ in moduli space. Suppose the occupancy of BPS states here is known. We will then explain how to construct a quiver that describes the theory at this point $u$.

To begin we split the BPS spectrum into two sets, the \emph{particles} and the \emph{antiparticles}.  We define particles to be those BPS states whose central charges lie in the upper half of the complex $\mathcal{Z}$ plane, and antiparticles those in the lower.  CPT invariance ensures that for each BPS particle of charge $\gamma$, there is an antiparticle of charge $-\gamma$. Thus the full BPS spectrum consists of the set of BPS particles plus their associated CPT conjugate antiparticles. We will use the occupancy of the particles to construct a quiver.

Among the particles, we choose a minimal basis set of hypermultiplets.  Since the lattice $\Gamma$ has rank $2r+f$, our basis will consist of $2r+f$ BPS hypermultiplets. Let us label their charges $\gamma_{i}$.  The particles in the basis set should be thought of as the elementary building blocks of the entire spectrum of BPS states.  As such they are required to form a positive integral basis for all occupied BPS particles in the lattice $\Gamma$.  This means that every charge $\gamma$ which supports a BPS particle satisfies
\begin{equation}
\gamma =\sum_{i=1}^{2r+f}n_{i}\gamma_{i}. \hspace{.2in}n_{i}\in \mathbb{Z}^{+} \label{nsum}
\end{equation}
We emphasize that the basis need not span $\Gamma$, but only the subset of occupied states in $\Gamma$. We will see in section \ref{bpsqm} that this equation can be interpreted as saying that the BPS particle with charge $\gamma$ can be viewed as a composite object built up from a set of elementary BPS states containing $n_{i}$ particles of charge $\gamma_{i}$.

It is important to notice that the requirement that a set of states form a positive integral basis for the entire spectrum of BPS particles is quite strong, and in particular uniquely fixes a basis when it exists.  To see this, we suppose that $\{\gamma_{i}\}$ and $\{\widetilde{\gamma}_{i}\}$ are two distinct bases.  Then there is a matrix $n_{ij}$ relating them
\begin{equation}
\widetilde{\gamma}_{i}=n_{ij}\gamma_{j}; \hspace{.5in} \gamma_{i}=(n^{-1})_{ij}\widetilde{\gamma_{j}}.
\end{equation}
However since both $\{\gamma_{i}\}$ and $\{\widetilde{\gamma}_{i}\}$ form positive integral bases, the matrix $n_{ij}$ and its inverse must have positive integral entries.  It is easy to see that this forces both matrices to be permutations.  Thus the two bases can differ only by a trivial relabeling.

Now, given the basis of hypermultiplets $\{\gamma_{i}\}$ there is a natural diagram, a \emph{quiver}, which encodes it.  This quiver is constructed as follows:
\begin{itemize}
\item For each element $\gamma_{i}$ in the basis, draw a node of the quiver.
\item For each pair of charges in the basis compute the electric-magnetic inner product $\gamma_{i}\circ \gamma_{j}$.  If $\gamma_{i}\circ \gamma_{j}>0,$ connect corresponding nodes $\gamma_{i}$ and $\gamma_{j}$ with $\gamma_{i}\circ \gamma_{j}$ arrows, each of which points from node $j$ to node $i$.
\end{itemize}

\begin{figure}[here!]
  \centering
  \subfloat[BPS Spectrum]{\label{fig:tri1}\includegraphics[width=0.65\textwidth]{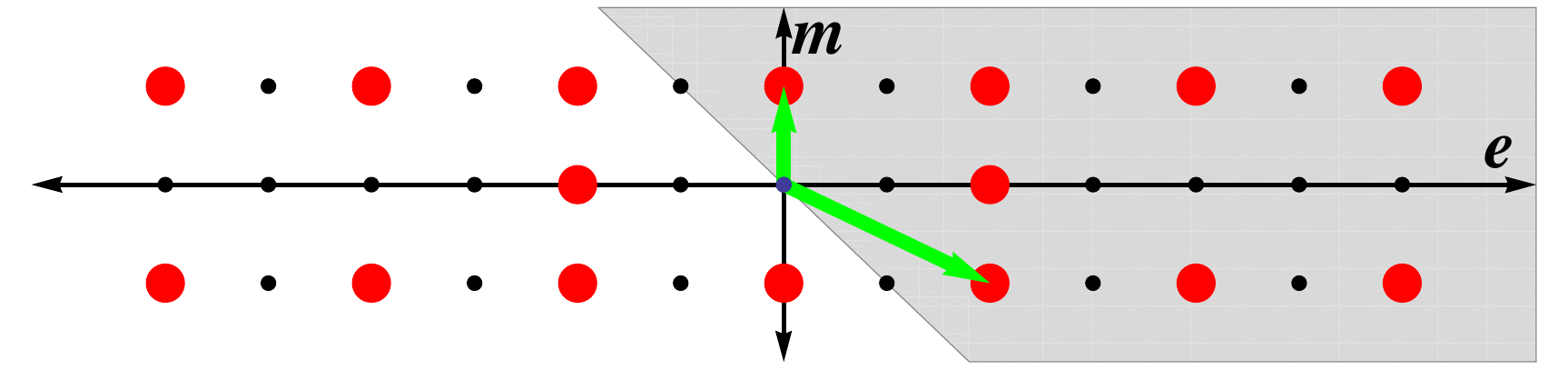}}   
  \hspace{.01in}   
  \subfloat[BPS Quiver]{\label{fig:tri2} \raisebox{0.23in}{
  \begin{xy}
  (0,70)*{\Circle}="a"; (0,65)*{(0,1)};
(30,70)*{\Circle}="c"; (30,65)*{(2,-1)};
  {\ar "a";"c" <1.5pt>}
    {\ar "a";"c" <-1.5pt>}
  \end{xy}
  }
  }
  \caption{The spectrum and BPS quiver of $SU(2)$ Yang-Mills.  In (a) the weak-coupling BPS spectrum, both particles and antiparticles, is plotted in the $(e,m)$ plane.  Red dots denote the lattice sites occupied by BPS states.  The green arrows show the basis of particles given by the monopole and dyon.  We have represented our choice of particle central charge half-plane by the grey region. In (b) the BPS quiver is extracted from this data.  It has one node for each basis vector, and the double arrow encodes the symplectic product.}
  \label{fig:su2}
\end{figure}
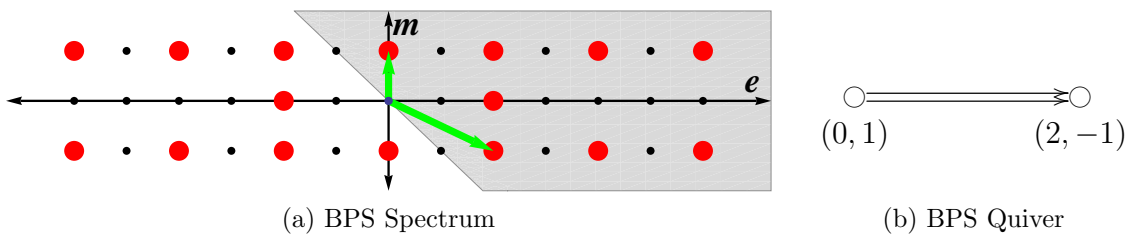

To illustrate this construction, we consider the simple case of pure $SU(2)$ gauge theory at a large value of the Coulomb branch modulus, where the theory is governed by semiclassical physics.  In terms of their associated electric and magnetic charges $(e,m)$, the occupied BPS states consist of:
\begin{equation}\begin{array}{c}
\mathrm{Vector \ multiplet \ } W-\mathrm{boson}: (2,0), \\ \mathrm{Hypermultiplet  \ dyons}: (2n,1), \hspace{.05in} (2n+2,-1) \hspace{.05in} n\geq 0.
\end{array}
\end{equation}
Choosing the particle half-plane represented in Fig. \ref{fig:tri1}, the unique basis is given by the monopole $(0,1)$ and the dyon $(2,-1)$.  The spectrum and the resulting quiver are then shown in Figure \ref{fig:su2}.

So, returning to the general story, we have given a map from BPS spectra to quivers.  At this stage, we pause to point out important subtleties in this procedure.  The first is that our identification of arrow being determined by the Dirac inner product glosses over the possibility of having arrows between nodes which point in opposite directions.  In fact, what the Dirac product truly captures is the net number of arrows.  It is a fortunate feature of all of the field theory examples discussed in this work, with the exception of section 6.2, the electric magnetic inner product accurately determines the arrows in the quiver.  Further analysis of this issue occurs in our discussion of superpotentials in section 3. 

A second important subtlety is that there exist field theories for which there is no BPS quiver whatsoever. To illustrate this, note that one assumption thus far was that we could find a basis of hypermultiplets in the upper half of the central charge plane. By linearity of the central charge function, this gives a constraint on the occupied subset of $\Gamma$. In particular, since the set $\{\gamma_{i}\}$ forms a basis, we have for an arbitrary BPS particle of charge $\gamma$,
\begin{equation}
\gamma=\sum_{i}n_{i}\gamma_{i} \Longrightarrow \mathcal{Z}_{u}(\gamma)=\sum_{i}n_{i}\mathcal{Z}_{u}(\gamma_{i}). \hspace{.2in} n_{i}\geq0 \label{zlin}
\end{equation}
Since $\mathcal{Z}(\gamma_{i})$ all lie in the upper half-plane, \eqref{zlin} implies that the central charges of all BPS particles lie in a cone in the upper half of the central charge plane, bounded by the left-most and right-most $\mathcal{Z}(\gamma_i);$ we denote this the \emph{cone of particles}.

One can see that many theories do not even have such a cone, and therefore don't have an associated BPS quiver. The simplest example is $\mathcal{N}=4$ Yang-Mills with gauge group $SU(2)$.  Because of S-duality, this theory has a spectrum of dyons with charges $(p,q)$, for $p$ and $q$ arbitrary coprime integers. It follows that the phases of the central charges of these dyons form a dense set in the unit circle in the central charge plane.  In particular, there is no cone of particles and hence no quiver. 

We can state the problem with $\mathcal{N}=4$ Yang-Mills from the $\mathcal{N}=2$ perspective: there is an adjoint hypermultiplet which is forced to be massless. The $\mathcal{N}=2^{*}$ theory, where the adjoint is given a mass, \emph{does} admit a BPS quiver, given in section \ref{su2examples}.  This situation is typical of gauge theories that become conformal when all mass deformations are turned off.  A conformal field theory has no single particle states at all, let alone BPS states.  A quiver description is therefore only possible when sufficiently many massive deformations of the theory exist and have been activated.

\subsubsection{Alternative Constructions of BPS Quivers} \label{altconst}
Thus far we have explained how BPS quivers provide a way of describing certain properties of the basis for the BPS spectates at a fixed point in moduli.  In the next section, we explain the reverse construction, that is, how to extract a BPS spectrum from a BPS quiver, and hence how a BPS quiver can be used as a convenient way for encoding the complete BPS spectrum.  However, the most important application of BPS quivers is that they can be used to deduce an \emph{unknown} BPS spectrum.  One reason this is so, is that our construction of BPS quivers is completely local in the Coulomb branch moduli space $\mathcal{U}$.  Given a point $u\in \mathcal{U}$ where the BPS spectrum is known, the quiver description of the spectrum is uniquely fixed if it exists.  But, as will be clear by the conclusion of section 3, once a quiver is determined for a single modulus $u$, the quiver description of the entire moduli space $\mathcal{U}$ is also fixed.  Thus, we may determine the quiver in say a region of weak coupling where the physics is under control, and then use it to calculate the BPS spectrum at strong coupling. 

Even more striking is the fact that BPS quivers can frequently be deduced by alternative geometric methods in various contexts in string theory, even when the BPS spectrum is unknown for any value of the moduli.  The quiver methods described in the following sections can then be used to determine the spectrum from scratch.  

The existing literature on the techniques used to extract BPS quivers is by now very vast, in the following we outline some of the various interrelated approaches:\footnote{See also \cite{Aspinwall09} and references therein for an excellent recent exposition of the mathematical structures used to describe to D-branes which in includes in particular the associated quiver representation theory.}

\begin{itemize}

\item Building on the original orbifold construction of quiver gauge theories of \cite{DM} refs. \cite{Dia97,Dia99,DFR1,DFR2}  provided the identification of the quiver nodes with a basis of BPS states obtained from fractional branes, these BPS quivers were further explored in \cite{FM00,Fiol}. 

\item The relation of the $4d$ quivers with the soliton spectrum in $2d$ \cite{CV92} was studied in various places, see for example \cite{HIV00,CFIKV01,FHHI02}, more recently this $2d/4d$ correspondence and the associated construction of BPS quivers was discussed in \cite{CNV}.

\item The toric methods of \cite{FHH00,Feng:2001xr} and the relation to dimer models \cite{HK05} were used in \cite{FHKVW05} to construct a large class of quivers, their construction using mirror symmetry was studied in \cite{Feng:2005gw}.

\item  Based on the geometric study of BPS states in SW theories pioneered in \cite{KLMVW} and further studied in \cite{SV,GMN09}, the BPS quivers can be obtained from triangulations of Riemann surfaces as described in \cite{CV11,ACCERV1} using the relation of triangulations and quivers of \cite{FST}.  Given a pair of M5-branes wrapping a Riemann surface $\mathcal{C}$, an ideal triangulation of $\mathcal{C}$ can be used to determine the BPS quiver.  There is reason to believe that these techniques can be generalized to larger numbers of M5-branes and we initiate this analysis in section 6.
\end{itemize}

\subsubsection{Quiver Quantum Mechanics: From BPS Quivers to BPS Spectra}
\label{bpsqm}

We now return to our general discussion of BPS quivers and explain how to deduce the full spectrum from the quiver.  Thus far the BPS quiver we have introduced is merely a way of encoding a basis of BPS states $\{\gamma_{i}\}$ for a given $\mathcal{N}=2$ theory. To construct a general BPS state, we must know, for a given charge
\begin{equation}
\gamma=\sum_{i}n_{i}\gamma_{i}
\end{equation}
whether any particles of this charge exist in the theory, and if so, determine their degeneracy and spins. We attack this question by viewing the hypothetical state with charge $\gamma$ as a quantum mechanical bound state of $n_{i}$ copies of each basis particle $\gamma_{i}$.  Since we seek a BPS particle, we introduce a four supercharge quantum mechanics problem and look for its supersymmetric ground states.  The precise quantum mechanics theory is constructed from the BPS quiver and the charge $\gamma$ in the following way: Let $i$ index nodes of the quiver, and $a$ index the arrows of the quiver. Then we introduce a gauge group for each node and bifundamental field $B^{a}_{ij}$ for each arrow pointing $i\rightarrow j,$
\begin{equation}
\mathrm{Gauge \  Group}= \prod_{\mathrm{nodes }\,i}U(n_{i}), \hspace{.5in} \mathrm{Matter}=\bigoplus_{\mathrm{arrows }\,a}B^{a}_{ij}.
\end{equation}
Thus, the BPS quiver, whose nodes and arrows were originally merely a presentation of a basis of hypermultiplets, now encodes the gauge groups and bifundamental matter of a quiver quantum mechanics.  

This prescription can be motivated most easily when the four-dimensional field theory is engineered in string theory.  In such a situation, BPS states are viewed as various supersymmetric bound states of D-branes.  Then the nodes of our quiver correspond to a collection of basic supersymmetric branes and the arrows are bifundamental fields that arise at brane intersections.  This also provides an elementary understanding of the appearance of \emph{non-abelian} gauge fields in the quantum mechanics: they are the usual non-abelian degrees of freedom that arise when branes coincide.  The quantum mechanics problem introduced above is then nothing but the worldvolume theory of a system of D-branes dimensionally reduced to 0+1 dimensions. 

Returning to our general analysis, to asses the existence of a BPS particle with charge $\gamma$, we look for supersymmetric ground states on the Higgs branch of this quiver theory.  These depend on two data which we must still specify:
\begin{itemize}
\item Fayet-Iliopoulos Terms

Since the gauge groups at each node are given by $U(n_{i})$, the overall $U(1)$ at each node can couple to an independent FI-term $\theta_{i}$.  These parameters are fixed by the central charges $\mathcal{Z}_{u}(\gamma_{i})$ of the constituent particles.  We state this identification in the case that all the central charges point in nearly the same direction in the complex plane.  Then let $\mathcal{Z}_{u}(\gamma)$ denote the central charge of a state with charge $\gamma$, and set 
\begin{equation}
\theta_{i}=|\mathcal{Z}_{u}(\gamma_{i})|\left(\phantom{\int}\hspace{-.2in}\arg(\mathcal{Z}_{u}(\gamma_{i}))-\arg(\mathcal{Z}_{u}(\gamma))\right).
\end{equation} 
For each node $i$ in the quiver there is then a D-term equation of motion
\begin{equation}
\sum_{\substack{\mathrm{arrows} \\ \mathrm{starting \ at \ }i}} \hspace{-.1in}|B^{a}_{ij}|^{2}-\sum_{\substack{\mathrm{arrows} \\ \mathrm{ending \ at \ }i}} \hspace{-.1in}|B^{a}_{ki}|^{2}=\theta_{i}. \label{dterm}
\end{equation}
When the central charges are not nearly aligned, the identification of the FI parameters is more involved, and for now the reader should assume that the moduli are such that this approximation is valid.\footnote{Alternatively one may tune the central charges to near alignment.  Since this involves no crossing of walls of marginal stability the spectrum is stable under this motion.} Later in section \ref{rep} we will see an elegant way of rephrasing our problem that completely avoids this issue.
\item Superpotentials

Whenever there are non-trivial oriented cycles in the BPS quiver, the quantum mechanics theory admits a non-trivial gauge invariant superpotential $\mathcal{W}$ which is a holomorphic function of the bifundamental fields.  Our procedure for producing a quiver does not fix a superpotential; it is an independent datum of our construction which must be computed by alternative means.   Later in section \ref{MUT} we will see general constraints on $\mathcal{W}$.  For now, we simply assume that $\mathcal{W}$ is given.  This superpotential yields F-term equations of motion
\begin{equation}
\frac{\partial \mathcal{W}}{\partial B^{a}_{ij}}=0.
\end{equation}
\end{itemize}

Having fully fixed the quantum mechanics, we now turn to the moduli space of supersymmetric ground states with charge $\gamma$,  $\mathcal{M}_{\gamma}$.\footnote{From now on, whenever we refer to supersymmetric ground states of the quiver quantum mechanics, we will always mean on the Higgs branch.  The Coulomb branch can also be studied and gives rise to equivalent results for BPS spectra. \cite{Denef}}  This space is simply the solution to the equations of motion described above, quotiented by the action of the unitary gauge groups.
\begin{equation}
\mathcal{M}_{\gamma}=\left\{B^{a}_{ij}\left | \frac{\partial \mathcal{W}}{\partial B^{a}_{ij}}=0, \sum_{\substack{\mathrm{arrows} \\ \mathrm{starting \ at \ }i}} \hspace{-.1in}|B^{a}_{ij}|^{2}-\sum_{\substack{\mathrm{arrows} \\ \mathrm{ending \ at \ }i}} \hspace{-.1in}|B^{a}_{ki}|^{2}=\theta_{i}\right.\right \}/\prod_{i}U(n_{i}). \label{mg}
\end{equation}
If $\mathcal{M}_{\gamma}$ is non-empty, then there exists a BPS particle in the spectrum with charge $\gamma$.  To determine spins and degeneracy from $\mathcal{M}_\gamma$, we examine the structure of its cohomology.  Specifically, since $\mathcal{M}_{\gamma}$ is the moduli space of a theory with four supercharges, it is a K\"{a}hler manifold, and as such its cohomology automatically forms representations of Lefschetz $SU(2)$.  For each such irreducible Lefschetz $SU(2)$ representation, we obtain a supersymmetric BPS multiplet.  The spacetime spin of a multiplet is then determined by tensoring the Lefschetz spin with an overall $\mathcal{N}=2$ hypermultiplet,
\begin{equation}
\mathrm{Spin}=\mathrm{Lefschetz}\otimes \left(\left[\frac{1}{2}\right]+2\left[0\right] \label{lef}\right).
\end{equation}

Equation \eqref{lef} can be intuitively understood by thinking about the worldvolume theory of a BPS particle.  This worldvolume theory supports four supercharges and hence has an R-symmetry group of $SU(2)$ which is none other than the Lefschetz $SU(2)$ of the moduli space.  On the other hand, the R-symmetry group of a brane, in this case our particle, can be identified with the group of rotations transverse to the worldvolume, which in turn controls the angular momentum of the state.  Thus the Lefschetz $SU(2)$ computes the orbital angular momentum of the state, and the overall shift by 1/2 in \eqref{lef} simply takes into account the intrinsic spin contribution.  

In practice the most important application of \eqref{lef} is to distinguish vector multiplets from hypermultiplets.  The latter are associated to Lefschetz multiplets of length zero, as would naturally occur if, say, $\mathcal{M}_{\gamma}$ were a point.  Meanwhile vector multiplets are associated to Lefschetz multiplets of length two, the canonical example of which is $\mathcal{M}_{\gamma}\cong \mathbb{P}^{1}$.  In complete generality the formula \eqref{lef} tells us that if $\mathcal{M}_{\gamma}$ has complex dimension $d$ then there is guaranteed to be a BPS multiplet of spin $\frac{d+1}{2}$ with charge $\gamma$ in the spectrum. Naive parameter counting gives the expected dimension of the $\mathcal{M}_\gamma$ as 
\begin{equation}
d=\sum_{B_{ij}^a} (n_i n_j)-\sum_{\mathrm{nodes}\,i}n_i^2-(\# \, \text{F-term constraints})+1.
\end{equation}
Here we have simply counted the degrees of freedom of the bifundamental fields, $B_{ij}^a,$ and subtracted the gauge degrees of freedom and the F-term constraints. The addition of 1 is for the overall diagonal gauge group $U(1)_d\subset \prod_i U(1)\subset \prod_i U(n_i).$ Since all fields are bifundamental, no field is charged under the simultaneous $U(1)$ rotation of all gauge groups, so this gauge degree of freedom is actually redundant. 

In summary, given a quiver we have defined a supersymmetric quantum mechanics problem, and the cohomology of the moduli spaces of grounds states of this quantum mechanics determines the occupancy of BPS states.

\subsection{Quiver Representations}
\label{rep}
While our supersymmetric quantum mechanics construction determines the BPS spectra as specified by a quiver, it is useful in practice to work in the language of quiver representation theory. Here the problem of determining the ground states of the supersymmetric quantum mechanics gets recast in a holomorphic framework. Our ability to rephrase the problem in terms of quiver representation theory arises from the fact that a supersymmetric moduli space of a theory with four supercharges, such as $\mathcal{M}_{\gamma}$, can be presented in two ways:
\begin{itemize}
\item As the solution to the F-term and D-term equations of motion modulo the action of the unitary gauge groups (this is what has been stated in \eqref{mg}).
\item As the solution to the F-term equations modulo the action of the complexified gauge group $\prod_{i}Gl(n_{i},\mathbb{C})$, augmented by a stability condition.
\end{itemize}
It is the second notion of $\mathcal{M}_\gamma$ that makes use of quiver representation theory.

To begin, we note that in a zero energy field configuration of supersymmetric quantum mechanics, the bifundamental fields are constants and hence their expectation values can be viewed as linear maps between vector spaces $\mathbb{C}^{n_i}$ associated to each node. These expectation values are constrained by the condition that they must solve the F-term equations of motion $\partial{\mathcal{W}}/\partial B^{a}_{ij}=0$.  A quiver representation is by definition precisely a choice of complex vector spaces $\mathbb{C}^{n_{i}}$ for each node, and linear maps $B^{a}_{ij}: \mathbb{C}^{n_{i}}\longrightarrow \mathbb{C}^{n_{j}}$ for each arrow in a quiver subject to the F-term equations.  So the data of a classical zero energy field configuration completely specifies a quiver representation (See \cite{Aspinwall09} and references therein).

Given a quiver representation $R$, defined by vector spaces $\mathbb{C}^{n_{i}}$ and maps $B^{a}_{ij}$ an important notion in the following will be the subrepresentations $S\subset R$.  A subrepresentation $S$  is defined by a choice of vector subspaces $\mathbb{C}^{m_{i}}\subset \mathbb{C}^{n_{i}}$ for each node and maps $b^{a}_{ij}: \mathbb{C}^{m_{i}}\longrightarrow \mathbb{C}^{m_{j}}$ for each arrow, such that all diagrams of the following form commute:
\begin{equation}
\begin{split}\xymatrix{
\mathbb{C}^{n_{i}}  \ar[r]^{B^{a}_{ij}} & \mathbb{C}^{n_{j}} \\
\mathbb{C}^{m_{i}}\ar[r]^{b^{a}_{ij}} \ar[u]& \mathbb{C}^{m_{j}} \ar[u]}\end{split}
\end{equation}

To complete our holomorphic description of the moduli space we must still specify a stability condition that ensures that a given quiver representation $R$ is related to a solution of the D-term equations in quiver quantum mechanics.  To motivate this, note that a quiver rep $R$ with vector spaces $\mathbb{C}^{n_i}$ is related to the description of a particle with charge $\gamma_R = \sum n_i \gamma_i$. Then heuristically, a subrepresentation $S$ of $R$ can be thought of as a bound state of smaller charge which may, in principle, form one of the constituents of a decay of a particle of charge $\gamma_R$.  To prohibit such a decay, we must restrict our attention to \emph{stable} quiver representations.  To define this notion of stability we let $\mathcal{Z}_{u}(R)$ denote the central charge of a representation,\footnote{When we speak of the central charge of a representation, we are always referring to the central charge of the bound state associated to that representation.}
\begin{equation}
\mathcal{Z}_{u}(R)\equiv \mathcal{Z}_u(\gamma_R)=\sum_{i}n_{i}\mathcal{Z}_{u}(\gamma_{i}).
\end{equation}
By construction the central charge vector lies in the cone of particles in the upper half of the central charge plane.  Then $R$ is called stable if for all subrepresentations $S$ other than $R$ and zero, one has
\begin{equation}
\arg(\mathcal{Z}_{u}(S))<\arg(\mathcal{Z}_{u}(R)). \label{stable}
\end{equation}
We will refer to any subrepresentation $S$ that violates this condition as a destabilizing subrepresentation. This condition is denoted $\Pi$-stability, and was studied in \cite{DFR1}.  We take this to be the requisite notion of stability at general points in moduli space.  One important consistency check on this choice is that when all the central charges are nearly aligned, the stability condition \eqref{stable} reduces to the D-term equations of motion presented earlier \cite{King,DFR1}. 

Given this notion of stability, we can now formulate the moduli space $\mathcal{M}_{\gamma}$ as set of stable quiver representations modulo the action of the complexified gauge group.
\begin{equation}
\mathcal{M}_{\gamma}=\left \{\phantom{\int}\hspace{-.2in}R =\{B^{a}_{ij}:\mathbb{C}^{n_{i}}\rightarrow \mathbb{C}^{n_{j}} \} \left |\phantom{\int} \hspace{-.15in} \frac{\partial \mathcal{W}}{\partial B^{a}_{ij}}=0, \  R \ \mathrm{is}\ \Pi-\mathrm{stable} \right. \right\}/\prod_{i}Gl(n_{i},\mathbb{C}).
\end{equation}
This is a completely holomorphic description of $\mathcal{M}_{\gamma}$, and in many examples is explicitly computable.

As a very elementary application, we note that the nodes of a quiver are always $\Pi$-stable reps. That is, consider $\gamma_j$ as the representation given by choosing $n_i=\delta_{ij}$. This is always stable since it has no non-trivial subrepresentations, and thus in particular no destabilizing subreps. Furthermore, since there is only one non-zero vector space, all maps must be chosen zero; thus the moduli space $\mathcal{M}_{\gamma_j}$ is given by a single point. We find that each node of a quiver gives a multiplicity one hypermultiplet BPS state.
\subsection{Walls of Marginal Stability and Examples of Quiver Representations} 
\label{wc}
The preceding discussion in this section has focused exclusively on utilizing BPS quivers to encode the spectrum of an $\mathcal{N}=2$ quantum field theory at a specific point $u$ on the Coulomb branch $\mathcal{U}$.  BPS states are stable under infinitesimal variations of the modulus, and thus our description can be viewed as local theory of BPS particles adequate on a patch in $\mathcal{U}$.  Of course we are interested in determining the spectrum across the entire moduli space, and this can also be achieved using the quiver.  

In the quiver representation theory problem, the moduli $u$ along with bare mass parameters and coupling constants enter the calculation through the central charge function $\mathcal{Z}_{u}$.  From the perspective of quiver representation theory, these are changes in the stability conditions.  For small deformations of the stability condition, the set of stable representations, and hence the BPS spectrum, is unchanged.  However at certain real codimension one loci in moduli space we encounter walls of marginal stability where a supersymmetric particle decays.    At the wall, the central charges of some representation $R$ and its subrep $S$ become aligned. On one side of the wall, $\arg\mathcal{Z}(S)<\arg\mathcal{Z}(R)$ so that $R$ stable, and hence some corresponding BPS particle exists. On the other side of the wall, the phases have crossed, and the stability condition has changed.  We will have  $\arg\mathcal{Z}(S)>\arg\mathcal{Z}(R),$ so the representation $R$ is no longer stable, and the associated particle has disappeared from the BPS spectrum.

It is a virtue of the description of the spectrum in terms of stable quiver representations that these wall-crossing processes are completely explicit. Indeed the BPS quiver gives us a way to calculate directly the BPS spectrum on either side of a wall.  One can then simply compare the answer on both sides, and see that properties such as the Kontsevich-Soibelman wall-crossing formula hold.  In this section we study these wall crossing phenomena in the context of the Argyres-Douglas conformal theories.

\subsubsection{$A_{2}$ Theory}
\label{twonodeex}
Let's begin with a simplest possible example which demonstrates wall-crossing. We will consider the Argyres-Douglas $A_2$ theory, whose quiver is given by two nodes connected by a single arrow \cite{CNV}.  We will denote by $\mathcal{Z}_{i}$ the central charges of the two basis particles,
\begin{equation}
\xy 
(-20,0)*+{1}*\cir<10pt>{}="a" ; (20,0)*+{2}*\cir<10pt>{}="b" 
\ar @{->} "a"; "b" 
 \endxy
 \end{equation}

No matter what the value of the central charges, the basis particles described by the nodes of the quiver are stable.  Thus the spectrum always contains at least two hypermultiplets.  Now let us search for a bound state involving $n_{1}$ particles of type $\gamma_{1}$ and $n_{2}$ particles of type $\gamma_{2}$.  According to the general theory developed in the previous sections we are to study a quiver representation of the following form
\begin{equation}
\begin{split}\xymatrix{
\mathbb{C}^{n_{1}}  \ar[r]^{B} & \mathbb{C}^{n_{2}} }\end{split}
\end{equation}
To determine stability we investigate subrepresentations.  Let's start with a subrepresentation of the form
\begin{equation}\begin{split}
\xymatrix{
\mathbb{C}^{n_{1}}  \ar[r]^{B} & \mathbb{C}^{n_{2}} \\
0\ar[r]^{0} \ar[u]& \mathbb{C}^{} \ar[u]}\end{split}
\end{equation}
There is no condition on the field $B$ for this diagram to commute; it is always a subrepresentation.  Thus, stability of our bound state requires
\begin{equation}
\arg(\mathcal{Z}_{2})<\arg(n_{1}\mathcal{Z}_{1}+n_{2}\mathcal{Z}_{2}) \Longrightarrow \arg(\mathcal{Z}_{2})<\arg(\mathcal{Z}_{1}). \label{2decay}
\end{equation}

Next we consider a similar decay involving the first basis particle
\begin{equation}
\begin{split}\xymatrix{
\mathbb{C}^{n_{1}}  \ar[r]^{B} & \mathbb{C}^{n_{2}} \\
\mathbb{C}\ar[r]^{0} \ar[u]& 0 \ar[u]}\end{split} \label{1decay}
\end{equation}
If this is a subrepresentation, then stability demands that $\arg(\mathcal{Z}_{1})<\arg(\mathcal{Z}_{2})$, so \eqref{2decay} cannot be satisfied.   Thus, to ensure the existence of a bound state we must forbid this subrepresentation, and hence we must choose $B$ so that the diagram in (\ref{1decay}) does not commute.  Thus $B$ should have no kernel, and in particular, we have $n_{1}\leq n_{2}$.

Finally we consider a decay involving the subrepresentation
\begin{equation}\begin{split}
\xymatrix{
\mathbb{C}^{n_{1}}  \ar[r]^{B} & \mathbb{C}^{n_{2}} \\
\mathbb{C}\ar[r]^{b} \ar[u]& \mathbb{C} \ar[u]}\end{split} \label{11decay}
\end{equation}
It is clear that $b$ can be chosen in such a way that this is always a subrepresentation.  Then stability demands that the central charges satisfy
\begin{equation}
\arg(\mathcal{Z}_{1}+\mathcal{Z}_{2})<\arg(n_{1}\mathcal{Z}_{1}+n_{2}\mathcal{Z}_{2}).
\end{equation}
However, given that $n_{1}\leq n_{2}$, and that $\mathcal{Z}_{2}$ has smaller phase than $\mathcal{Z}_{1}$, it is not possible to satisfy the above inequality.  It follows that the only possibility for a bound state is that \eqref{11decay} is not a subrepresentation, but an isomorphism of representations. So we only have the possibility of non-trivial moduli spaces for $n_1 = n_2 = 1$.

\begin{figure}[here!]
  \centering
  \frame{
  \subfloat[Chamber 1]{\label{a21}\includegraphics[width=0.45\textwidth]{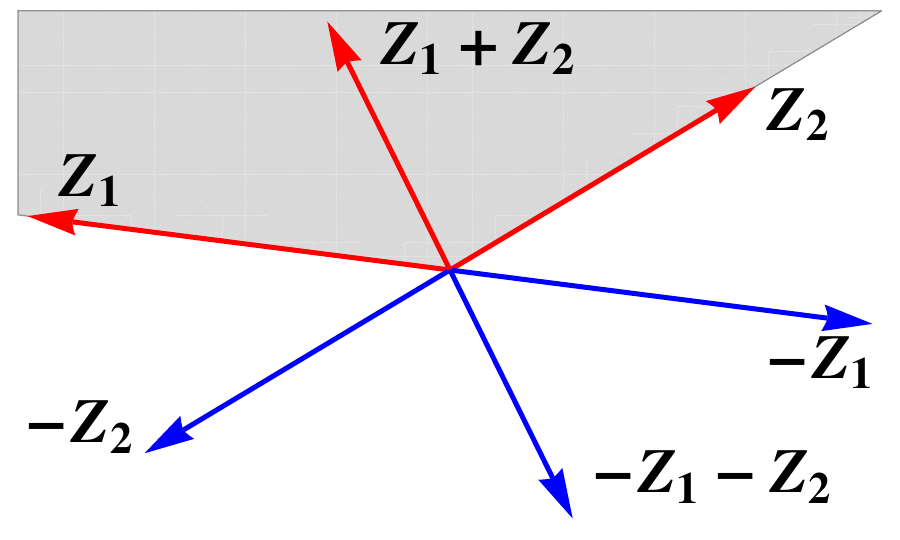}}}     
  \hspace{.01in}      
  \frame{     
  \subfloat[Chamber 2]{\label{a22}\includegraphics[width=0.45\textwidth]{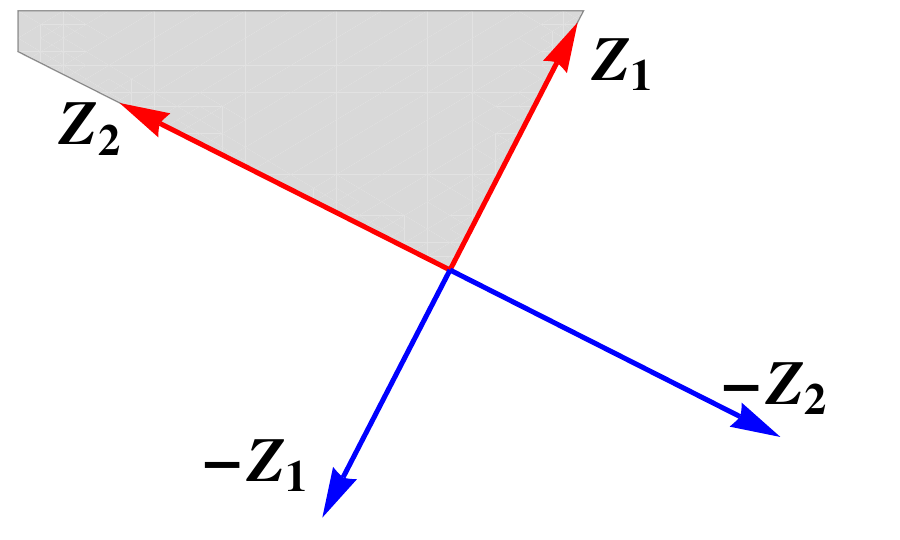}}}
  \caption{The chambers of the $A_{2}$ Argyres-Douglas theory.  The BPS spectrum is plotted in the central charge plane.  Particles are shown in red, antiparticles in blue.  The cone of particles is the shaded grey region.  In (a) the particles form a bound state.  In (b) the bound state is unstable and decays.}
  \label{fig:a2}
\end{figure}

In summary, when $\arg(\mathcal{Z}_{2})<\arg(\mathcal{Z}_{1})$ this theory supports a bound state with charge $\gamma_{1}+\gamma_{2}$.  The moduli space of representations of this charge is given by the quotient of a single non-zero complex number $B$ modulo the action of the complexified gauge group.  Clearly this moduli space is just a point, and so this representation describes a single hypermultiplet.  The complete spectrum for this example is depicted in Figure \ref{fig:a2}, and agrees with the known result for this theory \cite{SV}.
This basic 2-3 decay process is known in various contexts as a primitive decay \cite{DM07}.  In formalism of Kontevich and Soibelman this wall-crossing gives rise to the pentagon identity of quantum dilograthims.
\subsubsection{$A_{3}$ Theory}
\label{threenodeex}
As another example of quiver representation theory and wall-crossing we consider a quiver involving a non-trivial superpotential $\mathcal{W}$.  The quiver, known to be related to the $A_{3}$ Argyres-Douglas theory is given by

\begin{equation}
\xy 
(-20,0)*+{1}*\cir<10pt>{}="a" ; (20,0)*+{2}*\cir<10pt>{}="b" ;
(0,20) *+{3}*\cir<10pt>{}="c";
(0,-3) *{\alpha_1};
(13,12) *{\alpha_2};
(-13,12)*{\alpha_3};
\ar @{->} "b"; "a" ;
\ar @{->} "a"; "c" ;
\ar @{->} "c"; "b" ;
 \endxy
\end{equation}

We let $\alpha_{i}$ indicate the bifundamental field map exiting node $i$ and $\mathcal{Z}_{i}$ the central charge of node $i$.  The quiver is equipped with a superpotential 
\begin{equation}
\mathcal{W}=\alpha_{3}\alpha_{2}\alpha_{1}.
\end{equation}
Minimization of $\mathcal{W}$ implies that in any allowed field configuration all compositions of pairs of maps vanish
\begin{equation}
\alpha_{2}\circ \alpha_{1}=0, \hspace{.5in} \alpha_{3}\circ \alpha_{2}=0, \hspace{.5in} \alpha_{1}\circ \alpha_{3}=0. \label{falphas}
\end{equation}
We will show that this quiver has, up to relabeling the nodes, exactly two chambers with four or five BPS hypermultiplets respectively.   
 
 First, we note that as usual all of the node representation where the dimensions $n_{i}$ of the associated vector space are given by $n_{i}=\delta_{ij}$ for $j=1,2,3$ are stable and hence yield three hypermultiplets.  Further, when one of the $n_{i}$ vanishes, then two of the maps $\alpha$ must also vanish and the analysis reduces to the $A_{2}$ case considered in the previous section.  This yields two or one bound states depending on whether the phases of the $\mathcal{Z}_{i}$ are or are not cyclically ordered.  To conclude the analysis of this quiver, we now wish to illustrate that there are no further bound states that arise from representations
\begin{equation}
\xymatrix{
\mathbb{C}^{n_{1}} \ar[r]^{\alpha_{1}} & \mathbb{C}^{n_{2}} \ar[r]^{\alpha_{2}}  & \mathbb{C}^{n_{3}} \ar@/_2pc/[ll]^{\alpha_{3}} }\end{equation} 
with all $n_{i}$ non-zero.  

We begin by considering possible subrepresentations corresponding to node vectors, $(1,0,0)$, $(0,1,0)$, and $(0,0,1)$.  These are only subrepresentations when $\alpha_{i}$ has a kernel for $i=1,2,3$ respectively.  Clearly not all of these can be subreps simultaneously or else the representation would already be destabilized.  It follows that at least one of the $\alpha_{i}$, say $\alpha_{1}$ is injective and hence in particular $n_{1}\leq n_{2}$.

Now we apply the F-term equations \eqref{falphas}.  From the fact that $\alpha_{1}\circ \alpha_{3}=\alpha_{2}\circ \alpha_{1}=0$ and the fact that $\alpha_{1}$ is injective we learn that both $\alpha_{2}$ and $\alpha_{3}$ have non-vanishing kernels.  This means that both the node representations $(0,1,0)$ and $(0,0,1)$ are subreps so we deduce that $\mathcal{Z}_{1}$ must have largest phase for stability, and $\arg\mathcal{Z}_2,\arg\mathcal{Z}_3<\arg(n_{1}\mathcal{Z}_{1}+n_{2}\mathcal{Z}_{2}+n_{3}\mathcal{Z}_{3}).$

However now we consider a subrepresentation with dimension vector $(1,1,0)$.
\begin{equation}
\xymatrix{
\mathbb{C}^{n_{1}} \ar[r]^{\alpha_{1}} & \mathbb{C}^{n_{2}} \ar[r]^{\alpha_{2}}  & \mathbb{C}^{n_{3}} \ar@/_2pc/[ll]^{\alpha_{3}} \\
\mathbb{C} \ar[r]^{\beta_{1}} \ar[u]^{i}& \mathbb{C} \ar[u]^{j} \ar[r]^{\beta_{2}}  & 0 \ar[u]^{0} \ar@/^2pc/[ll]^{0}
}
\label{110sub}
\end{equation} 
This is a subrep exactly when the image of $\alpha_{1}$ meets the kernel of $\alpha_{2}$ non-trivially, which it does by the F-terms.  Thus we learn that 
\begin{equation}
\arg(\mathcal{Z}_{1}+\mathcal{Z}_{2})<\arg(n_{1}\mathcal{Z}_{1}+n_{2}\mathcal{Z}_{2}+n_{3}\mathcal{Z}_{3}).
\end{equation}
Given the conditions on the $\mathcal{Z}_{i}$ and the fact that $n_{1}\leq n_{2},$ the above is impossible.  

Thus we have arrived at a contradiction.  It follows that for this quiver with the given superpotential there are no states with all $n_{i}$ non-vanishing.  Note that this conclusion is altered when the superpotnetial is turned off.  In that case it is easy to check that the representation $(1,1,1)$ with all maps non-zero provides a stable hypermultiplet at all moduli. This completes our analysis of this quiver.
\section{Quiver Mutation and Duality}
\label{MUT}
We have seen how wall crossing is encoded into our quiver quantum mechanics picture. Walls of marginal stability correspond to hypersurfaces in which two central charges become aligned. The stability condition will differ on the two sides of this wall, and therefore there may be some representations which are stable on one side but not the other. There is in fact another type of hypersurface in moduli space that is strikingly relevant in our picture: hypersurfaces across which a fixed quiver quantum mechanics description of the BPS spectrum may break down entirely. Following \cite{KS} we will refer to these as walls of the second kind.

The situation is less dire than it may seem; we will be able to find another quiver description, valid on the other side of the wall. We will argue that the transformation of a quiver across a wall of the second kind is given by a canonical procedure, known as \emph{quiver mutation} which describes a quantum mechanical duality relating the ground state spectra of two distinct quivers.  Once the rule for transforming quivers at such walls is understood, we will be able to start with a quiver description at any point in moduli space and arrive at any other point by following an arbitrary path connecting them, doing the necessary mutations along the way.  Further, in section \ref{methodsection} we will revisit this procedure and see that the same transformation can be made on quivers at a fixed point in moduli space, and in this case the transformation will take us between quivers that describe the same physics. We will then immediately exploit this duality to circumvent the computations involved in solving the representation theory problem.

Recalling that the nodes of a quiver all correspond to particles, and must therefore have central charges which lie in the upper half-plane, we see what can go wrong. As we tune moduli, our central charge function changes, and as we cross some real co-dimension 1 subspace in $\mathcal{U},$ the central charge of one of the nodes may exit the half-plane.  This behavior defines the walls of the second kind.  They are the loci in moduli space (including as usual masses and couplings) where the central charge of a basis particle becomes real
\begin{equation}
\mathcal{Z}_{u}(\gamma_{i})\in \mathbb{R}.
\end{equation}

\begin{figure}[here!]
  \centering
  \frame{
  \subfloat[Spectrum pre-duality]{\label{fig:cone1}\includegraphics[width=0.4\textwidth]{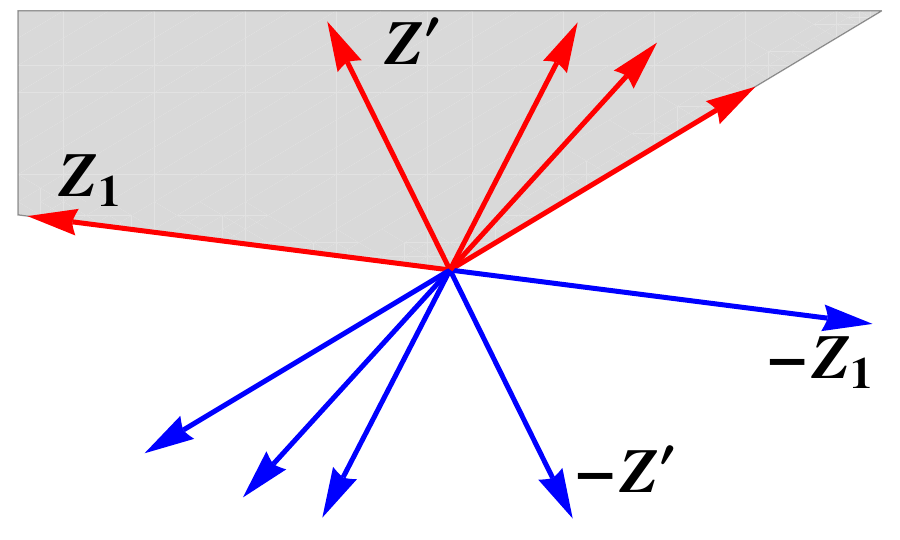}}}     
  \hspace{.01in}      
  \frame{     
  \subfloat[Spectrum post-duality]{\label{fig:cone2}\includegraphics[width=0.4\textwidth]{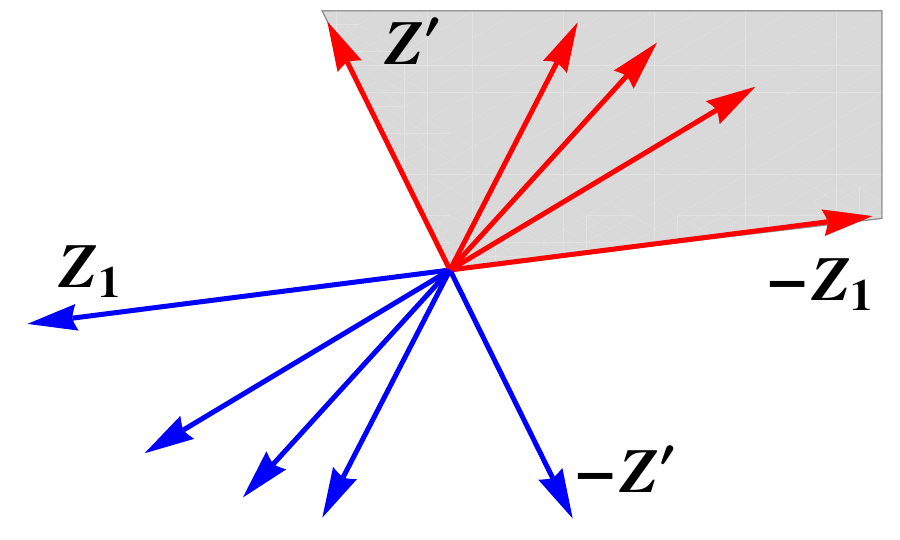}}}
  \caption{A discontinuity in the quiver description results in a quantum mechanical duality described by quiver mutation.  In both diagrams the BPS spectrum is plotted in the central charge plane.  Red lines denote particles while blue lines denote antiparticles.  The gray shaded region indicates the cone of particles.  In passing from (a) to (b) the particle with central charge $\mathcal{Z}_{1}$ changes its identity to an antiparticle.  The cone of particles jumps discontinuously and a new quiver description is required.}
  \label{fig:cones}
\end{figure}

Let us study the process of crossing a wall of the second kind in more detail. Consider the central charge configuration illustrated in Figure \ref{fig:cone1} where the BPS particles are described by the quiver $Q$.  As moduli are varied, the central charge of one of the basis elements, $\mathcal{Z}_1$ rotates out of the upper half-plane and we arrive at the new configuration illustrated in Figure \ref{fig:cone2}.

The first thing to notice about this process is that, since no central charges align, no walls of marginal stability are crossed, and hence the total BPS spectrum (consisting of both particles and antiparticles) is the same in Figures \ref{fig:cone1} and \ref{fig:cone2}. On the other hand, from the point of the quiver this process is discontinuous.  After $\mathcal{Z}_{1}$ has rotated out of the upper half of the central charge plane, it has changed its identity from a particle to an antiparticle.  Then the original basis of particles encoded by the quiver $Q$ is no longer an acceptable basis.  Specifically, in passing from Figure \ref{fig:cone1} to Figure \ref{fig:cone2}, the cone of particles has jumped discontinuously and as a result the original quiver description of the BPS spectrum is no longer valid.

\begin{figure}[here!]
  \centering
  \frame{
 \includegraphics[width=0.4\textwidth]{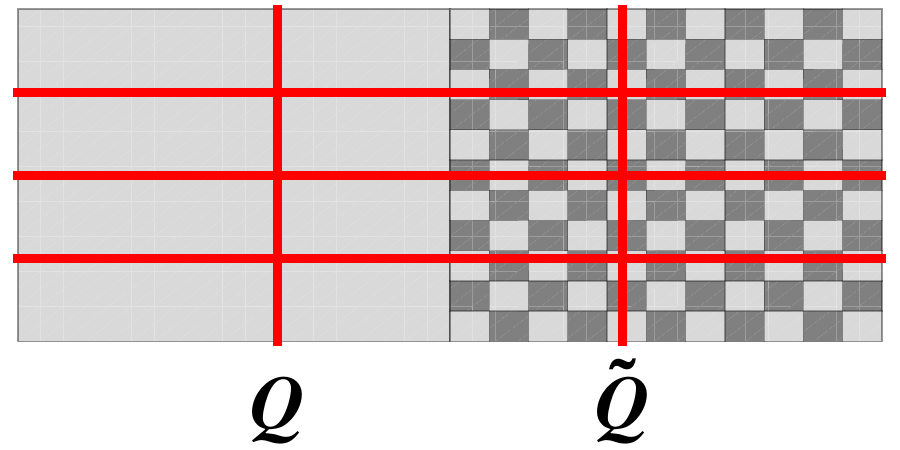}}     
  \caption{A cartoon of the moduli space and its relation to various BPS quiver descriptions.  The red lines denote walls of marginal stability where the BPS spectrum jumps.  The gray shaded region is the domain in moduli space where $Q$ describes the BPS spectrum.  The gray checkered region is the domain where $\widetilde{Q}$ describes the spectrum.  The two descriptions are glued together smoothly away from the walls of marginal stability.  Their interface is a wall of the second kind.}
  \label{fig:quiversew}
\end{figure}

To remedy this deficiency we must introduce a new quiver $\widetilde{Q}$ that encodes the BPS spectrum in the region of moduli space described by Figure \ref{fig:cone2}.  Since the total spectra of particles and antiparticles in $Q$ and $\widetilde{Q}$ are identical, the physical relation between them is that of a duality: they are equivalent descriptions of the same total spectrum of BPS states.  In the moduli space $\mathcal{U}$ the regions of validity of $Q$ and $\widetilde{Q}$ are sewn together smoothly along the loci where the central charge of an elementary basis particle is real.  This sewing is illustrated in Figure \ref{fig:quiversew}

In section \ref{mutationdefsection} we define the operation of mutation on a given quiver $Q$ to produces the quiver $\widetilde{Q}$, valid on the other side of the wall of the second kind.  In section \ref{methodsection} we explain how the existence of the mutation operation, when interpreted as duality between different quiver descriptions, leads to a powerful and striking method for determining BPS spectra.
\subsection{Quiver Mutation}
\label{mutationdefsection}

As the preceding discussion indicates, a global description of the BPS spectrum across the entire Coulomb branch will require many quivers all glued together in the fashion described above.  In this subsection we describe the algorithmic construction of this set of quivers by a graphical process known as quiver mutation.  In the following subsection we justify these rules using arguments from quiver representation theory.

To define mutation, let us suppose that node $\gamma_1$ is the BPS particle in the quiver whose central charge $\mathcal{Z}_{1}$ is rotating out of the half-plane.  We then seek to describe the dual quiver $\widetilde{Q}$ with corresponding nodes $\{\widetilde{\gamma}_i\}$.  Of course, since we have determined that a given spectrum of BPS particles admits at most one basis of BPS states, both $\widetilde{Q}$ and $\{\widetilde{\gamma}_i\}$ are uniquely fixed.  What's more, the quiver $\widetilde{Q}$ can be described in a simple graphical way starting from $Q$. \cite{Gabriel,BGP,HIV00,Beasley:2001zp,Feng:2001bn,CFIKV01,BD}.  The new basis is given by
\begin{eqnarray}\label{mutbasis}
\widetilde{\gamma}_1 & = & -\gamma_{1}\\
\widetilde{\gamma}_j  & = & 
  \begin{cases}
   \gamma_{j}+( \gamma_{j} \circ   \gamma_{1})\gamma_{1} & \text{if }  \gamma_{j} \circ   \gamma_{1} >0 \\
   \gamma_{j}      & \text{if }   \gamma_{j} \circ \gamma_{1} \leq0.
  \end{cases}
\end{eqnarray}
To construct $\widetilde{Q}$ graphically we follow the steps below:
\begin{enumerate}
\item The nodes of $\widetilde{Q}$ are in one-to-one correspondence with the nodes in $Q$.
\item The arrows of $\widetilde{Q}$, denoted $\widetilde{B}^{a}_{ij}$, are constructed from those of $Q$, denoted $B^{a}_{ij}$ as follows:
\begin{enumerate}
\item For each arrow $B^{a}_{ij}$ in $Q$ draw an arrow $\widetilde{B}^{a}_{ij}$ in $\widetilde{Q}$.
\item For each length two path of arrows passing through node $1$ in $Q$, draw a new arrow in $\widetilde{Q}$ connecting the initial and final node of the length two path
\begin{equation}
B^{a}_{i1}B^{b}_{1j}\longrightarrow \widetilde{B}^{c}_{ij}.
\end{equation}
\item Reverse the direction of all arrows in $\widetilde{Q}$ which have node 1 as one of their endpoints.
\begin{equation}
\widetilde{B}^{a}_{i1}\longrightarrow \widetilde{B}^{a}_{1i}; \hspace{.5in}\widetilde{B}^{a}_{1j}\longrightarrow \widetilde{B}^{a}_{j1}. 
\end{equation}
\end{enumerate}
\item The superpotential $\widetilde{\mathcal{W}}$ of $\widetilde{Q}$ is constructed from the superpotential $\mathcal{W}$ of $Q$ as follows:
\begin{enumerate}
\item Write the same superpotential $\mathcal{W}$.
\item For each length two path considered in step 2(b) replace in $\mathcal{W}$ all occurrences of the product $B^{a}_{i1}B^{b}_{1j}$ with the new arrow $\widetilde{B}^{c}_{ij}$.
\item For each length two path considered in step 2(b) $B^{a}_{i1}B^{b}_{1j}$ there is now a new length three cycle in the quiver $\widetilde{Q}$ formed by the new arrow created in step 2(b) and the reversed arrows in step 2(c)
\begin{equation}
\widetilde{B}^{a}_{1i}\widetilde{B}^{c}_{ij}\widetilde{B}^{b}_{j1}.
\end{equation}
Add to the superpotential all such three cycles.
\end{enumerate}
\end{enumerate}
As a simple example of this procedure we consider the $A_{3}$ quiver of section \ref{threenodeex} shown on the left and its mutation at node 1 shown on the right.

\noindent
  \begin{equation}\begin{array}{cc}
\begin{minipage}{.4\textwidth}
\[\xy
(-20,0)*+{1}*\cir<8pt>{}="a" ; (20,0)*+{2}*\cir<8pt>{}="b" ; (0,28.2)*+{3}*\cir<8pt>{}="c" ;
\ar @{->} "a"; "b" 
\ar @{->} "b"; "c"
\ar @{->} "c"; "a"
 \endxy
\]
\[
\mathcal{W}  = B_{12}B_{23}B_{31} 
\]
\end{minipage}&
\begin{minipage}{.4\textwidth}
\[
\xy 
(-20,0)*+{1}*\cir<8pt>{}="a" ; (20,0)*+{2}*\cir<8pt>{}="b" ; (0,28.2)*+{3}*\cir<8pt>{}="c" ;
\ar @{->} "b"; "a" 
\ar @{->} "b"; "c" <3pt>
\ar @{->} "c"; "b" <3pt>
\ar @{->} "a"; "c" 
 \endxy 
\]
\[\mathcal{W}  = \widetilde{B}_{32}\widetilde{B}_{23}+ \widetilde{B}_{32} \widetilde{B}_{21} \widetilde{B}_{13} 
\]\end{minipage}\end{array}
\end{equation}
\vspace{.1in}

As the above example illustrates, the process of quiver mutation in general creates cycles of length two in our new quiver.  From a physical perspective these are fields in the quiver quantum mechanics which admit a gauge invariant mass term.  In the example above such mass terms are present in the quadratic piece of the potential $\widetilde{B}_{32} \widetilde{B}_{23}$.  As is typical in physical theories, the massive fields decouple from the analysis of ground states and hence do not affect the BPS spectrum.  We may therefore integrate them out.  Thus to our list of quiver mutation rules we append the following final steps:
\begin{enumerate}
\item[4.]  For each two-cycle in $\widetilde{Q}$ for which a quadratic term appears in $\widetilde{W}$, delete the two associated arrows.
\item[5.] For each deleted arrow $\widetilde{B}^{a}_{ij}$ in step 4, solve the equation of motion
\begin{equation}
\frac{\partial{\widetilde{W}}}{\partial \widetilde{B}^{a}_{ij}}=0.
\end{equation}
Use the solution to eliminate $\widetilde{B}^{a}_{ij}$ from the potential.
\end{enumerate}

In the example illustrated above, the only two cycle has quadratic terms in the superpotential and is therefore deleted from the quiver.  This results in a vanishing superpotential and a quiver of the following form.
\begin{equation}
\xy 
(-40,0)*+{2}*\cir<8pt>{}="a" ; (0,0)*+{1}*\cir<8pt>{}="b" ; (40,0)*+{3}*\cir<8pt>{}="c" ;
\ar @{->} "a"; "b" 
\ar @{->} "b"; "c" 
 \endxy
\end{equation}

As a general rule, the study of BPS quivers is greatly complicated by the existence of pairs of opposite arrows whose associated fields cannot be integrated out from the superpotential.  When this is never the case, that is when the potential $\mathcal{W}$ is strong enough to integrate out to all opposite bifundamental fields after an arbitrary sequence of mutations, the potential is said to be \emph{non-degenerate}.  It is a fortunate simplification that for the vast majority of BPS quivers related to quantum field theories that we discuss in this paper the potential will turn out to be non-degenerate.  However exceptions to this general rule do arise.  For example in section \ref{t2quiv} we will see that the the quiver for the $\mathcal{T}_{2}$ theory defined by a free trifundamental half-hypermultiplet of a flavor group $SU(2)\times SU(2)\times SU(2)$ involves a quiver with canceling arrows and a potential which is too degenerate to integrate out all the associated by bifundamental fields.  In the following unless otherwise stated we avoid this complication and assume that all of our quivers involve non-degenerate superpotentials.  However, even when this is not the case one may still apply the mutation rules written above.  Mutation at a node supporting a pair of canceling arrows then results in adjoint fields at the mutated node.

\subsubsection{$A_{3}$ Revisited}
To put the above theory of quiver mutation in perspective, it is useful to consider the simplest example where the phenomenon of wall of the second kind occur.  This is the $A_{3}$ theory whose representation theory was investigated in section \ref{threenodeex}.  There are in fact four distinct quivers for the $A_{3}$ theory related by mutation.  These are given by
   \[
    \begin{xy}
    (0,-4)*{}; (0,0)*+{1}*\cir<8pt>{}="a";
    (40,-4)*{}; (40,0)*+{3}*\cir<8pt>{}="b";
    (20,-4)*{}; (20,0)*+{2}*\cir<8pt>{}="c";
   {\ar "a";"c"};{\ar "c";"b"}
    \end{xy}
    \]
       \[
    \begin{xy}
    (0,-4)*{}; (0,0)*+{1}*\cir<8pt>{}="a";
    (40,-4)*{}; (40,0)*+{3}*\cir<8pt>{}="b";
    (20,-4)*{}; (20,0)*+{2}*\cir<8pt>{}="c";
    {\ar "c";"a"};{\ar "c";"b"}
    \end{xy}
    \]
           \[
    \begin{xy}
    (0,-4)*{}; (0,0)*+{1}*\cir<8pt>{}="a";
    (40,-4)*{}; (40,0)*+{3}*\cir<8pt>{}="b";
    (20,-4)*{}; (20,0)*+{2}*\cir<8pt>{}="c";
    {\ar "a";"c"};{\ar "b";"c"}
    \end{xy}
    \]
   \[
    \begin{xy}
    (0,-4)*{}; (0,0)*+{1}*\cir<8pt>{}="a";
    (40,-4)*{}; (40,0)*+{3}*\cir<8pt>{}="b";
    (20,-4)*{}; (20,0)*+{2}*\cir<8pt>{}="c";
    {\ar @/^2pc/ "b";"a"};{\ar "c";"b"};{\ar "a";"c"}
    \end{xy}
    \]
    Let us name these four quivers respectively as $L$, $O$, $I$, and $C$.  The representation theory of the $C$ quiver was worked out in section \ref{threenodeex}.  In particular we determined that $C$ supports either $4$ or $5$ BPS states depending on moduli.  The representation theory of the other quivers is also readily calculated.  One finds that $L$ has 6 distinct chambers, while both $I$ and $O$ have 4.  If we dentote by $\theta_{i}$ the phase of $\mathcal{Z}_{i}$ and $\theta_{ij}$ the phase of $\mathcal{Z}_{i}+\mathcal{Z}_{j}$, then the complete list of chambers is given in table \ref{a3total}.

\begin{table}
\center
\begin{tabular}{|c|c|c|}
\hline
Chamber & Phase Conditions& Number of BPS States \\
\hline
$L_{1}$&$\theta_{3}>\theta_{2} >\theta_{1}$ &  3 \\
\hline
$L_{2}$&$\theta_{2}$ smallest, and $\theta_{1}, \theta_{3} >\theta_{12}$ &  4 \\
\hline
$L_{3}$&$\theta_{2}$ largest, and $ \theta_{23} >\theta_{1}, \theta_{3}$  &  4 \\
\hline
$L_{4}$&$\theta_{1}>\theta_{12} >\theta_{3}>\theta_{2}$ &  5 \\
\hline
$L_{5}$&$\theta_{2}>\theta_{1} >\theta_{23}>\theta_{3}$ &  5 \\
\hline
$L_{6}$&$\theta_{1}>\theta_{2} >\theta_{3}$ &  6 \\
\hline
$O_{1}$&$\theta_{2}$ smallest & 3\\
\hline
$O_{2}$&$\theta_{2}$ intermediate & 4 \\
\hline
$O_{3}$&$\theta_{2}$ largest, and $\theta_{12}<\theta_{3}$ or $\theta_{23}<\theta_{1}$& 5 \\
\hline
$O_{4}$&$\theta_{2}$ largest, and $\theta_{12}>\theta_{3}$ and $\theta_{23}>\theta_{1}$ & 6 \\
\hline
$I_{1}$&$\theta_{2}$ largest & 3\\
\hline
$I_{2}$&$\theta_{2}$ intermediate & 4 \\
\hline
$I_{3}$&$\theta_{2}$ smallest, and $\theta_{3}<\theta_{12}$ or $\theta_{1}<\theta_{23}$& 5 \\
\hline
$I_{4}$&$\theta_{2}$ smallest, and $\theta_{3}>\theta_{12}$ and $\theta_{1}>\theta_{23}$ &  6 \\
\hline
$C_{1}$&not cyclically ordered e.g. $\theta_{2}>\theta_{1} >\theta_{3}$ &  4 \\
\hline
$C_{2}$&cyclically ordered e.g.  $\theta_{1}>\theta_{2} >\theta_{3}$ &  5 \\
\hline
\end{tabular}
\caption{The chambers of the $A_{3}$ quivers before mutation equivalences are imposed. For each quiver labelled with node charges $Z_{i}$, $\theta_{i}$ denotes the argument of $Z_{i}$ while $\theta_{ij}$ denotes the argument of $Z_{i}+Z_{j}$.}
\label{a3total}
\end{table}
    
    In the global theory of $A_{3}$ these chambers are connected together across walls of the second kind where the quiver changes by a mutation. To understand mutations we then represent each chamber as a node in a graph and connect those mutation equivalent with directed arrows.  For example we define the expression
\begin{equation}
\xymatrix{
Q_{i} \ar[r] &\widetilde{Q}_{j} 
},
\label{qmutgraphex}
\end{equation}
to mean that mutation in chamber $i$ of quiver $Q$ on the leftmost boundary ray leads to chamber $j$ in the quiver $\widetilde{Q}$.  With these conventions the complete structure of walls of the second kind in the $A_{3}$ theory is encoded in the following diagrams.
\begin{equation}
\xymatrix{
L_{1} \ar[d]& \\
I_{1} \ar[r]& O_{1}\ar[ul]}
\hspace{.47in}
\xymatrix{
 & I_{2} \ar[dr] &  \\
L_{2} \ar[ur] \ar[dr] &C_{1} \ar[l] & \ar[l]L_{3} \\
 & O_{2} \ar[ur] & 
}
\hspace{.47in}
\xymatrix{
L_{5} \ar[r] &C_{2} \ar[r] & L_{4} \ar[d]\\
 I_{3}\ar[u]&  & \ar[ll]O_{3} 
}
\hspace{.47in}
\xymatrix{
L_{6} \ar[d]& \\
O_{4} \ar[r]& I_{4}\ar[ul]}\label{A3quivermutrels}
\end{equation} 
Where in the above, some chambers have two arrows leaving them because one can change the leftmost ray without crossing a wall.

\subsubsection{Justification of Mutation}

The previous subsection gives a straightforward recipe for producing, from a given quiver $Q$, all of its related duals by considering mutations at various nodes.  However we have not yet explained why this mutation rule is in fact correct.  In this subsection we fill in this gap.\footnote{The arguments in this section are somewhat technical and could be skipped in a first reading.}  Specifically our goal will be to derive the mutation rule, given the assumption that a quiver description $\widetilde{Q}$ exists after the transition illustrated by Figure \ref{fig:cones}.  

The basic point is that the new elementary basis particles $\widetilde{\gamma}_{i}$, are interpreted from the point of view of $Q$ as certain bound states of the original basis particles $\gamma_{i}$.  The key step is to identify which bound states.  

Consider again the cone geometry illustrated in Figure \ref{fig:cones}.  A special role is played by the two particles whose central charge rays form the boundary of the cone.  Such particles must always be included in the basis because, as their central charges are on the boundary of the cone, there is no way to generate these states by positive linear combinations of other rays in the cone.  Thus in Figure \ref{fig:cone2} the two states with central charges $\mathcal{Z}'$ and $-\mathcal{Z}_{1}$ must appear as nodes of the quiver $\widetilde{Q}$.  Of these, the latter is easy to identify as the antiparticle of the mutated node, $-\gamma_{1},$ and hence this charge must be in the new basis.  Meanwhile, in the following argument we will prove that the left-most ray, which we frequently refer to as the extremal ray, $\mathcal{Z}'$, is always a two particle bound state which may be identified explicitly.

To begin, we consider all connected length two subquivers of $Q$ which involve the node $\gamma_1$.  For a given node $\gamma_i$ there are $k_{i}$ arrows pointing either from $\gamma_i$ to $\gamma_1$ or from $\gamma_1$ to $\gamma_i$.
\begin{equation}
\xy 
(-20,0)*+{\gamma_1}*\cir<8pt>{}="a" ; (20,0)*+{\gamma_i}*\cir<8pt>{}="b" ;
(0,0)*+{\vdots}
\ar @/^2.5pc/ ^{B_{1}}"a"; "b"
\ar @/^1pc/ ^{B_{2}}"a"; "b"
\ar @/_2pc/ ^{B_{k_{i}}}"a"; "b"
 \endxy \hspace{.5in} 
 \mathrm{or}
 \hspace{.5in} 
 \xy 
(-20,0)*+{\gamma_i}*\cir<8pt>{}="a" ; (20,0)*+{\gamma_1}*\cir<8pt>{}="b" ;
(0,0)*+{\vdots}
\ar @/^2.5pc/ ^{B_{1}}"a"; "b"
\ar @/^1pc/ ^{B_{2}}"a"; "b"
\ar @/_2pc/ ^{B_{k_{i}}}"a"; "b"
 \endxy
 \label{sourcesink}
\end{equation}
Let us describe the leftmost bound state supported by these two node quivers.  In the case on the right of \eqref{sourcesink}, $\gamma_1$ appears as a sink.  Then, since $\mathcal{Z}(\gamma_1)$ has largest phase by hypothesis, $\gamma_1$ by itself is a destabilizing subrep of any possible bound state; thus no bound states can form.  

On the other hand, in the case on the left of \eqref{sourcesink}, where $\gamma_1$ appears as a source, bound states can exist.  We consider a general representation of the form
\begin{equation}
\xy 
(-23,0)*+{\mathbb{C}^{n}}*{}="c" ;(-20,0)*+{}*{}="a"; (20,0)*+{}*{}="b" ;(23,0)*+{\mathbb{C}^{m}}*{}="d" ;
(0,0)*+{\vdots}
\ar @/^2.5pc/ ^{B_{1}}"a"; "b" <3pt>
\ar @/^1pc/ ^{B_{2}}"a"; "b"
\ar @/_2pc/ ^{B_{k_{i}}}"a"; "b"<-3pt>
 \endxy
\end{equation} 

To make a bound state with largest possible phase we wish to make a representation where $n/m$ is as large as possible.  However, it is not difficult to see that the ratio $n/m$ is bounded.  Indeed, since $\mathcal{Z}(\gamma_1)$ has largest phase, there is a potentially destabilizing subrepresentation involving only the particle $\gamma_{1}$.  Such a subrepresentation is described by $k_{i}$ commutative diagrams of the form
\begin{equation}
\xymatrix{
\mathbb{C}^{n}  \ar[r]^{B_{j}} & \mathbb{C}^{m} \\
\mathbb{C}\ar[r]^{0} \ar[u]& 0 \ar[u]} 
\label{destablemut}
 \end{equation}
 In other words, the potential destabilizing subrepresentation is nothing but a non-zero vector which is simultaneously in the kernel of all of the maps $B_{j}$.  But then a simple dimension count shows that
 \begin{equation}
 \mathrm{dimension}\left(\bigcap_{j=1}^{k_{i}} \ker(B_{j})\right)\geq n-k_{i}m. \label{kernelcount}
 \end{equation}
And so in particular when the right-hand side of the above is positive, the subrepresentation \eqref{destablemut} exists and hence the bound state is unstable.  Thus we learn that stability requires
\begin{equation}
\frac{n}{m}\leq k_{i}.
\end{equation}

Finally, it is not difficult to find a stable representation $R$ which saturates the above bound.  Indeed let us take $n=k_{i}$ and $m=1$.  Then the maps $B_{j}$ are simply projections to a line.  The stability constraint that the $B_{j}$ have no common kernel implies that, up to gauge transformation, $B_{j}$ can be taken to be the dual vector to the $j$th basis element in the vector space attached to $\gamma_1$.   So defined, the representation $R$ is stable and has no moduli.  Thus it gives rise to a hypermultiplet with charge
\begin{equation}
\gamma_{i}+k_{i}\gamma_{1}.
\end{equation}

This completes the required analysis of quivers with two nodes.  To summarize, in the region of parameter space where $\mathcal{Z}(\gamma_1)$ has largest phase, we have determined the extremal bound state of all two-node subquivers involving $\gamma_1$.  The charges of the extremal  bound states are:   
\begin{itemize}
\item If $\gamma_{i}\circ \gamma_{1} <0$ then the extremal bound state is simply $\gamma_{i}$.
\item If $\gamma_{i}\circ \gamma_{1}>0$ then the extremal bound state is $\gamma_{i}+ (\gamma_{i}\circ \gamma_{1})\gamma_{1}$.
\end{itemize}

Now we claim that in the quiver $Q$ with an arbitrary number of nodes, one of the two particle bound states we have identified above will still be the left-most extremal ray after $\mathcal{Z}(\gamma_1)$ exits the upper half-plane.  To see this, we consider an arbitrary stable representation $R$ of $Q$.  We write the charge of $R$ as
\begin{equation}
\gamma_{R}=n\gamma_{1}+\sum_{\gamma_{i}\circ \gamma_{1}>0}m_{i}\gamma_{i}+\sum_{\gamma_{j}\circ \gamma_{1}\leq0}l_{j}\gamma_{j}
\end{equation} 
Let us focus in on the representation $R$ near the node $\gamma_1$.  There are now many nodes connected to the node 1 by various non-zero maps.  For those connections with $\gamma_{i}\circ \gamma_{1}\leq0,$ the node $\gamma_1$ appears as a sink, for those with $\gamma_{i}\circ \gamma_{1}>0,$ $\gamma_1$ appears as a source.

Our strategy is again to test whether $R$ is stable with respect to decays involving the subrepresentation $S$ with charge $\gamma_{1}$.  As in the two node case, in such a situation the connections where $\gamma_1$ is a sink are irrelevant.  On the other hand, if $S$ is really a subrepresentation then for each node link in the representation where node $1$ is a source, we have commutative diagrams of the form \eqref{destablemut}.

Given that $\mathcal{Z}(\gamma_1)$ has largest phase, stability of $R$ means that we must obstruct the existence of $S$.  As in the analysis of the two node quivers we see that $S$ will be a subrepresentation provided that the kernels of all maps exiting the node $\gamma_1$ have nonzero intersection.  However, just as in \eqref{kernelcount} we can see that this leads to an a priori bound on $n$, the amount of $\gamma_{1}$ contained in the representation $R$.  Explicitly we have
\begin{equation}
\mathrm{dimension}\left(\bigcap_{\gamma_{i}\circ \gamma_{1}>0}\bigcap_{j=1}^{k_{i}} \ker(B_{j})\right)\geq n-\sum_{\gamma_{i}\circ \gamma_{1}>0}k_{i}m_{i}.
\end{equation}
Hence to obstruct the existence of the subrepresentation $S$ we deduce the bound
\begin{equation}
n\leq \sum_{\gamma_{i}\circ \gamma_{1}>0}k_{i}m_{i}.
\end{equation}
But now we can directly see that $R$ cannot be extremal.  We have 
\begin{eqnarray}
\arg\left(\mathcal{Z}(R)\right) & = &\arg\left(nZ_{1}+\sum_{\gamma_{i}\circ \gamma_{1}>0}m_{i}Z_{i}+\sum_{\gamma_{j}\circ \gamma_{1}\leq 0}l_{j}Z_{j}\right)  \label{repineq}\\
& \leq &\arg\left(\sum_{\gamma_{i}\circ \gamma_{1}>0}m_{i}(k_{i}Z_{1}+Z_{i})+\sum_{\gamma_{j}\circ \gamma_{1}\leq 0}l_{j}Z_{j}\right). \nonumber
\end{eqnarray}
But the final expression in \eqref{repineq} is manifestly contained in the positive span of the two node extremal bound states, $k_i\gamma_1+\gamma_i$, that we identified in our analysis of two node quivers.  In particular, this means that $R$ cannot be a boundary ray and hence is not extremal.

Thus we deduce that the left-most ray after mutation is one of the two particle bound states that we have identified in our analysis of two node quivers.  Extremality then ensures that our new basis must include this two particle bound state.  But finally we need only notice that the central charges of all the two node extremal bound states that we have discovered are independent parameters.  Indeed letting the central charges vary in an arbitrary way, our conclusion is in fact that \emph{all} the two node bound states which we have determined must in fact be in the new basis.  In particular this means that the new basis of charges after mutation is completely fixed and we may write the transformation as follows:
\begin{eqnarray}
\widetilde{\gamma_{1}} & = & -\gamma_{1}\\
\widetilde{\gamma_{j}}  & = & 
  \begin{cases}
   \gamma_{j}+( \gamma_{j} \circ   \gamma_{1})\gamma_{1} & \text{if }  \gamma_{j} \circ   \gamma_{1} >0 \\
   \gamma_{j}      & \text{if }   \gamma_{j} \circ \gamma_{1} \leq0
  \end{cases}
\end{eqnarray}

As one can easily verify, the graphical quiver mutation rules described in the previous section are a direct consequence of computing the new BPS quiver $\widetilde{Q}$ from the symplectic products of the new basis of charges $\{\widetilde{\gamma}_{i}\}$.  This completes our argument justifying the mutation rules.
\subsection{The Mutation Method: BPS Spectra from Quiver Dualities}
\label{methodsection}
We saw above that at walls of the second kind, we were forced to change our quiver description because the central charge of some state exited the upper half of the complex half-plane, thereby turning from a particle to an antiparticle. We might also consider what happens if we fix a modulus $u\in \mathcal{U}$ and then consider a different definition of the particle half-plane, $\mathcal{H}$. If we imagine continuously changing our choice from one $\mathcal{H}$ to another, the situation is precisely the same as above; there is some parameter which we are tuning, and at some critical value the central charge of some state becomes such that it switches from particle to antiparticle. 

In this case, however, we are remaining at a fixed point in moduli space, and so all of these quivers describe precisely the same physics. That is, they are dual descriptions of the BPS spectrum. In fact, there is a whole class of quivers related to each other by duality at each point in moduli space. We will now exploit this fact to produce for us, in many cases, the entire spectrum for free.

First, let us reiterate that a single form of the quiver already in principle determines exactly which BPS states in the theory are occupied, including their spin and multiplicity. To find the answer, one can solve the representation theory of the quiver with superpotential, which amounts to the linear algebra problem described in section \ref{rep}. However, in practice this problem can become quite intractable. The mutation method we propose gets rid of all of the unsightly work required in solving the problem directly, and instead produces the spectrum using chains of dualities through different quiver descriptions of the theory.

Recall our first application of quiver rep theory in section \ref{rep}, where we checked that nodes of the quiver always correspond to multiplicity one hypermultiplets. This fact, together with an examination of which states are forced to be nodes for various choices of half-plane $\mathcal{H}$, is at the heart of what we call the mutation method. Imagine that for our initial choice of $\mathcal{H}$, with BPS basis $\{\gamma_i\}$, $\gamma_1$ is the node such that $\mathcal{Z}(\gamma_1)$ is left-most in $\mathcal{H}.$\footnote{From now on we will abuse verbiage slightly and simply say that ``$\gamma_1$ is left-most."} Say we then rotate our half-plane past it, and do the corresponding mutation to arrive at a new quiver description of the theory. This mutation includes an action on the charges of the quiver $\gamma_i$, as given in equation (3.2)-(3.3). Since this  new quiver is a description of the BPS states of the same theory, its nodes are also multiplicity one hypermultiplets. Consequently, we have discovered some subset of states in the 4d theory which we can say must exist. In particular, we generate some new BPS states of the form $-\gamma_1,\gamma_i+(\gamma_i\circ\gamma_1)\gamma_1.$ Of course, $-\gamma_1$ is just the antiparticle of the state $\gamma_1$, so this is no additional information. However, the states $\gamma_i + m_i\gamma_1$ are completely new. To discover these same states from the original quiver would have involved solving the non-trivial representation theory problem studied in the previous subsection. We are able to avoid this headache by observing that, because of duality, these states must be in the spectrum for consistency.

So we have found that duality will trivially produce some subset of the spectrum as nodes of various dual quivers. But in fact it does much more: in many cases, mutation produces the full spectrum in this way. Imagine we're in a chamber with finitely many BPS states, and pick an arbitrary state $\gamma$ which is a hypermultiplet of the 4d theory. Then we can rotate the half-plane $\mathcal{H}$ so that $\gamma$ is left-most. As usual, since the nodes of the quiver form a positive basis for states in $\mathcal{H}$, $\gamma$ must itself be a node. Therefore, if we start with any quiver description, and start rotating $\mathcal{H} \rightarrow e^{-i\theta}\mathcal{H}$ until $\gamma$ becomes left-most, we will go through a corresponding sequence of mutations, after which $\gamma$ will simply be a node of the quiver.

It is then easy to see how to systematically generate the spectrum in any finite chamber. We start with any quiver description which is valid at our given point in moduli space, and start rotating the half-plane. Since there are only finitely many states, we will only pass through finitely many mutations before we return to the original half-plane $\mathcal{H} \rightarrow e^{2\pi i}\mathcal{H}$.\footnote{Recall that for a given choice of $\mathcal{H}$, the quiver description is actually unique - there is a unique positive integral basis for the lattice of occupied BPS states, up to permutation. So we will also return to the original quiver up to permutation when $\mathcal{H}$ undergoes a full rotation.} The key point is every state in the chamber is left-most at some point during this rotation, so every state will indeed show up as a node of one of the dual quivers. Since rotating past a state corresponds to mutating on the node corresponding to that state, if we do the entire sequence of mutations and record each state we've mutated on, we will have exhausted all states in the chamber.

We can save a bit of work by making use of CPT: for any state $\gamma$ in the spectrum, $-\gamma$ is also occupied. So instead of taking $\mathcal{H} \rightarrow e^{2\pi i}\mathcal{H}$, we can just rotate half-way, $\mathcal{H} \rightarrow e^{i\pi}\mathcal{H}$, ending up at the quiver which describes all the antiparticles.\footnote{By a similar argument as above, the final quiver will have nodes $-\gamma_i$.} If we record every state $\gamma$ we mutate on as $\mathcal{H}$ is rotated, and then add all antiparticles $-\gamma$, we will have precisely the spectrum of the 4d theory. Note that we must repeat this procedure for each chamber, by doing mutations in some different order, as prescribed by the ordering of the phases of the central charges in that region of moduli space. As we discussed above any given quiver generally only covers some subset of moduli space; therefore, for different chambers, it will generally be necessary to apply this procedure to different mutation forms of the quiver.

Let's try an example. The representation theory for the Argyres-Douglas A3 theory was worked on in detail in section 2.3.2. We will see how to reproduce it with much less work in the present framework. We will assume that we are at a point in moduli space covered by the cyclic three node quiver. Imagine that $\gamma_1$ is leftmost. After the first mutation, the mutation that follows will depend on the ordering of $\gamma_3$ and $\gamma_1+\gamma_2$. Suppose that $\gamma_3$ is to the left. Then the particle half-plane, $\mathcal{H}$ and associated quiver before (i) and after (ii) the first mutation at $\gamma_1$ are
\begin{equation}\nonumber
\begin{array}{cc}
    \begin{minipage}[b]{0.5\linewidth}
    \centering
    \frame{\includegraphics[width=3in]{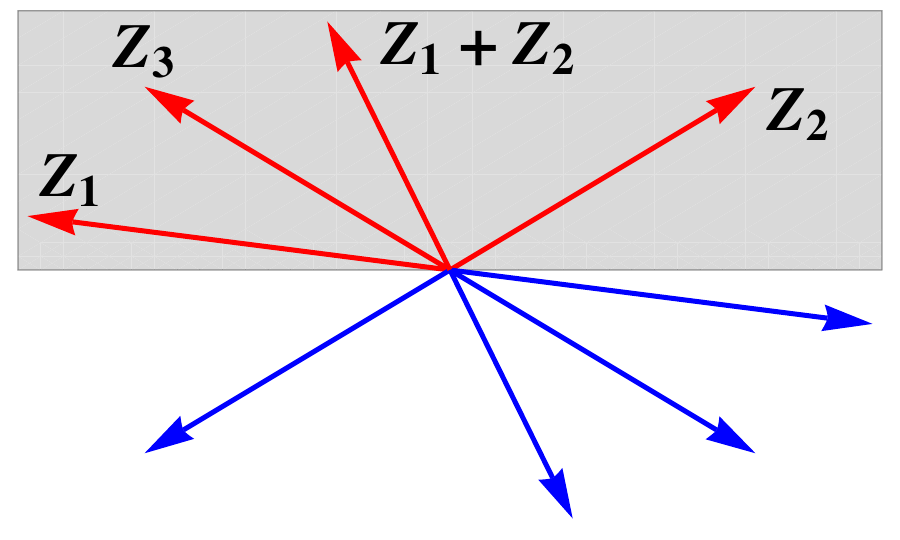}}
    \centering
    \[
    \begin{xy}
    (0,-4)*{\gamma_1}; (0,0)*{\CIRCLE}="a";
    (40,-4)*{\gamma_2}; (40,0)*{\Circle}="b";
    (20,-4)*{\gamma_3}; (20,0)*{\Circle}="c";
    {\ar @/^2pc/ "a";"b"};{\ar "b";"c"};{\ar "c";"a"}
    \end{xy}
    \]
        \small (i)
    \end{minipage} &
    \begin{minipage}[b]{0.5\linewidth}
    \centering
       \frame{\includegraphics[width=3in]{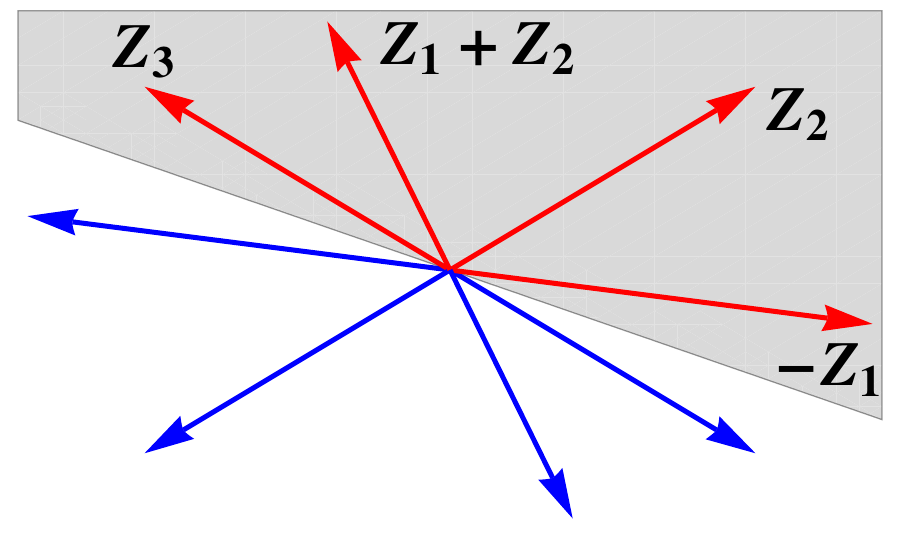}}
       \centering
  \[
    \begin{xy}
    (0,-4)*{-\gamma_1}; (0,0)*{\Circle}="a";
    (40,-4)*{\gamma_1+\gamma_2}; (40,0)*{\Circle}="b";
    (20,-4)*{\gamma_3}; (20,0)*{\CIRCLE}="c";
    {\ar @/_2pc/ "b";"a"};{\ar "a";"c"}
    \end{xy}
    \]
    \small (ii)
    \end{minipage}
    \end{array}
    \end{equation}
In the above diagrams, we denote the left-most particle state in each quiver, which indicates the next node to be mutated, by drawing the corresponding node in black, $\CIRCLE$. Now since the $\gamma_i$ were in the original half-plane $\mathcal{H}$ to begin with, it must be that $\gamma_1+\gamma_2$ is to the left of $-\gamma_1$ and $-\gamma_3$ in the current half-plane. This is true in general: one never mutates on negative nodes in going through a $\pi$-rotation of $\mathcal{H}$ from a quiver to its antiparticle quiver. The remaining mutations are completely fixed, and we find (iii,iv,v)
    \begin{equation}\nonumber
\begin{array}{cc}
    \begin{minipage}[b]{0.5\linewidth}
    \centering
    \frame{\includegraphics[width=3in]{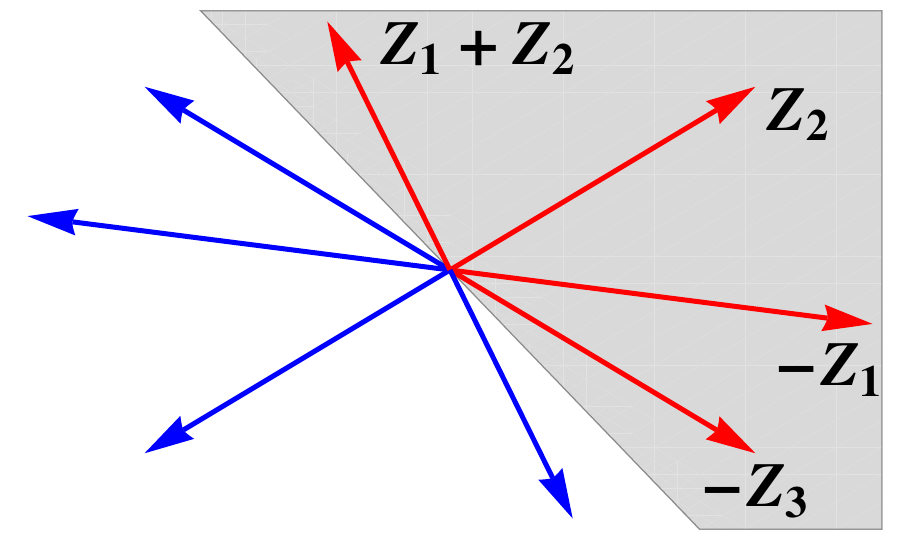}}
    \centering
    \[
    \begin{xy}
    (0,-4)*{-\gamma_1}; (0,0)*{\Circle}="a";
    (40,-4)*{\gamma_1+\gamma_2}; (40,0)*{\CIRCLE}="b";
    (20,-4)*{-\gamma_3}; (20,0)*{\Circle}="c";
    {\ar @/_2pc/"b";"a"};{\ar "c";"a"}
    \end{xy}
    \]
        \small (iii)
    \end{minipage} &
    \begin{minipage}[b]{0.5\linewidth}
    \centering
       \frame{\includegraphics[width=3in]{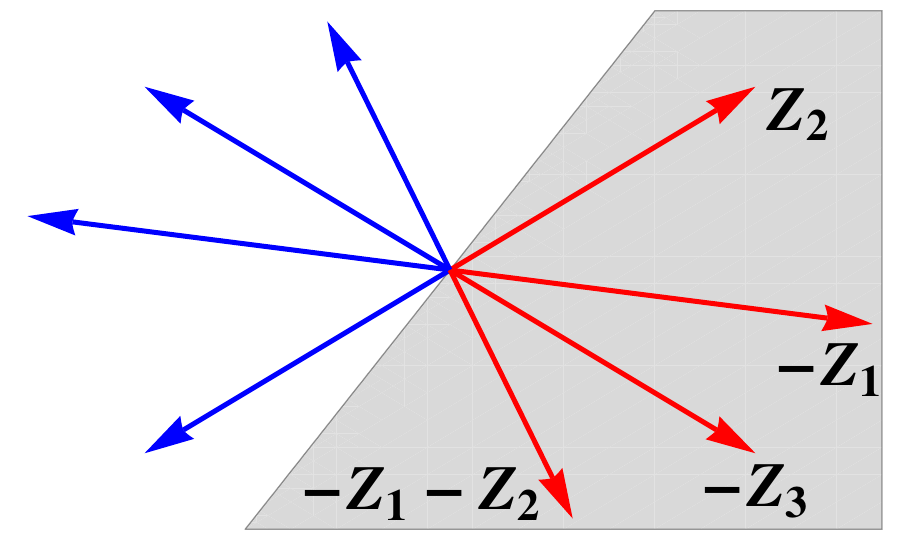}}
       \centering
    \[
    \begin{xy}
    (0,-4)*{\gamma_2}; (0,0)*{\CIRCLE}="a";
    (40,-4)*{-\gamma_1-\gamma_2}; (40,0)*{\Circle}="b";
    (20,-4)*{-\gamma_3}; (20,0)*{\Circle}="c";
    {\ar @/^2pc/"a";"b"};{\ar "c";"a"}
    \end{xy}
    \]
    \small (iv)
    \end{minipage}
    \end{array}
    \end{equation}  
    \begin{equation}\nonumber
    \begin{minipage}[b]{0.5\linewidth}
        \centering
  \frame{\includegraphics[height=2in]{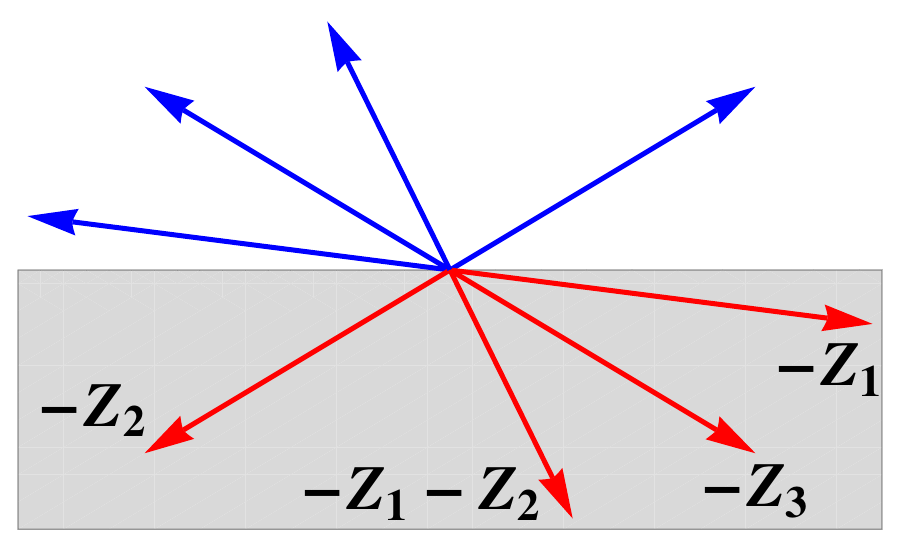}}
    \centering
    \[
    \begin{xy}
    (0,-4)*{-\gamma_2}; (0,0)*{\Circle}="a";
    (40,-4)*{-\gamma_1}; (40,0)*{\Circle}="b";
    (20,-4)*{-\gamma_3}; (20,0)*{\Circle}="c";
    {\ar @/_2pc/"b";"a"};{\ar "a";"c"};{\ar "c";"b"};
    \end{xy}
    \]
    \small (v)
    \end{minipage}
    \end{equation}
So we've arrived at the antiparticle quiver, which at the level of quiver without charges is the same, because the antisymmetric product is not affected by an overall sign on charges.\footnote{If you try to label nodes and keep track of them, which the drawings may subliminally suggest you do, in general you will return to $(-1)\times$permutation.} Therefore we've discovered a chamber with the states $\gamma_1$, $\gamma_2$, $\gamma_3$ and $\gamma_1 + \gamma_2$. This indeed agrees with one of the chambers found in 2.3.2. All of the chambers can similarly be mapped out, without ever doing the linear algebra analysis.

We pause here to emphasize two important points. The first is to recall that a quiver from the mutation class generically only covers a subset of moduli space. Therefore to map out all chambers, one must carry forth the above with the starting quiver being any one of the quivers in the mutation class. The second point is that, using the above method, one will not find any chamber covered by the cyclic quiver which contains the state $\gamma_1+\gamma_2+\gamma_3$. In the analysis of section 2.3.2, it was found that the $\gamma_1+\gamma_2+\gamma_3$ state was there in the quiver without superpotential, but killed when the (unique) non-degenerate superpotential was included. Thus we see that this mutation method knows about the associated non-degenerate superpotential indirectly. This is expected, because a non-degenerate superpotential is required for the mutation rule written above to be sensible.

There are some simple non-trivial statements which we can immediately make based on this method. One is that any finite chamber can only contain hypermultiplets, with multiplicity one. The argument here is simply that any state in a finite chamber can be made into a node of some dual quiver, and nodes, as we've mentioned, can never correspond to higher spin objects or higher multiplicity hypers. Therefore, it would be inconsistent with duality to ever have a higher spin or higher multiplicity object in a finite chamber.

Now let's consider infinite chambers. An additional layer of complexity, as compared to the finite case, is that two dual quiver descriptions may be separated by an infinite sequence of mutations. This is because, as we rotate between two choices of $\mathcal{H}$, we will generically have infinitely many BPS states which rotate out to the left. Our method above depended on our ability to keep track of the sequence of mutations which happens as $\mathcal{H}\rightarrow e^{i\pi}\mathcal{H}$. Now the infinitude of states in some sense blocks us from competing this sequence of mutations. For example, if we start with a given quiver description, we can't explore beyond the closest accumulation ray in the $\mathcal{Z}$-plane. Because of this difficulty, we can't make a similarly definite statement about the method as it applies to infinite chambers. Indeed, for certain theories, such as $\mathcal{N} = 2^*$ $SU(2)$ (the mass deformed $\mathcal{N}=4$ theory), it appears that the method isn't sophisticated enough to exhaust the spectrum.\footnote{Of course we can always produce some arbitrarily large subset of states of the theory by mutating until exhaustion (of the mutator, that is).}

However, as we will see in several examples, infinite chambers may also be understood by this method. Infinitude of the chamber is often due to higher spin objects, and we can often make progress by being just a bit clever. Note that any higher spin object must in fact be an accumulation ray of states in the central charge plane: If it weren't, we could rotate $\mathcal{H}$ so that it was left-most, and as above, in this dual quiver description our higher spin state would be a node. Of course this is a contradiction - nodes are always multiplicity one hypers. Higher multiplicity hypers must similarly be accumulation rays, a fact which may be less intuitive outside of this framework.

Before going on to examples and applications, we make some additional technical notes about the actual implementation of the mutation method. As we have described it here, we choose a point of the physical moduli space, compute central charges at that point, and mutate on the nodes in the order given by the ordering of phases of the central charges, as we tune $\mathcal{H}\rightarrow\mathcal{H}_\pi.$ Instead, when exploring the possible BPS spectra, it is sometimes more practical to simply mutate on the nodes in any order, and then check two things: (1) that the ordering chosen is consistent, and (2) that the ordering chosen is realized somewhere in physical moduli space. By consistent, we mean that there exists some choice of central charges $\mathcal{Z}(\gamma_i)$ that correspond to the ordering chosen. As it turns out, there is no need to check the first point: as long as we mutate only on nodes whose charges are given by positive linear combinations of the original $\gamma_i$, then the ordering is consistent. Of course, we expect to only mutate on positive nodes since we are only rotating by $\pi$ through the particle half-plane, and all particles should be given by positive integer linear combinations of the initial $\gamma_i.$ Note that the only condition for consistency is that $\arg\mathcal{Z}(\gamma_1+\gamma_2)$ lie between $\arg\mathcal{Z}(\gamma_1),\arg\mathcal{Z}(\gamma_2).$ In fact, the mutation method protects us from making inconsistent choices. Fix $\arg\mathcal{Z}(\gamma_1)>\arg\mathcal{Z}(\gamma_2),$ and suppose we have already mutated past $\gamma_1,$ but not yet $\gamma_2.$ Thus $-\gamma_1$ is in the positive integral span of the mutated quiver basis. Suppose both $\gamma_1+\gamma_2$ and $\gamma_2$ to appear as nodes; this is an immediate contradiction with the fact that the nodes form a basis, since now $\gamma_2$ is both a basis element and a non-trivial linear combination of basis elements $(\gamma_1+\gamma_2)+(-\gamma_1).$ So only one of these can appear as nodes and be mutated on next. If it is $\gamma_1+\gamma_2,$ there we are safe, and there is no inconsistency. If it is $\gamma_2,$ let's mutate past so that both $-\gamma_1, -\gamma_2$ are in the positive integral span of the mutated quiver basis; now it is impossible for $\gamma_1+\gamma_2$ to appear as a node of the quiver, or else we can construct 0 as a non-trivial linear combination of basis elements $\gamma_1+\gamma_2+(-\gamma_1)+(-\gamma_2).$

Therefore we can apply the mutation method by simply mutating on the positive nodes in any order we like, until we arrive at a quiver with all nodes labelled by negative charges, indicating that we have completed the rotation $\mathcal{H}\rightarrow\mathcal{H}_\pi.$ It remains to be checked whether the ordering we have applied is actually physically realized in moduli space. We can dispense of this final check when the physical moduli space has  complex dimension equal to the number of nodes. Then as we move in moduli space, it is possible to tune all central charges of nodes however we wish. These theories are known as \emph{complete} theories, studied and classified in \cite{CV11}. In a companion paper \cite{ACCERV1} we studied the application of these techniques to the class of complete theories. In the more general case of non-complete theories, existence of the desired changer in the physical moduli space must be checked by hand.

\subsubsection{Quiver Mutation and Quantum Monodromy}

The mutation method outlined in the previous section can be extended to compute not only the BPS spectrum, but also the full Kontsevich-Soibelman (KS) quantum monodromy operator itself \cite{KS, CV11, CDZ}.  In this section we briefly discuss these techniques.

To implement the KS formalism one first introduces the quantum torus algebra.  Let $i$ index the nodes of the quiver, as discussed in detail in previous sections, these nodes integrally generate the lattice of BPS charges.  Then the quantum torus algebra is defined by:
\begin{itemize}
\item A generator $Y_{i}$ for each node of the quiver.
\item Commutation relations between the generators.
\begin{equation}
Y_{i}Y_{j}=q^{-\gamma_{i}\circ \gamma_{j}}Y_{j}Y_{i},
\end{equation}
where in the above, $q$ is a parameter.
\end{itemize}
Given a general charge $\gamma=\sum_{i}n_{i}\gamma_{i}$ we introduce the operator $Y_{\gamma}$ as a normal ordered product of the corresponding generators:
\begin{equation}
Y_{\gamma}\equiv N[ Y_{1}^{n_{1} }\cdot Y_{2}^{n_{2}}\cdots Y_{m}^{n_{m}}].
\end{equation}
The KS framework gives a characterization of the BPS spectrum in terms of a certain operator $M(q)$ which acts on the quantum torus algebra and is constructed as a product of certain quantum dilogarithm operators, $\Psi(Y_{\gamma},q)$ built form the $Y_{\gamma}$.  These operators act naturally on the quantum torus algebra by conjugation
\begin{equation}
Y_{\alpha}\rightarrow \Psi(Y_{\gamma},q) Y_{\alpha} \Psi(Y_{\gamma},q)^{-1}.
\end{equation}
Meanwhile, the operation of quiver mutation studied in the previous sections also acts on the algebra through its action on the charges at various nodes.  We let $\mu_{k}$ denote the operation on the charge lattice induced by quiver mutation at the $k$-th node.  The induced action on the generators $Y_{i}$ is then given in parallel to equations (3.2)-(3.3) as
\begin{equation}
\mu_{k}(Y_{i})=\begin{cases} Y_{k}^{-1} &\mathrm{if} \  i=k \\
Y_{i} & \mathrm{if} \ \gamma_{i}\circ \gamma_{k}>0 \\
Y_{\gamma_{i}+(\gamma_{k}\circ \gamma_{k})\gamma_{k}} & \mathrm{if} \  \gamma_{i}\circ \gamma_{k}<0
\end{cases}
\end{equation}
We can combine the action of conjugation by the quantum dilogarithm with quiver mutation to produce a quantum mutation operator which acts on the torus algebra
\begin{equation}
\mathcal{Q}_{k}=\mathrm{Ad}(\Psi(Y_{k},q))\circ \mu_{k}.
\end{equation}
The quantum mutation operator is the natural generalization of quiver mutation to the torus algebra.  Furthermore, just as ordinary quiver mutations, like those studied in the previous section, allow us to easily determine the BPS spectrum, the quantum mutation operator allows us to write the full quantum monodromy operator $M(q)$.  Specifically, in a chamber consisting of finitely many BPS states there exists a sequence of mutations which acts as the identity (up to a permutation of nodes) on the quiver $Q$
\begin{equation}
\label{muteq}
\mu_{k(s)}\cdots\mu_{k(2)}\mu_{k(1)}Q=Q.
\end{equation}
A key feature of this sequence is that it is phase ordered; the state $k(1)$ is left-most, the state $k(2)$ is next to left-most and so on.  Associated to this sequence is an ordered product of quantum mutation operators
\begin{equation}
\mathcal{Q}_{k(s)}\cdots\mathcal{Q}_{k(2)} \mathcal{Q}_{k(1)}.
\end{equation}
The above operator can be expressed in terms of the adjoint action of a single operator which is none other than the desired operator $M(q)$.  As a consequence of the fact that the original sequence of mutations in equation  \eqref{muteq} is phase ordered, the operator $M(q)$ has the desired expression in terms of a phase ordered product over the BPS states of quantum dilogarithm operators \cite{Keller08,CNV,Keller11}.  In this way we recover the full KS monodromy operator from ordered mutation sequences.

\section{$SU(2)$ Gauge Theories}
\label{su2examples}
We begin our study of examples with $SU(2)$ gauge theories. This is a natural starting point, as the BPS spectra of several of these theories have been worked out from different points of view \cite{BF1,BF2,BF3,GMN09}. We will reproduce those results straight-forwardly from the mutation method. These examples serve as non-trivial confirmation of our framework, and also as a demonstration of the power of the mutation method.


\subsection{Pure $SU(2)$}
The quiver for pure $SU(2)$ gauge theory has been worked out in various papers \cite{Fiol,Denef,DGS,CNV,CV11}. Here we will content ourselves to fix it based on the known $SU(2)$ spectrum, as was done in section \ref{quiverqm}, and then check that the mutation method produces the correct spectrum.

The quiver we are studying is given by
  \begin{equation}
  \begin{xy}
  (0,0)*{\Circle}="a"; (-7,-5)*{\gamma_1=(0,1)};
(30,0)*{\Circle}="c"; (37,-5)*{\gamma_2=(2,-1)};
  {\ar "a";"c" <1.5pt>}
    {\ar "a";"c" <-1.5pt>}
  \end{xy}
  \end{equation}
\noindent
The strong coupling chamber is given by $\arg\mathcal{Z}(\gamma_2)>\arg\mathcal{Z}(\gamma_1).$
As we rotate $\mathcal{H}$, we have the following sequence of mutations

\begin{equation}\begin{array}{ccc}

\begin{minipage}[b]{0.28\textwidth}
 \centering
 \[ 
 \begin{xy}
  (0,0)*{\Circle}="a"; (0,-5)*{\gamma_1};
  (30,0)*{\CIRCLE}="c"; (30,-5)*{\gamma_2};
  {\ar "a";"c" <1.5pt>}
  {\ar "a";"c" <-1.5pt>}
  \end{xy}
  \]
  \small{(i)}
\end{minipage}
 &
\begin{minipage}[b]{0.28\textwidth}
  \centering
  \[
  \begin{xy}
  (0,0)*{\CIRCLE}="a"; (0,-5)*{\gamma_1};
  (30,0)*{\Circle}="c"; (30,-5)*{-\gamma_2};
  {\ar "c";"a" <1.5pt>}
    {\ar "c";"a" <-1.5pt>}
  \end{xy}
  \]
  \small{(ii)}
\end{minipage}
 &
\begin{minipage}[b]{0.28\textwidth}
  \centering
  \[
  \begin{xy}
  (0,0)*{\Circle}="a"; (0,-5)*{-\gamma_1};
  (30,0)*{\Circle}="c"; (32,-5)*{-\gamma_2};
  {\ar "a";"c" <1.5pt>}
  {\ar "a";"c" <-1.5pt>}
  \end{xy}
  \]
  \small{(iii)}
\end{minipage}
\end{array}
\end{equation}
\noindent We see that we end with the antiparticle quiver, and that the only states in this chamber are $\gamma_1$ and $\gamma_2$. This agrees with the well known result that only the monopole and dyon are stable at strong coupling.

We can move on to do the same analysis at weak coupling, where $\arg\mathcal{Z}(\gamma_1)>\arg\mathcal{Z}(\gamma_2).$
\begin{equation}\begin{array}{ccc}\vspace{7mm} 
\begin{minipage}[b]{0.28\textwidth}
 \centering
 \[ 
\begin{xy}
  (0,0)*{\CIRCLE}="a"; (0,-5)*{\gamma_1};
(30,0)*{\Circle}="c"; (30,-5)*{\gamma_2};
  {\ar "a";"c" <1.5pt>}
    {\ar "a";"c" <-1.5pt>}
  \end{xy}
  \]
  \small{(i)}
\end{minipage}
 &
\begin{minipage}[b]{0.28\textwidth}
  \centering
  \[
 \begin{xy}
  (0,0)*{\Circle}="a"; (0,-5)*{-\gamma_1};
(30,0)*{\CIRCLE}="c"; (26,-5)*{2\gamma_1+\gamma_2};
  {\ar "c";"a" <1.5pt>}
    {\ar "c";"a" <-1.5pt>}
  \end{xy}
  \]
  \small{(ii)}
\end{minipage}
 &
\begin{minipage}[b]{0.28\textwidth}
  \centering
  \[
  \begin{xy}
  (0,0)*{\CIRCLE}="a"; (2,-5)*{3\gamma_1+2\gamma_2};
(30,0)*{\Circle}="c"; (28,-5)*{-2\gamma_1-\gamma_2};
  {\ar "a";"c" <1.5pt>}
    {\ar "a";"c" <-1.5pt>}
  \end{xy}
  \]
  \small{(iii)}
  \end{minipage}
\\  \raisebox{15mm}{$\cdots$} & \begin{minipage}[b]{0.28\textwidth}
  \centering
  \[
  \begin{xy}
  (0,0)*{\CIRCLE}="a"; (5,5)*{(k+1)\gamma_1+k\gamma_2};
(30,0)*{\Circle}="c"; (25,-5)*{-k\gamma_1-(k-1)\gamma_2};
  {\ar "a";"c" <1.5pt>}
    {\ar "a";"c" <-1.5pt>}
  \end{xy}
  \]
  \small{(k+1)}
  \end{minipage}
    & \raisebox{15mm}{$\cdots$}
\end{array}
\end{equation}
\noindent It is quite clear that we are in an infinite chamber. The entire sequence we'll find is obvious: we will have $(k+1)\gamma_1 + k\gamma_2$ for $k\ge 0$, with charge $(2k,1)$. In the $\mathcal{Z}$ plane these limit to the ray $\alpha \mathcal{Z}(\gamma_1 + \gamma_2)$. Notice that the $(e,m)$ charge of $\gamma_1+\gamma_2$ is $(2,0).$ We're finding the expected accumulation ray associated with the vector, the $W$ boson, in the weak coupling spectrum. In terms of rotating the half-plane, $W$ is protected from being a node because it is an accumulation ray of hypermultiplet dyons. In terms of the mutations, the ``quiver with $W$ as a node" is infinitely many mutations away in the space of dualities, preventing a contradiction. As mentioned above, this accumulation ray is blocking us from exploring the states lying in the rest of the central charge plane. We expect to only find one vector in the pure $SU(2)$ theory, but we have not yet found all the dyons. We would expect another set of dyons, $(2k,-1)$ which decompose as $k\gamma_1+(k+1)\gamma_2$ for $k\ge 0.$ These would all lie to the right of the $W$ boson, $\gamma_1+\gamma_2;$ thus we need some method for exploring that region of the $\mathcal{Z}$-plane.

In this case, and in any case where there is only a single accumulation ray, we can get around this problem easily. To do so, we recall that our mutation rule came from rotating the half-plane clockwise, $\mathcal{H}\rightarrow e^{-i\theta}\mathcal{H}$. We'll refer to this as left-mutation, because it is associated with states rotating out of the left of $\mathcal{H}$. There should of course be a similar mutation rule corresponding to rotating the half-plane counter-clockwise instead, $\mathcal{H}\rightarrow e^{i\theta}\mathcal{H}$, which we will call right-mutation. Both of these rules can be expressed as an action of a linear operator on the set of charges $\gamma_i$ which label the nodes of the quiver. If we call the usual left-mutation action on charges $M_L$, and the right-mutation action $M_R$, then we should have the obvious relations

\begin{equation}
M_L M_R = M_R M_L = Id_{\{\gamma\}}
\end{equation}

One can check that the transformation which satisfies the above identities (for $\gamma_1$ rotating out of $\mathcal{H}$) is simply

\begin{eqnarray}\label{mutbasis}
\widetilde{\gamma}_1 & = & -\gamma_{1}\\
\widetilde{\gamma}_j  & = & 
  \begin{cases}
   \gamma_{j}+( \gamma_{1} \circ   \gamma_{j})\gamma_{1} & \text{if }  \gamma_{1} \circ   \gamma_{j} >0 \\
   \gamma_{j}      & \text{if }   \gamma_{1} \circ \gamma_{j} \le 0.
  \end{cases}
\end{eqnarray}

Pictorially, mutation to the left (on node 1) acts non-trivially on those nodes which 1 points to, while right mutation acts non-trivially on nodes which point to 1. With this new rule in hand, we can start with the quiver and begin mutating ``to the right". Then we'll explore the BPS states starting from the right side of $\mathcal{H}$, as these are the ones leaving the half-plane. If there is only a single accumulation ray in $\mathcal{H}$, left and right-mutation together will allow us to explore both sides of it, filling out the entire $\mathcal{Z}$-plane except for the ray of the accumulation point.

Let's apply right mutation starting from the original $SU(2)$ quiver to find the remaining states. Now we use $\otimes$ to indicate the right-most node which will be right-mutated next.
\begin{equation}\begin{array}{ccc}\vspace{7mm} 
\begin{minipage}[b]{0.28\textwidth}
 \centering
 \[ 
\begin{xy}
  (0,0)*{\Circle}="c"; (0,-5)*{\gamma_1};
(30,0)*{\otimes}="a"; (30,-5)*{\gamma_2};
  {\ar "c";"a" <1.5pt>}
    {\ar "c";"a" <-1.5pt>}
  \end{xy}
  \]
  \small{(i)}
\end{minipage}
 &
\begin{minipage}[b]{0.28\textwidth}
  \centering
  \[
 \begin{xy}
  (0,0)*{\otimes}="c"; (2,-5)*{\gamma_1+2\gamma_2};
(30,0)*{\Circle}="a"; (30,-5)*{-\gamma_2};
  {\ar "a";"c" <1.5pt>}
    {\ar "a";"c" <-1.5pt>}
  \end{xy}
  \]
  \small{(ii)}
\end{minipage}
 &
\begin{minipage}[b]{0.28\textwidth}
  \centering
  \[
  \begin{xy}
  (0,0)*{\Circle}="c"; (2,-5)*{-\gamma_1-2\gamma_2};
(30,0)*{\otimes}="a"; (28,-5)*{2\gamma_1+3\gamma_2};
  {\ar "c";"a" <1.5pt>}
    {\ar "c";"a" <-1.5pt>}
  \end{xy}
  \]
  \small{(iii)}
  \end{minipage}
\\  \raisebox{15mm}{$\cdots$} & \begin{minipage}[b]{0.28\textwidth}
  \centering
  \[
  \begin{xy}
  (0,0)*{\otimes}="c"; (5,5)*{k\gamma_1+(k+1)\gamma_2};
(30,0)*{\Circle}="a"; (25,-5)*{-(k-1)\gamma_1-k\gamma_2};
  {\ar "a";"c" <1.5pt>}
    {\ar "a";"c" <-1.5pt>}
  \end{xy}
  \]
  \small{(k+1)}
  \end{minipage}
    & \raisebox{15mm}{$\cdots$}
\end{array}
\end{equation}
\noindent We have generated the states $k\gamma_1 + (k+1)\gamma_2 = (2k, -1)$. So mutation to the right obtains the dyons that we didn't see before, namely the ones lying on the other side of the vector. Since these states limit to the same ray in the $\mathcal{Z}$ plane, at $\mathcal{Z}(\gamma_1+\gamma_2)$, we have understood the stability of all states except those lying on this ray. To complete the analysis, in principle one should do the representation theory for states along the ray $\gamma_1+\gamma_2.$ At a generic choice of parameters, the only particles that may exist along this ray are of the form $n(\gamma_1+\gamma_2).$ \footnote{\label{nongeneric} This statement heavily relies on the fact that this theory is complete. If the central charges of nodes cannot be varied independently, and the theory is thus incomplete, then there are non-trivial relations satisfied by the central charges of nodes at all points of parameter space. For example, there may be a relation of the form $\gamma_k=\gamma_1+\gamma_2,$ satisfied for all parameter choices. Then the general particle at the ray $\mathcal{Z}(\gamma_1+\gamma_2)$ is of the form $n(\gamma_1+\gamma_2)+m\gamma_k.$ We will see how this may come about in section \ref{nf=1}.} It turns out that there is indeed a single vector present with the expected charge. This seems slightly obnoxious, because we still have to do some representation theory, but keep in mind that the work has been drastically reduced in that we only have to check for representations along this ray. 

\begin{figure}
\centering
\includegraphics[height=3in]{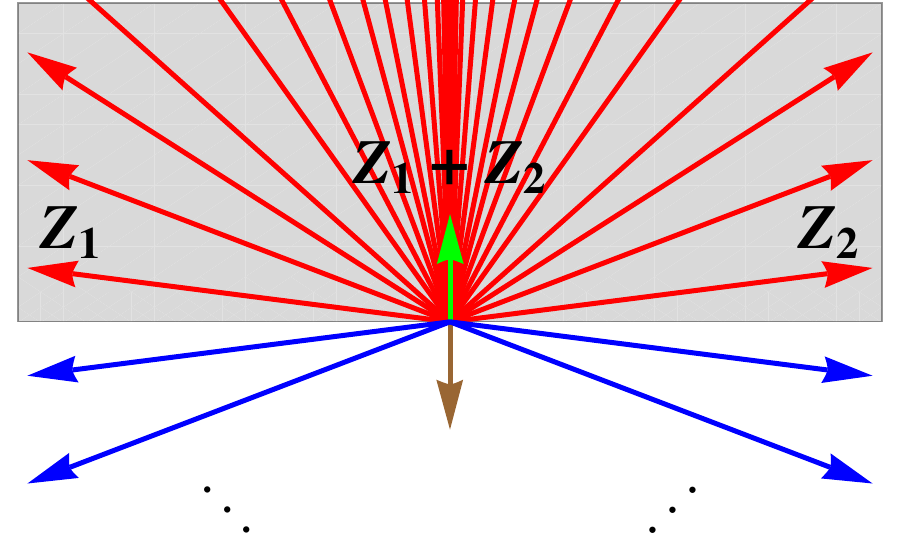}
\caption{The BPS spectrum of pure $SU(2)$ gauge theory, plotted in the central charge $\mathcal{Z}$-plane. The spectrum contains a vector state with charge $\mathcal{Z}_1+\mathcal{Z}_2$ (plotted in green), which is forced to occur in the $\mathcal{Z}$-plane at an accumulation ray of hypermultiplet states. On either side of the vector state, there is an infinite sequence of dyons whose central charges asymptotically approach the ray on which the vector lies. The mutation method is able to capture the full spectrum of the theory by rotating the half-plane to the left (yielding particles on the left of the vector particle) and the to right (yielding particles on the right of the vector particle).}
\label{godfigure}
\end{figure}

To summarize, we have found the strong coupling $SU(2)$ spectrum by a completely trivial application of the mutation method. For the weak coupling chamber, we introduced \emph{right-mutation} to be able to explore the central charge $\mathcal{Z}$-plane on both sides of the accumulation ray at the $W$ boson. Here we found, as expected, the $W$ boson and the infinite tower of dyons. In Figure \ref{godfigure}, we draw the spectrum in the $\mathcal{Z}$-plane to clarify how the mutation method is capable of obtaining all states of the theory.  The well-known resulting spectra are summarized in the table below:
\[
\begin{tabular}{rl||rl}
\multicolumn{2}{c||}{\bf{Strong coupling}} & \multicolumn{2}{c}{\bf{Weak coupling}} \\ \hline
Monopole: &$(0,1)$ &Positive dyons:& $(2n, 1)$ \\
Dyon: &$(2,-1)$&Negative dyons:& $(2n+2, -1)$\\ 
& & $W$ boson:& $(2,0,0)$
\end{tabular}
\]


\subsection{Adding matter}\label{su2matter}
The quiver of $SU(2)\;N_f=1$ was deduced using general considerations in \cite{CV11}. Here we simply recall their reasoning. We expect $2r+f=3$ nodes of the quiver. First we note that we can tune the mass of the quark to infinity. Then the massive quark fields should decouple from the theory, leaving the BPS states of pure $SU(2)$. This suggest that the quiver should consist of the pure $SU(2)$ quiver (with the usual monopole and dyon charges) along with an additional node. In the decoupled limit, there should be additional states with $(e,m)$ charges $(\pm 1,0);$ the third node should correspond to one of these two charges. However, we need to make the correct choice for third node that allows \emph{both} of these new states to be generated by positive linear combinations of the nodes. If we take $(1,0),$ all nodes of the quiver have positive electric charge, and the state $(-1,0)$ cannot be generated; the correct choice is then $(-1,0),$ which can be combined with the $W$ boson $(2,0)$ of the $SU(2)$ subquiver to form $(1,0)$. Computing electric-magnetic inner products, we find the following quiver:
 \begin{equation}\label{nf=1}\begin{split}  \begin{xy}
  (-10,-5)*{\gamma_1=(0,1)}; (15,20)*{\Circle}="a"; (15,25)*{\gamma_3=(-1,0)}; (0,0)*{\Circle}="b";
(30,0)*{\Circle}="c"; (40,-5)*{\gamma_2=(2,-1)};
  {\ar "a";"b"};{\ar "c";"a"};  {\ar "b";"c" <1.5pt>}
    {\ar "b";"c" <-1.5pt>}
  \end{xy}\end{split}
  \end{equation}

We can repeat this argument to add as many additional flavors as we like; the result is to produce $N_f$ copies of the node $\gamma_3$ with different flavor charges. 
\begin{equation}\label{arbnf}\begin{split}
  \begin{xy}
(0,20)*+{\gamma_4}="a";
(-15,20)*+{\gamma_3}="d";
(45,20)*+{\gamma_{N_f}}="f";
  (0,0)*+{\gamma_1}="b";
(30,0)*+{\gamma_2}="c";
(22.5,20)*{\dots};
  {\ar "a";"b"};{\ar "c";"a"};  {\ar "b";"c" <1.5pt>};
    {\ar "f";"b"};{\ar "c";"f"};  {\ar "d";"b"};{\ar "c";"d"};
    {\ar "b";"c" <-1.5pt>}
  \end{xy}\end{split}
\end{equation}

Alternatively, we can add hypermultiplet matter charged under other representations of the gauge group. If instead of a fundamental $\mathbf{2}$ of $SU(2)$ we consider a $\mathbf{j}$ rep of $SU(2),$ we find that $\gamma_3$ has charge $(-j,0).$ Generalizing our analysis above, we conclude that if a quiver description of this theory exists, it is given by a similar quiver with $j$ arrows $\gamma_3\rightarrow\gamma_1,\gamma_2\rightarrow\gamma_3$
\begin{equation}\label{spinj}\begin{split}
  \begin{xy}
  (-10,-5)*{\gamma_1=(0,1)}; (15,20)*{\Circle}="a"; (15,25)*{\gamma_3=(-j,0)}; (0,0)*{\Circle}="b";
(30,0)*{\Circle}="c"; (40,-5)*{\gamma_2=(2,-1)};
(26,10)*{j};
(5,10)*{j};
  {\ar "a";"b"};{\ar "c";"a"};  {\ar "b";"c" <1.5pt>}
    {\ar "b";"c" <-1.5pt>}
  \end{xy}\end{split}
 \end{equation}
Certainly this quiver can generate the full $\mathbf{j}$ representation, raising the electric charge by adding the $W$ boson. However, for $\mathbf{j}\ne \mathbf{2},$ it is possible that the quiver generates some additional representations of the gauge group. Indeed, it turns out that such a quiver will correspond to $SU(2)$ with a full $\otimes^j \mathbf{2}$ representation of the gauge group.  We will see an explicit occurrence of this in section \ref{n=2*}, where for $j=2$ the quiver above (\ref{spinj}) produces the matter representation $\mathbf{3} \oplus \mathbf{1}.$


\subsection{Massless $N_f=1$}\label{nf=1}
Recall that a single quiver from the mutation class generally does not cover all of moduli space. If we start with a valid quiver description and move in moduli space, it may be that at some point the central charges $\mathcal{Z}(\gamma_i)$ no longer lie in a common half-plane. We deduced the $SU(2)$ with matter quivers in the decoupling limit of infinite quark mass, so there is no reason to expect it to cover the chamber with the bare mass of the quark set to zero. Actually, one can easily see that the massless chamber should have $\mathcal{Z}(\gamma_1+\gamma_2-2\gamma_3)=0$ for the charges given in (\ref{nf=1}). Thus we have $\mathcal{Z}(\gamma_3) = -\frac{1}{2}\mathcal{Z}(\gamma_1)+\mathcal{Z}(\gamma_2))$. There is no way that the three central charges can lie in a single half-plane. 

It is easy to remedy this situation by properly applying mutations. Imagine beginning at a point of parameter space where the quiver above is valid. Then we consider tuning parameters until we reach the desired point. As we do this, we should keep track of any states leaving the half-plane, and perform the appropriate mutations. This sounds as though it involves detailed knowledge of the moduli space geometry, but that turns out to be completely unnecessary. There is no need to restrict our path to the physical parameter space; instead we are free to move throughout full space of central charges for the theory. In other words, we are free to pretend that the theory is complete as we tune parameters.\footnote{This theory actually \emph{is} complete; however, in any other non-complete examples, the same approach is valid.} This drastically simplifies the procedure. Now we may start with a valid quiver at a certain choice of parameters, and then tune the central charges one-by-one to produce the arrangement at the desired endpoint in parameter space.

For the $N_f=1$ quiver (\ref{nf=1}), let's keep $\gamma_1,\gamma_2$ fixed in the central charge plane, and tune $\gamma_3$ from its initial value within the half-plane by rotating it to the right. It will exit on the right, inducing a right-mutation on $\gamma_3.$ We should continue rotating $\gamma_3$ all the way to $\mathcal{Z}(\gamma_3)=-\mathcal{Z}(\gamma_1+\gamma_2),$ and keep track of mutations of the charges of the mutated quiver. In this case, no additional mutations occur. This gives\footnote{The monopole and dyon acquire flavor charges \cite{SW2}, which we now include in the charge labels.}
  \begin{equation}\begin{split}
  \begin{xy}
  (0,0)*{\Circle}="a"; (-10,-5)*{\gamma_1=(0,1,1/2)};
  (15,20)*{\Circle}="c"; (15,25)*{\gamma_3=(1,0,-1)};
  (30,0)*{\Circle}="b"; (40,-5)*{\gamma_2=(1,-1,1/2)};
  {\ar "a";"b"};{\ar "a";"c"};
  {\ar "c";"b"};
  \end{xy}\end{split}
  \end{equation}
The flavor group for $N_f=1$ is $SO(2)\cong U(1)$, so we label the charges of our states by their $U(1)$ charge $f$; the nodes then correspond to the electromagnetic and flavor charges $(e,m,f)$ as given above. At zero bare mass, the central charge function only depends on the electric and magnetic charges of the states, so the third node is constrained as $\mathcal{Z}(\gamma_3) = \mathcal{Z}(\gamma_1) + \mathcal{Z}(\gamma_2)$. Thus, just as in the pure $SU(2)$ theory, there are only two distinct chambers, one with $\text{arg }\mathcal{Z}(\gamma_1)> \text{arg }\mathcal{Z}(\gamma_2)$, and the other with $\text{arg }\mathcal{Z}(\gamma_2)> \text{arg }\mathcal{Z}(\gamma_1)$. This will turn out to be a feature of all the massless examples we consider.

  Let's start by exploring the chamber with $\mathcal{Z}(\gamma_2)$ ahead of $\mathcal{Z}(\gamma_1)$. We start by mutating on $\gamma_{2}$, after which we have the nodes $\gamma_1$, $\gamma_3$ and $-\gamma_2$. $\gamma_3$ is now left most, so we must mutate on it next, and so on.
\begin{equation}\begin{array}{cc}
    \begin{minipage}[b]{0.4\linewidth}
    \centering
    \[
    \begin{xy}
    (0,0)*{\Circle}="a"; (-5,0)*{\gamma_1};
    (15,20)*{\Circle}="c"; (15,25)*{\gamma_3};
    (30,0)*{\CIRCLE}="b"; (35,0)*{\gamma_2};
    {\ar "a";"b"};{\ar "a";"c"};
    {\ar "c";"b"};
    \end{xy}
    \]
    \small (i)
    \end{minipage}
 &
    \begin{minipage}[b]{0.4\linewidth}
    \centering
    \[
    \begin{xy}
    (0,0)*{\Circle}="a"; (-5,0)*{\gamma_1};
    (15,20)*{\CIRCLE}="c"; (15,25)*{\gamma_3};
    (30,0)*{\Circle}="b"; (36,0)*{-\gamma_2};
    {\ar "b";"a"};{\ar "a";"c"};
    {\ar "b";"c"};
    \end{xy}
    \]
    \small (ii)
    \end{minipage}
   \\
    \begin{minipage}[b]{0.4\linewidth}
    \centering
    \[
    \begin{xy}
    (0,0)*{\CIRCLE}="a"; (-5,0)*{\gamma_1};
    (15,20)*{\Circle}="c"; (15,25)*{-\gamma_3};
    (30,0)*{\Circle}="b"; (36,0)*{-\gamma_2};
    {\ar "b";"a"};{\ar "c";"a"};
    {\ar "c";"b"};
    \end{xy}
    \]
    \small (iii)
    \end{minipage}
   &
    \begin{minipage}[b]{0.4\linewidth}
    \centering
    \[
    \begin{xy}
    (0,0)*{\Circle}="a"; (-6,0)*{-\gamma_1};
    (15,20)*{\Circle}="c"; (15,25)*{-\gamma_3};
    (30,0)*{\Circle}="b"; (36,0)*{-\gamma_2};
    {\ar "a";"b"};{\ar "a";"c"};
    {\ar "c";"b"};
    \end{xy}
    \]
    \small (iv)
    \end{minipage}
    \end{array}
\end{equation}
  We see the only states in this chamber were the nodes of the original quiver and their antiparticles. We've discovered the strong coupling chamber of the $N_f = 1$ theory, whose spectrum indeed coincides with these hypermultiplets.

  Now let's explore the other chamber. Here we take $\mathcal{Z}(\gamma_1)$ ahead of $\mathcal{Z}(\gamma_2)$. We have the sequence
\begin{equation}\begin{array}{cc}
    \begin{minipage}[b]{0.4\linewidth}
    \centering
\[
  \begin{xy}
  (0,0)*{\CIRCLE}="a"; (-5,-5)*{\gamma_1};
  (15,20)*{\Circle}="c"; (15,25)*{\gamma_3};
  (30,0)*{\Circle}="b"; (35,-5)*{\gamma_2};
  {\ar "a";"b"};{\ar "a";"c"};
  {\ar "c";"b"};
  \end{xy}
  \]
  \small (i)
    \end{minipage}
    &
    \begin{minipage}[b]{0.4\linewidth}
    \centering
  \[
  \begin{xy}
  (0,0)*{\Circle}="a"; (-5,-5)*{-\gamma_1};
  (15,20)*{\CIRCLE}="c"; (15,25)*{\gamma_1+\gamma_3};
  (30,0)*{\Circle}="b"; (30,-5)*{\gamma_1+\gamma_2};
  {\ar "b";"a"};{\ar "c";"a"};
  {\ar "c";"b"};
  \end{xy}
  \]
  \small (ii)
    \end{minipage}
    \\
    \begin{minipage}[b]{0.4\linewidth}
    \centering
  \[
  \begin{xy}
  (0,0)*{\Circle}="a"; (-5,-5)*{\gamma_3};
  (15,20)*{\Circle}="c"; (15,25)*{-\gamma_1-\gamma_3};
  (30,0)*{\CIRCLE}="b"; (25,-5)*{2\gamma_1+\gamma_2+\gamma_3};
  {\ar "b";"a"};{\ar "a";"c"};
  {\ar "b";"c"};
  \end{xy}
  \]
  \small (iii)
    \end{minipage}
&
    \begin{minipage}[b]{0.4\linewidth}
    \centering
  \[
  \begin{xy}
  (0,0)*{\CIRCLE}="a"; (-5,-5)*{\gamma_1+\gamma_2+2\gamma_3};
  (15,20)*{\Circle}="c"; (15,25)*{\gamma_1+\gamma_2};
  (30,0)*{\Circle}="b"; (30,-5)*{-2\gamma_1-\gamma_2-\gamma_3};
  {\ar "a";"b"};{\ar "a";"c"};
  {\ar "c";"b"};
  \end{xy}
  \]
  \small (iv)
    \end{minipage}
    \end{array}
    \end{equation}
  We're clearly in an infinite chamber. Continuing in this way, we see our spectrum includes the states
  \begin{align*}
  (n+1)\gamma_1+n(\gamma_2 + \gamma_3) &= (2n,1,1/2)\\
  (n+1)(\gamma_1+\gamma_3)+n\gamma_2 &= (2n+1,1,-1/2)
  \end{align*}
As in the weak chamber of the pure $SU(2)$ theory, we are seeing the accumulation ray which should contain the $W$ boson of the theory. Here we are actually getting twice as many hypermultiplets as in pure $SU(2)$ since we have states of both even and odd electric charge. We will identify the odd electric charge states as quark-dyon bound states.

As before, let's start with the original quiver and mutate to the right to study the BPS states on the other side of the accumulation ray. This generates the states
  \begin{align*}
  n(\gamma_1+\gamma_3)+(n+1)\gamma_2 &= (2n+1,-1,1/2)\\
  n\gamma_1 + (n+1)(\gamma_2 +\gamma_3) &= (2n+2,-1,-1/2)
  \end{align*}
This sequence of states also accumulate at the same ray in the central charge plane; between these two sequences of infinities, the only central charges that can appear are proportional to $\mathcal{Z}(\gamma_1+\gamma_2)$. We might again expect that these dyons limit to a single vector in the central charge plane. We could attempt to test this hypothesis by actually doing the representation theory along this ray, but instead let's appeal to some physical reasoning to see why this is indeed wrong. Namely, we're in the weak coupling chamber of the $N_f = 1$ theory. We would expect that this theory indeed contains BPS states corresponding to the fundamental quark hypermultiplet, and at zero bare mass the central charge of this hyper lies directly at the same BPS phase as the $W$ boson. This is precisely the non-generic situation we hinted at in footnote \ref{nongeneric}.

Actually, given the non-genericity, something special has happened in this example. This quarks, given by $\gamma_3$ and by $\gamma_1+\gamma_2$, appeared as nodes after a finite sequence of mutations. Note that we never mutated on these quark nodes, because the nodes we mutate on are left-most (or right-most) and being on an accumulation ray, the quark can never be made left-most (or right-most). Instead, they simply appeared as one of the other ``interior" nodes in some of the dual quiver descriptions of the theory. This doesn't have to happen, and indeed won't happen in the undeformed $N_f=2,3$ cases below. We simply got lucky. If we hadn't seen the quark this way, we would have had to find it by hand. In either case, how can we be sure there are no other hypermultiplets lying on top of the vector, which aren't showing up as interior nodes elsewhere? One should consider a slightly deformed $N_f=1$ theory with $m\ne 0$ and check that there are no additional hypermultiplets (aside from those predicted by wall-crossing formulae). In this way, one can check that there are no additional hypermultiplets coinciding with the vector when $m\rightarrow 0$.  In principle, it is irrelevant whether or not the deformation we take is physically realized - thus, even in a non-complete theory, the same strategy works for understanding the particles along an accumulation ray. Alternatively, of course, one could always directly use quiver representation theory to rule out other states with that BPS phase.
 
Putting everything together, we find the following possible spectra for massless $N_f=1$.
\[
\begin{tabular}{rl||rl}
\multicolumn{2}{c||}{\bf{Strong coupling}} & \multicolumn{2}{c}{\bf{Weak coupling}} \\ \hline
 Quark:& $(1,0,-1)$ & Quarks:&$(1,0,\pm 1)$\\
Monopole:& $(0,1,1/2)$&Positive dyons:& $(2n,1,1/2) $  \\
Dyon:& $(1,-1,1/2)$&Negative dyons:& $(2n+2,-1, -1/2)$ \\&& Quark-dyons:& $(2n+1,\mp 1,\pm 1/2)$\\ && $W$ boson: & $(2,0,0)$
\end{tabular}
\]
where $n$ ranges over integers $n\ge 0.$
Along with their antiparticles, this collection agrees with the well known weak coupling spectrum of massless $SU(2)\; N_f=1$ (\cite{BF2}).


\subsection{Massive $N_f=1$}\label{massivespec}
For just one flavor, it is not too difficult to actually find all possible spectra of the theory with $m\ne 0$. It turns out that the acyclic quiver used in the previous subsection covers all chambers. Unfortunately, there is a great deal of redundancy in the full chamber spectrum - there are many distinct regions of moduli space that give the same spectrum due to dualities. By duality here, we mean the following: the spectrum depends only on the quiver and the central charges decorating the nodes, but not on the actual charge $(e,m,f)$ labels themselves. Thus, there may be widely separated regions of moduli space that happen to have the same quiver and associated central charges, but different charge labels; consistency of this framework requires that such regions actually have spectra that are equivalent up to some appropriate $Sp(2r,\mathbb{Z})$ relabeling of charges. Here we will simply list the possible spectra, without choosing a particular point in moduli space or duality frame; the downside is that as a result we cannot give the charges of the states, since charge labels require a choice of duality frame.

\begin{itemize}
\item Minimal chamber: 3 nodes are the only BPS states.
\item 4 state chamber: 3 nodes and 1 bound state hypermultiplet.
\item 5 state chamber: 3 nodes and 2 bound state hypermultiplet.
\item Weak coupling chambers, labelled by $k$. These consist of:
\begin{itemize}
\item 2 quark hypermultiplets,
\item $W$ boson vector multiplet,
\item Infinite tower of dyon hypermultiplets, 
\item $k$ additional quark-dyon bound state hypermultiplets, for $0\le k \le \infty.$
\end{itemize}
\end{itemize}
This list exhausts the BPS spectra that can be supported by quivers in this mutation class.
Embedded in this result are the two massless chambers, which correspond to the 3 state minimal chamber and the $k=\infty$ weak coupling chamber. It is a relatively straight-forward  exercise to find all these chambers beginning with the minimal massless spectrum, by repeated application of the pentagon and $SU(2)$ wall-crossing identities.


\subsection{Massless $N_f = 2$}
The relevant quiver for massless $N_f = 2$ follows from analogous mutations of the decoupling limit quiver (\ref{arbnf}) in section \ref{su2matter}. Here we find
  \begin{equation}
  \begin{xy}
  (0,0)*{\Circle}="d"; (-22,0)*{\gamma_1=(1,-1,0,-1/2)};
  (15,-15)*{\Circle}="a"; (20,-20)*{\gamma_4=(0,1,-1/2,0)};
  (30,0)*{\Circle}="b"; (50,0)*{\gamma_2=(1,-1,0,1/2)};
  (15,15)*{\Circle}="c"; (20,20)*{\gamma_3=(0,1,1/2,0)};
  {\ar "a";"d"};{\ar "a";"b"};
  {\ar "c";"d"};{\ar "c";"b"};
  \end{xy}
  \end{equation}
The flavor group is now $Spin(4)\cong SU(2)\times SU(2)$, and we
will denote our states by $(e,m,f_1,f_2)$, where $f_i$ are the charges
under the $U(1)$ contained in the $i$th $SU(2)$ factor. We see that there
are only two distinct values for the central charge between the four
nodes when the bare masses vanish. This means that there will again only be two chambers, given by the relative ordering of $\mathcal{Z}(\gamma_1)=\mathcal{Z}(\gamma_2), \mathcal{Z}(\gamma_3)=\mathcal{Z}(\gamma_4).$

There is a small added subtlety that was absent for $N_f = 1$. Namely, we technically can't rotate the central charge of a single node out of the half plane by itself. All mutations will happen for two nodes simultaneously. Also, as mentioned above, we don't get lucky in this example - the quarks don't show up as interior nodes of any of the quivers as we start mutating. If we mass deform the theory, however, the central charge of the quarks no longer coincides with the vector, and we will see them appear after a finite number of mutations. This tells us that there are the quark hypermultiplets lying on top of the vector when $m\rightarrow 0$, but no extra states. For simplicity, we will work out the $m=0$ point and quote this result.

For strong coupling, we first mutate on nodes $1$ and $2$, and find

\begin{equation}\begin{array}{c}
    \begin{minipage}[b]{0.4\linewidth}
    \centering
  \[
  \begin{xy}
  (15,-15)*{\Circle}="a"; (15,-20)*{\gamma_4};
  (30,0)*{\CIRCLE}="b"; (35,0)*{\gamma_2};
  (15,15)*{\Circle}="c"; (15,20)*{\gamma_3};
  (0,0)*{\CIRCLE}="d"; (-5,0)*{\gamma_1};
  {\ar "a";"d"};{\ar "a";"b"};
  {\ar "c";"d"};{\ar "c";"b"};
  \end{xy}
  \]
  \small (i)
    \end{minipage}
    \hspace{0.25cm}
    \begin{minipage}[b]{0.4\linewidth}
    \centering
  \[
  \begin{xy}
  (15,-15)*{\CIRCLE}="a"; (15,-20)*{\gamma_4};
  (30,0)*{\Circle}="b"; (36,0)*{-\gamma_2};
  (15,15)*{\CIRCLE}="c"; (15,20)*{\gamma_3};
  (0,0)*{\Circle}="d"; (-6,0)*{-\gamma_1};
  {\ar "d";"a"};{\ar "b";"a"};
  {\ar "d";"c"};{\ar "b";"c"};
  \end{xy}
  \]
  \small(ii)
    \end{minipage}
    \\
\begin{minipage}[b]{0.4\linewidth}
    \centering
    \[
  \begin{xy}
  (15,-15)*{\Circle}="a"; (15,-20)*{-\gamma_4};
  (30,0)*{\Circle}="b"; (36,0)*{-\gamma_2};
  (15,15)*{\Circle}="c"; (15,20)*{-\gamma_3};
  (0,0)*{\Circle}="d"; (-6,0)*{-\gamma_1};
  {\ar "a";"d"};{\ar "a";"b"};
  {\ar "c";"d"};{\ar "c";"b"};
  \end{xy}
  \]
  \small (iii)
\end{minipage}
\end{array}
\end{equation}

Thus we see that this chamber contains no bound states, and the only
states are hypermultiplets contributed by the nodes. We
have one hypermultiplet of electromagnetic charge $(1,-1)$ in the
$(\mathbf{1},\mathbf{2})$ rep of $SU(2)\times SU(2)$, and one of charge
$(0,1)$ in the $(\mathbf{2},\mathbf{1})$.

The other chamber is of course more interesting. We have the following
sequence of mutations:

\begin{equation}\begin{array}{c}
    \begin{minipage}[b]{0.4\linewidth}
    \centering
  \[
  \begin{xy}
  (15,-15)*{\CIRCLE}="a"; (15,-20)*{\gamma_4};
  (30,0)*{\Circle}="b"; (35,0)*{\gamma_2};
  (15,15)*{\CIRCLE}="c"; (15,20)*{\gamma_3};
  (0,0)*{\Circle}="d"; (-5,0)*{\gamma_1};
  {\ar "a";"d"};{\ar "a";"b"};
  {\ar "c";"d"};{\ar "c";"b"};
  \end{xy}
  \]
  \small{(i)}
    \end{minipage}
    \hspace{0.25cm}
    \begin{minipage}[b]{0.4\linewidth}
    \centering
  \[
  \begin{xy}
  (15,-15)*{\Circle}="a"; (13,-20)*{-\gamma_4};
  (30,0)*{\CIRCLE}="b"; (45,0)*{\gamma_2+\gamma_3+\gamma_4};
  (15,15)*{\Circle}="c"; (13,20)*{-\gamma_3};
  (0,0)*{\CIRCLE}="d"; (-14,0)*{\gamma_1+\gamma_3+\gamma_4};
  {\ar "d";"a"};{\ar "b";"a"};
  {\ar "d";"c"};{\ar "b";"c"};
  \end{xy}
  \]
  \small{(ii)}
    \end{minipage}
\\
\begin{minipage}[b]{0.4\linewidth}
\centering
  \[
  \begin{xy}
  (15,-15)*{\CIRCLE}="a"; (15,-20)*{\gamma_1+\gamma_2+2\gamma_3+\gamma_4};
  (30,0)*{\Circle}="b"; (45,0)*{-\gamma_2-\gamma_3-\gamma_4};
  (15,15)*{\CIRCLE}="c"; (15,20)*{\gamma_1+\gamma_2+\gamma_3+2\gamma_4};
  (0,0)*{\Circle}="d"; (-16,0)*{-\gamma_1-\gamma_3-\gamma_4};
  {\ar "a";"d"};{\ar "a";"b"};
  {\ar "c";"d"};{\ar "c";"b"};
  \end{xy}
  \]
  \small{(iii)}
\end{minipage}
\end{array}\end{equation}
Continuing in this way, we generate the states
\begin{align*}
n (\gamma_1+\gamma_2+\gamma_4)+(n+1)\gamma_3&= (2n,1,1/2,0)\\
(n+1) \gamma_4 + n(\gamma_1 + \gamma_2 + \gamma_3) &= (2n,1,-1/2,0)\\
(n+1)(\gamma_1+\gamma_3+\gamma_4)+n\gamma_2 &= (2n+1,1,0,-1/2)\\
(n+1)(\gamma_2+\gamma_3+\gamma_4)+n\gamma_1 &= (2n+1,1,0,1/2).
\end{align*}
\noindent
On the other hand, mutating to the right gives the states
\begin{align*}
n(\gamma_1+\gamma_3+\gamma_4)+(n+1) \gamma_2 &= (2n+1,-1,0,1/2)\\
n( \gamma_2 + \gamma_3 + \gamma_4) + (n+1)\gamma_1 &= (2n+1,-1,0,-1/2)\\
(n+1)(\gamma_1+\gamma_2+\gamma_4)+n\gamma_3 &= (2n+2,-1,-1/2,0)\\
n\gamma_4+(n+1)(\gamma_1+\gamma_2+\gamma_3) &= (2n+2,-1,1/2,0).
\end{align*}
These fill out dyons $(2n,\pm 1)$ in the
$(\mathbf{2},\mathbf{1})$ and quark-dyons $(2n+1,\pm 1)$ in the
$(\mathbf{1},\mathbf{2})$. Trapped between the
two infinite sequences we have the vector boson $\gamma_1+\gamma_2+\gamma_3+\gamma_4
= (2,0,0,0)$, which we identify as the $W$. The quarks also lie at the same BPS phase, and are given by $\gamma_2+\gamma_4,\gamma_1+\gamma_4,\gamma_2+\gamma_3,\gamma_1+\gamma_3.$

The two spectra are tabulated below, where we now assemble the states into representations of the full $SU(2)\times SU(2)$ with charges given as $(e,m)_{\mathbf{f_1},\mathbf{f_2}}$:
\[
\begin{tabular}{rl||rl}
\multicolumn{2}{c||}{\bf{Strong coupling}} & \multicolumn{2}{c}{\bf{Weak coupling}} \\ \hline
 Monopole:& $(0,1)_{\mathbf{2},\mathbf{1}}$ & Quarks:& $(1,0)_{\mathbf{2},\mathbf{2}}$\\
Dyon:& $(1,-1)_{\mathbf{1},\mathbf{2}}$& Positive dyons:& $(2n,1)_{\mathbf{2},\mathbf{1}}$  \\
& &Negative dyons:& $(2n+2,-1)_{\mathbf{2},\mathbf{1}}$\\&& Quark-dyons: & $(2n+1,\pm 1)_{\mathbf{1},\mathbf{2}}$\\ & &$W$ boson:& $(2,0)_{\mathbf{1},\mathbf{1}}$
\end{tabular}
\]
This agrees with the well known weak coupling spectrum of the $SU(2)$
$N_f=2$ theory.


\subsection{Massless $N_f = 3$}
The $N_f = 3$ quiver is given, after mutations to reach the massless chamber, as
  \begin{equation}
  \begin{xy}
  (0,0)*{\Circle}="a"; (-18,0)*{\gamma_5=(0,1,1,0,0)};
  (15,15)*{\Circle}="b"; (15,20)*{\gamma_2=(0,1,-1,1,0)};
  (30,0)*{\Circle}="c"; (45,5)*{\gamma_3=(0,1,0,-1,1)};
  (15,-15)*{\Circle}="d"; (15,-20)*{\gamma_4= (0,1,0,0,-1)};
  (15,0)*{\Circle}="e"; (33,-3)*{\gamma_1=(1,-2,0,0,0)};
  {\ar "a";"e"};{\ar "b";"e"};
  {\ar "c";"e"};{\ar "d";"e"};
  \end{xy}
  \end{equation}
The flavor group is $SO(6)\cong SU(4)$ and the nodes of the quiver make up
a monopole of electric/magnetic charge $(0,1)$ in the $\mathbf{4}$ of
$SU(4)$, and a dyon of charge $(1,-2)$ in the $\mathbf{1}$. We have labelled the flavor charges as $(e,m,q_1,q_2,q_3)$, where $q_i$ are the
eigenvalues under the respective generators of the Cartan of $SU(4)$. The central charge
degeneracy we experienced in the $N_f = 2$ case is again present, among $\gamma_i$ for $2\le i \le 5.$ Half the spectrum will come as sets of $4$ simultaneous mutations.

There are again two chambers, one with $\text{arg }\mathcal{Z}(\gamma_5)>\text{arg }\mathcal{Z}(\gamma_1)$, and the other with
$\text{arg }\mathcal{Z}(\gamma_1)>\text{arg }\mathcal{Z}(\gamma_5)$. The second chamber is strong coupling, and just includes
the particles that correspond to the original nodes of the quiver. In the other chamber, the mutations generate the spectrum
\begin{align*}
\gamma_i+n (\gamma_2+\gamma_3+ \gamma_4 + \gamma_5) + 2n \gamma_1 &= (2n,1,1,0,0)\\
(n+1)(\gamma_2+\gamma_3+\gamma_4+\gamma_5) + (2n+1)\gamma_1 &= (2n+1,2,0,0,0)\\
-\gamma_i+(n+1) (\gamma_2+\gamma_3+ \gamma_4 + \gamma_5) + (2n+1) \gamma_1 &= (2n+1,1,-1,0,0)
\end{align*}
The states in which $\gamma_i$ appears are repeated for $1\le i\le 4$. Thus we see that we have a magnetic charge $2$ dyon that is a singlet under flavor $SU(4)$, as well as magnetic charge $1$ dyons in the $\mathbf{\bar{4}}$ and quark-dyons in the $\mathbf{4}$. 

As usual, the mutations to the right will fill out the dyons on the other side of the accumulation ray. Right mutation generates:
\begin{align*}
n(\gamma_2+\gamma_3+\gamma_4+\gamma_5) + (2n+1)\gamma_1 &= (2n+1,-2,0,0,0)\\
\gamma_i + n (\gamma_2\gamma_3+ \gamma_4 + \gamma_5) + (2n+1) \gamma_1 &= (2n+1,-1,1,0,0)\\
-\gamma_i+(n+1) (\gamma_2+\gamma_3+ \gamma_4 + \gamma_5) + (2n+2) \gamma_1 &= (2n+2,-1,-1,0,0)
\end{align*}
The vector $W$ boson, is at  an accumulation ray, and the subtlety about generating the quarks is the same as in the $N_f=2$ case. Here the quarks are given by $\gamma_1+\gamma_i+\gamma_j,$ where $2\le i<j\le 5.$ 
\[
\begin{tabular}{rl||rl}
\multicolumn{2}{c||}{\bf{Strong coupling}} & \multicolumn{2}{c}{\bf{Weak Coupling}} \\ \hline
 Monopole:& $(0,1)_{\mathbf{4}}$ & Quarks:& $(1,0)_{\mathbf{6}}$\\
Dyon:& $(1,-2)_{\mathbf{1}}$& Positive dyons:& $(2n,1)_{\mathbf{4}}$\\
&& Negative dyons:& $(2n+2,-1)_{\mathbf{\bar{4}}}$ \\&& $m=2$ dyons:& $(2n+1,\pm 2)_\mathbf{1}$ \\ &&Quark-dyons: & $(2n+1,- 1)_{\mathbf{4}}$\\&&& $(2n+1,1)_{\mathbf{\bar{4}}}$\\ && $W$ boson:& $(2,0)_{\mathbf{1}}$
\end{tabular}
\]


\subsection{$N_f=4$}
For $N_f=4$ the massless theory is conformal; mass deformations break conformality. The quiver in the decoupling $m\rightarrow \infty$ limit is given as\footnote{Our analysis will break the $SO(8)$ flavor symmetry, so we supress all flavor data.}
\begin{equation}\begin{xy}
(30,0) *{\Circle} ="4",
(0,0) *{\Circle} ="5",
(-15,0) *+{\gamma_1=(2,-1)},
(45,0) *+{\gamma_2=(0,1)},
(-5,-25) *+{\gamma_5=(-1,0)},
(-5,25) *+{\gamma_3=(-1,0)},
(35,-25) *+{\gamma_6=(-1,0)},
(35,25) *+{\gamma_4=(-1,0)},
(-5,-20) *{\Circle} ="7",
(35,-20) *{\Circle} ="8",
(-5,20) *{\Circle} ="9",
(35,20) *{\Circle} ="10",
"5", {\ar"4"<1.5pt>},
"5", {\ar"4"<-1.5pt>},
"4", {\ar"7"},
"7", {\ar"5"},
"4", {\ar"8"},
"8", {\ar"5"},
"4", {\ar"9"},
"9", {\ar"5"},
"4", {\ar"10"},
"10", {\ar"5"},
\end{xy}\end{equation}
There are many additional subtleties in this BPS spectrum because it corresponds to a massive deformation of the conformal theory. In particular, there is no quiver that describes the $m\rightarrow 0$ limit; if we try to follow the strategy employed in the asymptotically free cases to trace the quiver from $m=\infty$ to $m=0,$ we find that any path goes through infinitely many mutations, preventing us from identifying a quiver for the $m=0$ chamber.

Nonetheless, we may take a finite mass and find various chambers in which the mutation method can successfully compute BPS spectra. The following is an example of a finite chamber of this theory, with the BPS states listed in decreasing order of BPS phase:
\begin{equation}
\gamma_3,\gamma_4,\gamma_{2},\gamma_1+\gamma_3+\gamma_4,\gamma_2+\gamma_5,\gamma_2+\gamma_6,\gamma_1+\gamma_3,\gamma_1+\gamma_4,\gamma_2+\gamma_5+\gamma_6,\gamma_1,\gamma_5,\gamma_6.
\end{equation}
This theory is complete, so, as previously discussed, this chamber must occur in physical moduli space.

In principle, the BPS spectrum can be worked out in all of moduli space by applying the KS wall crossing formula to this chamber. However, the spectrum in some regions of moduli space becomes extremely complicated. To give a general sense of this, we will describe some wall crossings in this theory, which were first studied in \cite{GMN09}.

Focus on the first three states, $\gamma_3,\gamma_4,\gamma_2.$ If we move $\gamma_2$ all the way to the left, we will produce $\gamma_2,\gamma_2+\gamma_3,\gamma_2+\gamma_4,\gamma_2+\gamma_3+\gamma_4,\gamma_3,\gamma_4.$ Separating the rest of the spectrum into similar consecutive sets of three, analogous wall crossings will produce a spectrum of 24 states.
\begin{align}
\notag &\gamma_2,\gamma_2+\gamma_3,\gamma_2+\gamma_4,\gamma_2+\gamma_3+\gamma_4,\gamma_3,\gamma_4,\\
\notag &\gamma_2+\gamma_5,\gamma_2+\gamma_6,\gamma_1+2\gamma_2+\gamma_3+\gamma_4+\gamma_5+\gamma_6,\gamma_1+\gamma_2+\gamma_3+\gamma_4+\gamma_6,\\ 
\notag & \hspace{.5in}\gamma_1+\gamma_2+\gamma_3+\gamma_4+\gamma_5,\gamma_1+\gamma_3+\gamma_4,\\
\notag & \gamma_2+\gamma_5+\gamma_6,\gamma_1+\gamma_2+\gamma_3+\gamma_5+\gamma_6,\gamma_1+\gamma_2+\gamma_4+\gamma_5+\gamma_6,\\
\notag &\hspace{0.5in} 2\gamma_1+\gamma_2+\gamma_3+\gamma_4+\gamma_5+\gamma_6,\gamma_1+\gamma_3,\gamma_1+\gamma_4, \\
& \gamma_5,\gamma_6,\gamma_1+\gamma_5+\gamma_6,\gamma_1+\gamma_5,\gamma_1+\gamma_6,\gamma_1.
\end{align}
Now we can produce various vectors by crossing states between the four sets of six; for example, $(\gamma_1+\gamma_3+\gamma_4)\circ(\gamma_2+\gamma_5+\gamma_6)=-2,$ so exchanging them will produce a tower of dyons and a vector $\gamma_1+\gamma_2+\gamma_3+\gamma_4+\gamma_5+\gamma_6=(-2,0)$, by the $SU(2)$ wall crossing identity. Similarly, exchanging $\gamma_3,\gamma_4$ with $\gamma_2+\gamma_5,\gamma_2+\gamma_6=$ will produce a vector $2\gamma_2+\gamma_3+\gamma_4+\gamma_5+\gamma_6=(-4,2)$ along with two dyon towers and four additional hypers; this is just the wall crossing of massless $SU(2),N_f=2.$ Two more vectors will be generated by this procedure, $2\gamma_1+\gamma_3+\gamma_4+\gamma_5+\gamma_6=(0,-2)$ and $\gamma_1-\gamma_2=(2,-2).$\footnote{To obtain this last vector, we must rotate the half-plane, allowing $\gamma_2$ to exit and mutating on $\gamma_2$ in the quiver.} These four vectors have non-trivial electric-magnetic inner products, and so additional wall crossing of the vectors will produce some highly complicated spectrum with infinitely many vectors.

One would expect such wild BPS behavior in the massless conformal limit, where conformal dualities produce some infinite set of vectors dual to the familiar $W$ boson. It is interesting to observe that this complicated structure begins to emerge even with finite mass, in regions of moduli space where the quiver description is perfectly valid.


\subsection{$\mathcal{N}=2^*$}\label{n=2*}
The $\mathcal{N}=2^*$ theory is a massive deformation of conformal $\mathcal{N}=4,$ where we give the adjoint hypermultiplet some non-zero mass. Alternatively, it is simply a gauge theory with a massive hypermultiplet charged under the adjoint of the gauge group. For $SU(2)$ this is given, following the discussion in section \ref{su2matter}, by the following quiver:

  \begin{equation}\label{n=2*quiv}\begin{split}
  \begin{xy}
  (0,0)*{\Circle}="a"; (-10,-4)*{\gamma_1=(0,1)};
  (15,20)*{\Circle}="c"; (15,24)*{\gamma_3=(-2,0)};
  (30,0)*{\Circle}="b"; (40,-4)*{\gamma_2=(2,-1)};
  {\ar "a";"b"<1.5pt>};{\ar "c";"a"<1.5pt>};
  {\ar "b";"c"<1.5pt>};  {\ar "a";"b"<-1.5pt>};{\ar "c";"a"<-1.5pt>};
  {\ar "b";"c"<-1.5pt>};
  \end{xy}\end{split}
  \end{equation}
  
As indicated in section \ref{su2matter}, this quiver indeed turns out to generate matter content of the full $\mathbf{2}\otimes\mathbf{2}=\mathbf{3}\oplus\mathbf{1}.$ Thus it gives the $\mathcal{N}=2^*$ theory plus an uncharged singlet hypermultiplet. In \cite{ACCERV1}, this quiver was obtained in studying the rank two Gaiotto theory on a torus with one puncture. We can understand this matter content from the point of view of \cite{Gaiotto}. We start with a pair of pants, corresponding to a half-hypermultiplet charged as a trifundamental under three $SU(2)$ flavor groups, represented by the three boundary components. Glueing together two boundary components of the pair of pants identifies the two $SU(2)$'s and gauges them. To form the punctured torus, we glue two legs together, producing an $SU(2)$ gauge group, and matter content $\mathbf{2}\otimes\mathbf{2}=\mathbf{3}\oplus\mathbf{1}.$
  
This fact can be checked from the BPS spectrum as follows. Consider the rep $\gamma_1+\gamma_2+\gamma_3$ of this quiver. This rep has charge $(0,0)$ meaning that it is a pure flavor state. For $\mathcal{N}=2^*$ we would expect such a hypermultiplet, corresponding to the state inside the $\mathbf{3}$ that is uncharged under the $U(1)\subset SU(2);$ if we add an uncoupled singlet, we would then expect this site of the charge lattice to be occupied by two BPS particles. Quiver representation theory finds the latter situation, as we now demonstrate.

The superpotential for this quiver was worked out in \cite{ACCERV1}. The result was
\begin{equation}
\mathcal{W}=X_{12}X_{23}X_{31}+Y_{12}Y_{23}Y_{31}+X_{12}Y_{23}X_{31}Y_{12}X_{23}Y_{31}.
\end{equation}
Here, $X_{ij},Y_{ij}$ correspond to the two maps between nodes $i,j$ in the representation. The resulting F-terms are of the form
\begin{align}
X_{23}X_{31}+Y_{23}X_{31}Y_{12}X_{23}Y_{31}&=0,\\
X_{12}X_{23}+Y_{12}X_{23}Y_{31}X_{12}Y_{23}&=0,\\
X_{31}X_{12}+Y_{31}X_{12}Y_{23}X_{31}Y_{12}&=0,\\
Y_{23}Y_{31}+X_{23}Y_{31}X_{12}Y_{23}X_{31}&=0,\\
Y_{12}Y_{23}+X_{12}Y_{23}X_{31}Y_{12}X_{23}&=0,\\
Y_{31}Y_{12}+X_{31}Y_{12}X_{23}Y_{31}X_{12}&=0.
\end{align} 
We are studying the rep $\gamma_1+\gamma_2+\gamma_3$, so all gauge groups are $U(1)$, and the bifundamental fields here are simply $1
\times 1$ matrices. In this example, we can solve the full equations by just truncating to the quadratic pieces and solving those, since setting the quadratic pieces to zero also sets the quintic terms to zero.\footnote{There is also a solution given by nontrivial cancellation between the quadratic and quintic terms. However, the resulting moduli space is non-compact, so its cohomology contains no normalizable forms, and as such it does not contribute to the particle spectrum}.
\begin{align}
X_{23}X_{31}&=0\\
X_{12}X_{23}&=0\\
X_{31}X_{12}&=0\\
Y_{23}Y_{31}&=0\\
Y_{12}Y_{23}&=0\\
Y_{31}Y_{12}&=0
\end{align} 
These will set two of the $X$'s and two of the $Y$'s equal to zero. We will focus on the two non-zero fields, $X_i,Y_j,$ with $i,j\in \{(12),(23),(31)\}.$ Before going on, we pause to consider what the possible moduli spaces may be. For any choice of $i,j,$ there is enough gauge symmetry to set both $X_i,Y_j$ to one; thus the moduli space is at most 9 points, one for each choice of $(i,j).$ Some of these points will be eliminated by the stability analysis. Note that $\Pi$-stability does not distinguish between $X,Y,$ so if $X_i,Y_j\ne0$ is stable, then $X_j,Y_i\ne0$ is also stable. We will show below that the stability analysis always yields a moduli space of 2 points.

The simplest way to proceed is a case-by-case analysis of the possible orderings of central charges. For each choice of orderings, we will consider the following cases of $(i,j)$: (a) $(12,23)$, (b) $(23,31)$, (c) $(31,12),$ (d) $(12,12),$ (e) $(23,23),$ (f) $(31,31).$  There are three more cases obtained by exchanging $(i,j).$ A simple study of commutative diagrams shows that, for (a) the subreps are $\gamma_3,\gamma_2+\gamma_3.$ By cyclic symmetry, (b) has subreps  $\gamma_1,\gamma_3+\gamma_1,$ and for (c),  $\gamma_2,\gamma_1+\gamma_2.$ For (d) we find subreps $\gamma_2,\gamma_3,\gamma_2+\gamma_3,\gamma_1+\gamma_2;$ (e) and (f) have subreps given by cyclic symmetry. We can choose $\gamma_1$ to be the left-most node without loss of generality. Automatically, (e) and (f) are unstable due to the subrep $\gamma_1$ which has $\arg\mathcal{Z}(\gamma_1)>\arg\mathcal{Z}(\gamma_1+\gamma_2+\gamma_3).$ Suppose $\arg\mathcal{Z}(\gamma_1)>\arg\mathcal{Z}(\gamma_2)>\arg\mathcal{Z}(\gamma_3).$ Then rep (b) is destabilized by subrep $\gamma_1,$ and reps (c,d) are destabilized by subrep $\gamma_1+\gamma_2.$ Rep (a), on the other hand, is stable since its subreps have $\arg\mathcal{Z}(\gamma_1+\gamma_2+\gamma_3)>\arg\mathcal{Z}(\gamma_2+\gamma_3)>\arg\mathcal{Z}(\gamma_3).$ So here the moduli space is 2 points, $X_{12},Y_{23}\ne 0$ and $X_{23},Y_{12}\ne 0.$ Next, we consider $\arg\mathcal{Z}(\gamma_1)>\arg\mathcal{Z}(\gamma_1+\gamma_2)>\arg\mathcal{Z}(\gamma_3)>\arg\mathcal{Z}(\gamma_2).$ Rep (a) is again stable, while rep (b) is destabilized by $\gamma_1$ and reps (c,d) are destabilized by $\gamma_1+\gamma_2.$ The final case we must study is $\arg\mathcal{Z}(\gamma_1)>\arg\mathcal{Z}(\gamma_3)>\arg\mathcal{Z}(\gamma_1+\gamma_2)>\arg\mathcal{Z}(\gamma_2).$ Now we find that rep (c) is stable, while reps (a,d) are destabilized by $\gamma_3$ and rep (b) is destabilized by $\gamma_1.$ The conclusion is that the moduli space of the rep $(\gamma_1+\gamma_2+\gamma_3)$ is simply two points for any choice of parameters. Therefore, at all values in the parameter space of this theory, we find \emph{two} hypermultiplets with no electric-magnetic charge. This confirms that the quiver is describing the Gaiotto construction, $\mathcal{N}=2^*$ plus a single uncharged hypermultiplet.

The spectrum of this theory is extremely intricate for any chamber of the moduli space. We will demonstrate the existence of at least two vector particles for any choice of central charges. Without loss of generality, we take $\gamma_1$ to be leftmost. Then we should consider two cases. If $\arg \mathcal{Z}(\gamma_1)>\arg\mathcal{Z}(\gamma_2)>\arg\mathcal{Z}(\gamma_3),$ then the $\Pi$-stability analysis yields $\gamma_1+\gamma_2=(2,0)$ and $\gamma_1+\gamma_3=(-2,1)$ as stable vector particles. Alternatively, if $\arg \mathcal{Z}(\gamma_1)>\arg\mathcal{Z}(\gamma_3)>\arg\mathcal{Z}(\gamma_2),$ then  $\gamma_1+\gamma_2$ is a stable vector particle, along with either $(n+1)\gamma_1+n\gamma_2+\gamma_3$ or $n\gamma_1+(n+1)\gamma_2+\gamma_3$ for some choice of $n.$ In any of the cases, the two vector particles identified have non-zero electric-magnetic inner product. Consequently, the stable vector states could form a highly complicated spectrum of bound states.  The presence of multiple accumulation rays (one at each vector) obstructs the mutation method as defined from producing an unambiguous result for the spectrum. We can use left and right mutation to identify some set of dyons, along with the left-most and right-most vector states; however, the region of the $\mathcal{Z}$-plane between the two vectors could be arbitrarily wild. It would be interesting to try to develop an extension of the algorithm capable of computing the spectrum for this theory.


\subsection{Flavor Symmetries and Gauging}\label{su2gauge}
The above $SU(2)$ examples involve a well-known $SO(2N_f)$ flavor symmetry at the massless point of parameter space. In fact, the quivers used in the analysis all display quite suggestive symmetries themselves. In this section we will study the relationship between global symmetries of the physical theory and discrete symmetries of the quiver. This will turn out to provide a powerful tool for constructing quivers for new theories by gauging global symmetries.

Suppose a physical theory has some known global symmetry. Generally speaking, turning on various deformations of the theory will break the global symmetry, so here we consider studying the theory at the precise point of parameter space that preserves the full global symmetry of interest. Of course, the BPS spectrum should reflect this symmetry. The first question we wish to explore is how this symmetry should be encoded in the BPS quiver. 

It is possible that every state in the BPS spectrum might be singlet under the global symmetry; then it would be very difficult to find evidence for the symmetry in either the quiver or the full BPS spectrum. So we should refine the question a bit. Let us restrict to a global $SU(n)$ symmetry, and further, let us study the case in which there is some BPS hypermultiplet in the fundamental of $SU(n).$  In this case we can give a very straightforward answer to the question. The full fundamental multiplet of BPS states must have identical central charges. We simply choose our quiver half-plane so that this multiplet is left-most in the $\mathcal{Z}$-plane.\footnote{This choice of half-plane will be impossible when the phase of central charge of the fundamental of hypermultiplets occurs at some accumulation ray of BPS states. In fact, this exact situation occurs in the case of $SU(2),\,\mathcal{N}=2^*.$ This theory has an enhanced $SU(2)$ flavor symmetry at the massless point. However, we are never able to see the symmetry in the quiver (which has a single mutation form, given in section (\ref{n=2*quiv})). The massless theory is conformal, and the spectrum is dense; hence there is no half-plane that admits a positive integer basis. Barring this complication, there exists a half-plane that yields a mutation form of the quiver which explicitly presents the symmetry.} Since they carry distinct flavor charges spanning the weight space, all $n$ states of fundamental must occur in the quiver.\footnote{The weight space is only $n-1$-dimensional, so one may worry that only $n-1$ of the states appear. However, the weights obey $\sum_i f_i=0$ so that the last weight is given by a \emph{negative} integer linear combination of the others. As long as the multiplet carries some non-zero electric-magnetic charge, the last state be linearly independent from the others. Then, to fill out the $n$ states of the fundamental, all $n$ states must appear in the quiver.} These states of course have different global charges, but identical electric-magnetic charges. Since the quiver is only sensitive to electric-magnetic charges, we will find $n$ identical nodes in the resulting quiver, and thus an $S_n$ permutation symmetry that exchanges these identical nodes.

The above $SU(2)$ examples with massless matter illustrate this fact. For $N_f=2,$ we have an $SO(4)=SU(2)\times SU(2)$ flavor symmetry, which manifests as two $S_2$ discrete symmetries in the quiver, given by exchanging $\gamma_1,\gamma_2$ and $\gamma_3,\gamma_4.$ For $N_f=3$, we have an $SO(6)=SU(4)$ symmetry, manifested as an $S_4$ on $\gamma_1,\gamma_2,\gamma_3,\gamma_4.$ For $N_f=4$, there should be a full $SO(8)$ flavor symmetry; however, it is only preserved at the massless conformal point, where we have no quiver description. For any mass deformation, the maximal symmetry is $SU(4),$ which corresponds to the obvious $S_4$ acting on $\gamma_3,\gamma_4,\gamma_5,\gamma_6.$

Alternatively, suppose we start with a quiver containing $n$ identical nodes and an $S_n$ symmetry. If we assign identical charges to these nodes, the resulting BPS spectra will be forced to organize into representations of $SU(n),$ because the quiver representation theory does not distinguish among these $n$ identical nodes. The nodes themselves will form a multiplet in the fundamental representation, while bound states involving combinations of the identical nodes will form various tensor representations. Unfortunately, we cannot conclude from this that the full theory preserves this symmetry - perhaps is it is preserved by the BPS states, but broken by some non-BPS states. Nonetheless, if we are expecting an $SU(n)$ global symmetry, it is quite natural to identify it with this discrete symmetry of the quiver.

From these observations, we can suggest a powerful rule for constructing quivers of new theories by gauging global symmetries of a theory with a known quiver. For now, let us focus on gauging a global $SU(2)$ symmetry that is manifested as an $S_2$ symmetry in the quiver acting on a pair of identical nodes. We will extend to general $SU(n)$ after we have discussed quivers of more general gauge theories. Physically, to gauge a symmetry, we add gauge degrees of freedom and couple them appropriately to the matter already present in the theory. At the level of the quiver, the procedure is quite analogous. We should add two nodes of an $SU(2)$ subquiver to add the gauge degrees of freedom. Then we must couple to the existing pair of identical nodes to this subquiver to form a fundamental of the $SU(2).$ Recall that when we added a flavor to $SU(2)$, we added only one state of the doublet fundamental representation, because bound states would generate the second. Here we must do the same thing - we delete one of the nodes, and connect the other to the $SU(2)$ subquiver in an oriented triangle. The deleted state will now be generated by a bound state with the $SU(2)$ nodes.

To give an example, we can consider gauging one of the $SU(2)$ flavor symmetries of $SU(2),\,N_f=2,$ which exchanges $\gamma_1,\gamma_2.$
  \begin{equation}
  \begin{xy}
  (0,0)*+{\gamma_1}="d";
  (10,10)*+{\gamma_3}="a";
  (20,0)*+{\gamma_2}="b";
  (10,-10)*+{\gamma_4}="c";
  {\ar "a";"d"};{\ar "a";"b"};
  {\ar "c";"d"};{\ar "c";"b"};
  (30,0) *{\stackrel{\mathrm{gauge}}{\Longrightarrow}};
  (40,10)*+{\gamma_3}="e";
  (50,0)*+{\gamma_2}="f";
  (40,-10)*+{\gamma_4}="g";
  (60,-10)*+{b}="1";
  (60,10)*+{c}="2";
  {\ar "e";"f"};
  {\ar "g";"f"};
  {\ar "1";"2" <1.5pt>};{\ar "1";"2" <-1.5pt>};
  {\ar "2";"f"};{\ar "f";"1"}
  \end{xy}
  \end{equation}
We have added an $SU(2)$ subquiver $b,c$ and charged the flavor node $\gamma_2$ under it; now we have two $SU(2)$ gauge groups with a bifundamental matter field. In this case, we can actually see the weak coupling description of the resulting theory from the quiver, if we apply some mutations. Mutating on $\gamma_1,\gamma_2,b,c$ in that order produces
  \begin{equation}
  \begin{xy}
  (40,10)*+{\Circle}="e";
  (50,0)*+{\Circle}="f";
  (40,-10)*+{\Circle}="g";
  (60,-10)*+{\Circle}="1";
  (60,10)*+{\Circle}="2";
  {\ar "e";"f"};
  {\ar "f";"g"};
  {\ar "1";"2" <1.5pt>};{\ar "1";"2" <-1.5pt>};
  {\ar "g";"e" <1.5pt>};{\ar "g";"e" <-1.5pt>};
  {\ar "2";"f"};{\ar "f";"1"}
  \end{xy}
  \end{equation}
in which there are two $SU(2)$ subquivers, each coupled to the same node as a fundamental matter state, producing a bifundamental.

This gauging procedure can be understood very nicely from the perspective of the Gaiotto curve \cite{Gaiotto}. That work studied the conformal $\mathcal{N}=2$ theories that arise from wrapping stacks of $n$ M5-branes on some punctured Riemann curve known as the Gaiotto curve; $n$ is denoted the \emph{rank} of the theory. The punctures correspond to mass deformations of the theory; an exactly conformal theory would have all punctures turned off. In the case of two M5-branes, the resulting theories have gauge group $SU(2)^k.$ We briefly recall the map between the Gaiotto curve and the weak coupling gauge theory description for the rank 2 case. Each puncture of the Gaiotto curve corresponds to an $SU(2)$ flavor symmetry. Such Riemann surfaces may be glued together at punctures by opening a hole at each puncture and glueing the two together with a tube. This results in gauging the diagonal subgroup of the $SU(2)$'s corresponding to the punctures. The sphere with three punctures corresponds to a half-hypermultiplet trifundamental under the three $SU(2)$'s associated to the three punctures. Then from the pair-of-pants decomposition of a Riemann surface, we can break any surface into some number of three-punctured spheres connected up in some way.  From this, we may determine a weak coupling description of any such theory. Since the pair-of-pants decomposition is non-unique, there may be many different weak coupling descriptions; these are precisely the $\mathcal{N}=2$ dualities studied in \cite{Gaiotto}. For our purposes, we simply want to note that this glueing procedure can be translated to the quiver gauging rule at the level of the quiver, if we can identify the appropriate $S_2$ symmetries in some mutation forms of each quiver. Then the quiver of the glued surface is precisely the quiver obtained by simultaneously gauging the $S_2$ symmetries in the two quivers. That is, we add an $SU(2)$ subquiver, remove one of each pair of identical nodes in the two quivers, and couple both of the remaining nodes to the same $SU(2)$ subquiver.

As another example, consider glueing the $SU(2),\,N_f=4$ quiver to itself other by gauging the diagonal subgroup $SU(2)_d\subset SU(2)\times SU(2)\subset SU(4).$ The original quiver presents $S_2\times S_2\subset S_4$ symmetries given by exchanging $\gamma_3,\gamma_4$ and $\gamma_5,\gamma_6$ respectively. The gauging procedure looks as follows
\begin{equation}\begin{xy}
(20,0) *+{\gamma_2} ="4",
(0,0) *+{\gamma_1} ="5",
(5,-15) *+{\gamma_3} ="7",
(15,-15) *+{\gamma_4} ="8",
(5,15) *+{\gamma_5} ="9",
(15,15) *+{\gamma_6} ="10",
"5", {\ar"4"<1.5pt>},
"5", {\ar"4"<-1.5pt>},
"4", {\ar"7"},
"7", {\ar"5"},
"4", {\ar"8"},
"8", {\ar"5"},
"4", {\ar"9"},
"9", {\ar"5"},
"4", {\ar"10"},
"10", {\ar"5"},
(30,0) *{\stackrel{\mathrm{gauge}}{\Longrightarrow}},
(40,0)*+{\gamma_1}="15",
(60,0)*+{\gamma_2}="14",
(80,0)*+{b}="17",
(100,0)*+{c}="16",
(70,15)*+{\gamma_5}="11",
(70,-15)*+{\gamma_3}="12",
"15", {\ar"14"<1.5pt>},
"15", {\ar"14"<-1.5pt>},
"17", {\ar"16"<1.5pt>},
"17", {\ar"16"<-1.5pt>},
"16",{\ar"12"},
"12",{\ar"17"},
"16",{\ar"11"},
"11",{\ar"17"},
"14",{\ar"12"},
"12",{\ar"15"},
"14",{\ar"11"},
"11",{\ar"15"},
\end{xy}\end{equation}

For these rank 2 theories, there is actually a more systematic way to generate quivers for all surfaces via triangulations from special lagrangian flows, as developed in \cite{CV11,ACCERV1}. The quiver gauging rule just described can in fact be understood from this triangulation view point, as explained in \cite{CV11}. For example, the theory $SU(2),\,N_f=4$ corresponds to a sphere with 4 punctures; the gauged quiver shown above is known from that analysis to correspond to a torus with 2 punctures, which is precisely the surface produced after glueing two punctures from the 4-punctured sphere. Notice that, since the resulting surface contains 2 punctures, we would expect there to be two more $SU(2)$'s available for gauging. In fact, a mutation sequence can produce one $S_2$ in the quiver, but there is no way to produce two such symmetries. The analysis from the triangulation perspective shows that we can produce all but one $S_2$ in the quiver; that is, we can realize one fewer $S_2$ than the total number of punctures. Actually, there is a very good reason that we are unable to gauge the last $SU(2).$ If we did so, we would remove all punctures from the surface, and produce a quiver for a punctureless surface. However, a punctureless surface supports an exactly conformal theory - all mass deformations have been turned off. Hence the BPS spectrum would exhibit some duality, and in general be dense  in the central charge plane, obstructing the existence of a quiver. Thus for consistency, it is necessary that we not be able to gauge the $SU(2)$ symmetry of a once-punctured surface. Nonetheless, we can be able to build up a quiver for any surface with at least one puncture, and these all agree with the quivers obtained from triangulations. For higher rank theories, the analog of the triangulation approach is not known; however, the gauging rules will allow us to construct quivers for a large class of theories whose quiver descriptions were previously unknown.


\section{$SU(N)$ Gauge Theories and Beyond}
In this section we apply the formalism discussed in the previous section to the examples of non-abelian ADE Yang-Mills theories with matter.

  
\subsection{Construction of $SU(N)$ Quivers} \label{sunweak}
Quivers for pure $SU(N)$ gauge theory were constructed in \cite{CNV} via the $2d/4d$ correspondence studied there. These BPS quivers have also been studied previously in \cite{Fiol}. That work identified as nodes of the quiver a set of fractional branes in an orbifold phase of the geometries used in the type IIA geometric engineering \cite{KKV,KaMV}.\footnote{Fractional branes as a basis of BPS quivers were studied in \cite{Dia99,DFR1,DFR2}. Their charges for $SU(N)$ were identified from a boundary CFT analysis in \cite{Lerche00}. BPS particles with magnetic and electric charge in the IIA geometric engineering context correspond to even branes wrapped on cycles of the geometry. The fractional branes are identified with the monopoles and dyons which can become massless somewhere in moduli space, equivalently these states correspond to the vanishing cycles in the homology lattice of the Seiberg-Witten curves of these theories found in refs. \cite{KLYT,AF,KLT}. See also \cite{Lerche96} and references therein.} 

 Here we will provide a purely $4d$ motivation for that result, and use it to extend the proposal to $SU(N)$ gauge theory with arbitrary matter.  First we fix some notation. We have been using $(e,m)$ for electric and magnetic charges. Electric charges will naturally be associated to weights of the gauge group, and magnetic charges associated to roots. We denote simple roots $\alpha_i$ and fundamental weights $\omega_i;$ the appropriate inner product is given by $\alpha_i\cdot\omega_j=\delta_{ij}.$

By the $2r+f$ counting, the quiver should consist of $2(N-1)$ nodes. Let us consider the mutation form of the quiver that covers the decoupling limits in which each $W$ boson associated to a simple root $\alpha_i$ separately becomes infinitely massive. In order to separately decouple these vectors, the $N-1$ simple root $W$ bosons must be disjointly supported as reps of the quiver. Since the reps supported on only one node cannot give vectors, and we only have $2(N-1)$ nodes, each $W$ boson must be supported on two distinct nodes. So we have two nodes $b_i,c_i,$ forming an $SU(2)$ subquiver associated to each simple root. Then we simply need to choose charge assignments within the $SU(2)$ subquivers. In order to obtain the associated $W$ boson, the two nodes should have the charges of a consecutive pair of dyons, $((n_i+1)\alpha_i,-\alpha_i),(-n_i\alpha_i,\alpha_i).$ The most obvious choice is just $n_i=0$, the appropriate monopole and dyon for each simple root. If we make this choice, the result is precisely the quiver computed by \cite{CNV} using the 2d/4d correspondence:
\begin{equation}\label{sunweakquiv}\begin{split}\begin{xy} 0;<1pt,0pt>:<0pt,-1pt>:: 
(150,3)*+{\;\;} ="12",
(150,48)*+{\;\;}="13",
(200,4)*+{\;\;} ="14",
(200,48)*+{\;\;}="15",
(175,25) *+{\dots},
(250,0) *+{c_{N-1}} ="0",
(250,50) *+{b_{N-1}} ="1",
(100,0) *+{c_2} ="2",
(100,50) *+{b_2} ="3",
(50,0) *+{c_1} ="4",
(50,50) *+{b_1} ="5",
"1", {\ar"0" <1.5pt>},
"1", {\ar"0" <-1.5pt>},
"3", {\ar"2" <1.5pt>},
"3", {\ar"2" <-1.5pt>},
"2", {\ar"5"},
"4", {\ar"3"},
"2",{\ar"13"},
"12",{\ar"3"},
"14",{\ar"1"},
"0",{\ar"15"},
"5", {\ar"4"<1.5pt>},
"5", {\ar"4"<-1.5pt>},
\end{xy}\end{split}\end{equation}
where $b_i=(0,\alpha_i)$ and $c_i=(\alpha_i,-\alpha_i).$

The $SU(N)$ quivers we have deduced contain closed oriented cycles; thus the quiver requires a superpotential to be specified. The orbifold construction of \cite{Fiol} produces this superpotential by reducing the superpotential of the $\mathcal{N}=4$ theory.\footnote{The quiver (and superpotential) discussed on \cite{Fiol} is actually related by some mutations to the quiver we study here.}  Explicitly, the appropriate superpotential is given as, 
\begin{equation}\begin{split}\begin{xy} 0;<1pt,0pt>:<0pt,-1pt>:: 
(80,80) *+{c_2} ="2",
(80,00) *+{b_2} ="3",
(0,0) *+{c_1} ="4",
(0,80) *+{b_1} ="5",
(160,0)*+{\;}="6",
(160,80)*+{\;}="7",
(240,0)*+{\;}="8",
(240,80)*+{\;}="9",
(320,0) *+{c_{N-1}} ="10",
(320,80) *+{b_{N-1}} ="11",
(-10,40) *+{X_1},
(10,40) *+{Y_1},
(90,40) *+{Y_2},
(70,40) *+{X_2},
(340,40) *+{Y_{N-1}},
(300,40) *+{X_{N-1}},
(120,70)*+{\phi_{2}},
(120,-10)*+{\phi'_2},
(280,70)*+{\phi_{N-2}},
(280,-10)*+{\phi'_{N-2}},
(200,40)*+{\dots},
(40,70)*+{\phi'_{1}},
(40,-10)*+{\phi_1},
"3", {\ar"2" <1.5pt>},
"3", {\ar"2" <-1.5pt>},
"2", {\ar"5"},
"4", {\ar"3"},
"5", {\ar"4"<1.5pt>},
"5", {\ar"4"<-1.5pt>},
"6",{\ar"3"},
"2",{\ar"7"},
"11", {\ar"10"<1.5pt>},
"11", {\ar"10"<-1.5pt>},
"10",{\ar"8"},
"9",{\ar"11"}
\end{xy}\end{split}\end{equation}
with \begin{equation}\label{SUNpot}
\mathcal{W}=\sum_{i=1}^{N-2} X_i\phi'_i X_{i+1}\phi_i-Y_i\phi'_i Y_{i+1}\phi_{i}.
\end{equation} 

Before going on, we will demonstrate a weak-coupling check on this superpotential. The quivers given above explicitly display $W$ bosons associated to the simple roots; the ordering $\arg\mathcal{Z}(b_i)>\arg\mathcal{Z}(c_i)$ ensures that there will be a $W$ boson associated to the $i$th simple root. However, at weak coupling we would expect massive vector $W$ bosons associated to all roots of the $SU(N)$ algebra, due to Higgsing of the gauge bosons. The set of massive vectors should fill out exactly one adjoint of the $SU(N)$, except for the Cartan elements, which remain massless. 

Let us see how these additional vectors come about by first considering $SU(3).$  We seek a vector state corresponding to a representaion with dimension vector $(1,1,1,1)$.  The superpotential is then
\begin{equation}\mathcal{W}=X_1\phi' X_2\phi-Y_1\phi' Y_2\phi,\end{equation}
and the resulting F-terms are
\begin{align}
\label{fterm1}\phi\phi'X_2=\phi\phi'Y_2=\phi\phi' X_1=\phi\phi'Y_1=0\, ,\\
\label{fterm2}\phi(X_1X_2-Y_1Y_2)=\phi'(X_1X_2-Y_1Y_2)=0\,.
\end{align}
If both $\phi,\phi'$ are zero,  the rep is given by $X_i,Y_i$, and falls apart into the direct sum of two subreps, $b_1+c_1,b_2+c_2.$ Such a situation is described as a decomposable representation; decomposable reps are never stable, since one of the two subreps must be to the left of decomposable rep in the $\mathcal{Z}$-plane. If $\phi,\phi'$ are both nonzero, then $X_i,Y_i$ are all zero by (\ref{fterm1}), and again the rep is decomposable. We are left with two cases, $\phi=0,\phi'\ne0$ and vice versa. Having set one of the $\phi$'s to zero, there is one more equation in (\ref{fterm2}) that must be satisfied: $X_1 X_2=Y_1 Y_2.$ Naive dimension counting gives us $6-2-3=1,$ so we have a vector. Gauge fixing sets $\phi \text{ (or }\phi')=X_1=Y_1=1;$ then the actual moduli space is parameterized by $X_2=1/Y_2,$ which forms $\mathbb{P}_1$. Lefschetz $SU(2)$ gives exactly one vector of this charge, and no hypers. It remains to check the stability conditions. For $\phi=0,$ there are subreps $c_1,b_1+c_1,b_1+c_1+c_2;$ these are \emph{not} destabilizing precisely when, in addition to the weak coupling conditions, we also have $\arg\mathcal{Z}(b_1+c_1)<\arg\mathcal{Z}(b_2+c_2).$ On the other hand, when $\arg\mathcal{Z}(b_1+c_1)>\arg\mathcal{Z}(b_2+c_2),$ then $c_1+b_1$ is certainly a destabilizing subrep. Similarly, $\phi_2=0$ is stable precisely for $\arg\mathcal{Z}(b_1+c_1)>\arg\mathcal{Z}(b_2+c_2).$ Therefore, at any region in weak coupling, we find precisely one $W$ boson of the desired charge.

Now we consider arbitrary $SU(N).$ By embedding the $SU(3)$ quiver as a subquiver of an arbitrary $SU(N)$ quiver, we see that the specified superpotential (\ref{SUNpot}) guarantees that exactly one $W$ boson vector with charge $(\alpha_i+\alpha_{i+1},0)$ appears at weak coupling. It remains to check the $W$ bosons associated to the rest of the roots, which have charges $\left(\sum_{i=j}^{j+k} \alpha_i,0\right)$ for any $k>1.$ As representations, these are given by $\sum_{i=j}^{j+k} b_i+c_i.$ It is clear that, for this analysis, we can simply focus on the subquiver formed by $b_i,c_i$ for $j\le i \le j+k;$ all other nodes (and maps involving them) are set to zero in this rep, and consequently, any superpotential terms from them are trivial. Thus we can simply study the rep $v=\sum_{i=1}^{k} b_i+c_i$ of the $SU(k+1)$ quiver and superpotential as shown above.

The F-terms are now a bit more subtle. 
\begin{align}
\label{ftermk1}\phi_{i-1}\phi'_{i-1}X_{i-1}+\phi_{i}\phi'_{i}X_{i+1}&=\phi_{i-1}\phi'_{i-1}Y_{i-1}+\phi_{i}\phi'_{i}Y_{i+1}=0 \,,\\
\label{ftermk2}\phi_i(X_iX_{i+1}-Y_iY_{i+1})&=\phi'_i(X_iX_{i+1}-Y_iY_{i+1})\,.
\end{align}
Again, not both $\phi_i,\phi'_i$ can be zero, or else the rep is decomposable. However, it seems that perhaps both $\phi_i, \phi'_i$ may be nonzero; since (\ref{ftermk1}) now has two terms, this no longer forces the rep to become decomposable. Nonetheless, we can dispose of this possibility by stability. If both $\phi_i,\phi'_i$ are nonzero, then either both $\phi_{i-1},\phi'_{i-1}$ are nonzero or $X_{i+1},Y_{i+1}$ are zero due to (\ref{ftermk1}). By induction, we will find that $X_j,Y_j$ are zero for some $j.$ This situation cannot be $\Pi$-stable; because $X_j,Y_j$ vanish, we have two subreps, $b_j$ (which is now effectively a sink in the quiver), and $v-c_j,$ the subrep where we set to zero $c_j,$ (which is now an effective source in the quiver). It must be the case that one of these is destabilizing. If $\arg\mathcal{Z}(c_j)>\arg\mathcal{Z}(v)$, then we have $\arg\mathcal{Z}(b_j)>\arg\mathcal{Z}(c_j)>\arg\mathcal{Z}(v)$ so that $b_j$ is destabilizing; otherwise $\arg\mathcal{Z}(v-c_j)>\arg\mathcal{Z}(v)>\arg\mathcal{Z}(c_j),$ so that $v-c_j$ is destabilizing.

Having dealt with this subtlety, we can continue with the analysis. The remaining case is that exactly one of $\phi_i, \phi'_i$ is nonzero for each $i;$ this gives $2^k$ possibilities. First, we check the dimension of the parameter space: we start with $4k-2$ maps and $2k-1$ gauge symmetries; we have set $k-1$ maps to zero, and we have $k-1$ remaining constraints (\ref{ftermk2}); thus $(4k-2)-(2k-1)-(k-1)-(k-1)=1.$ We may gauge fix $\phi_i \text{ (or } \phi'_i)=X_i=Y_i=1$ for $1 \le i < N-1;$ then the moduli space is $\mathbb{P}^1$ parametrized by $X_{N-1}=1/Y_{N-1}.$  Thus we have $2^k$ vector states. Using stability, we will find that precisely one of these vectors is stable for any region of weak coupling. To see this, fix $j$ and choose $\phi_j\ne 0.$ Because of this choice, there is a subrep $\sum_{i=j+1}^{k} b_i+c_i,$ which is destabilizing when $\arg\mathcal{Z}\left(\sum_{i=j+1}^k b_i+c_i\right)>\arg\mathcal{Z}(v)>\arg\mathcal{Z}\left(\sum_{i=1}^{j} b_i+c_i\right).$ If we had chosen $\phi'_j\ne 0,$ we would have a subrep $\sum_{i=1}^j (b_i+c_i)$ which is destabilizing in exactly the opposite situation, $\arg\mathcal{Z}\left(\sum_{i=1}^{j} b_i+c_i\right)>\arg\mathcal{Z}(v)>\arg\mathcal{Z}\left(\sum_{i=j+1}^k b_i+c_i\right).$\footnote{There are some additional subreps that should be considered, but ultimately play no role. For example, if $\phi_j\ne0,\phi'_m\ne0$ for $j<m$, then there is a subrep $\sum_{i=j+1}^m b_i+c_i,$ which is destabilizing when $\arg\mathcal{Z}\left(\sum_{i=j+1}^m b_i+c_i\right)>\mathcal{Z}(v).$ Suppose that neither subreps described above are destabilizing; then  $\arg\mathcal{Z}\left(\sum_{i=j+1}^k b_i+c_i\right)<\arg\mathcal{Z}(v)$ and $\arg\mathcal{Z}\left(\sum_{i=1}^{m} b_i+c_i\right)<\arg\mathcal{Z}(v).$ Summing these inequalities, we find $\arg\mathcal{Z}\left(\sum_{i=j+1}^m b_i+c_i\right)<\mathcal{Z}(v),$ so that this new subrep cannot be destabilizing. Further, if $c_i+b_i$ is a subrep, then so is $c_i,$ but this again gives no additional destabilizing constraints since $\arg\mathcal{Z}(b_i)>\arg\mathcal{Z}(b_i+c_i)>\arg\mathcal{Z}(c_i).$} So we have arrived at the desired conclusion, namely, that we obtain precisely one vector for each root of $SU(N).$ With a bit more work it is possible to see that, up to field redefinitions, this is the unique superpotential at quartic order that properly produces exactly one set of $W$ bosons. In principle this leaves the possibility of higher order terms in the superpotential, but the derivation of \cite{Fiol} shows that indeed no such terms arise.


\subsection{General ADE-type Gauge Group}\label{ADEtype}
Some brief comments will allow us to extend the above analysis to arbitrary ADE-type (ie simply-laced) gauge group $G.$ At weak coupling, we would again expect to be able to decouple the $\text{rank } G$ distinct $SU(2)$ subgroups, again with one corresponding to each simple root of the algebra. Then we would again find an $SU(2)$ subquiver for each simple root $\alpha_i.$ If we again make the ansatz of fixing charges $(0,\alpha_i), (\alpha_i,-\alpha_i),$ then we find that, for each line in the Dynkin diagram (ie $\alpha_i\cdot\alpha_j=-1$), we must connect the respective $SU(2)$ subquivers as
\begin{equation}\begin{split}\begin{xy} 0;<1pt,0pt>:<0pt,-1pt>:: 
(80,80) *+{c_j} ="2",
(80,00) *+{b_j} ="3",
(0,0) *+{c_i} ="4",
(0,80) *+{b_i} ="5",
(-10,40) *+{X_i},
(10,40) *+{Y_i},
(90,40) *+{Y_j},
(70,40) *+{X_j},
(40,70)*+{\phi_{ij}'},
(40,-10)*+{\phi_{ij}},
"3", {\ar"2" <1.5pt>},
"3", {\ar"2" <-1.5pt>},
"2", {\ar"5"},
"4", {\ar"3"},
"5", {\ar"4"<1.5pt>},
"5", {\ar"4"<-1.5pt>},
\end{xy}\end{split}\end{equation}
with the quartic superpotential $\mathcal{W}=X_i\phi_{ij}' X_j\phi_{ij}-Y_i\phi_{ij}' Y_j\phi_{ij}.$ 

Thus there is a straightforward graphical prescription for constructing a quiver for pure SYM with simply-laced gauge group $G,$ starting from the Dynkin diagram of G. For every node $i$ of the Dynkin diagram, we draw and $SU(2)$ subquiver with nodes $b_i,c_i;$ for every line in the Dynkin diagram given $i-j$ we connect the $SU(2)$ subquivers as above, with the quartic superpotential. This is exactly the quiver $\widehat{A}_1\boxtimes G,$ which was found to describe these theories via $2d/4d$ in \cite{CV11}. The superpotential guarantees the existence of some subset of the $W$ bosons, namely those contained in any $SU(N)$ subquiver of the full $G$ quiver; studying the full root system of $W$ bosons becomes quite complicated, and we omit the analysis here. While the quartic terms must be present in the superpotential, there may or may not be some additional higher order terms. For clarity, we draw the Dynkin diagrams along with resulting quivers for $D_4, E_6.$

\begin{equation}\begin{xy} 0;<1pt,0pt>:<0pt,-1pt>:: 
(-165,25) *+{\alpha_2}*\cir<8pt>{}="10",
(-215,25) *+{\alpha_1}*\cir<8pt>{}="11",
(-140,-15.3) *+{\alpha_4}*\cir<8pt>{}="12",
(-140,68.3) *+{\alpha_3}*\cir<8pt>{}="13",
"10", {\ar@{-}"11"},
"10", {\ar@{-}"12"},
"10", {\ar@{-}"13"},
(100,0) *+{c_2} ="2",
(100,50) *+{b_2} ="3",
(50,0) *+{c_1} ="4",
(50,50) *+{b_1} ="5",
(135,93.3) *+{b_3}="7",
(135,43.3) *+{c_3}="6",
(135,6.7) *+{b_4}="9",
(135,-43.3) *+{c_4}="8",
"3", {\ar"2" <1.5pt>},
"3", {\ar"2" <-1.5pt>},
"2", {\ar"5"},
"4", {\ar"3"},
"5", {\ar"4"<1.5pt>},
"5", {\ar"4"<-1.5pt>},
"9", {\ar"8" <1.5pt>},
"9", {\ar"8" <-1.5pt>},
"7", {\ar"6" <1.5pt>},
"7", {\ar"6" <-1.5pt>},
"2", {\ar"7"},
"6", {\ar"3"},
"2", {\ar"9"},
"8", {\ar"3"},
\end{xy}\end{equation}

\begin{equation}\begin{xy} 0;<1pt,0pt>:<0pt,-1pt>:: 
(-0,25) *+{\alpha_5}*\cir<8pt>{}="14",
(-50,25) *+{\alpha_4}*\cir<8pt>{}="13",
(-100,25) *+{\alpha_3}*\cir<8pt>{}="12",
(-100,-25)*+{\alpha_6}*\cir<8pt>{}="15",
(-150,25)*+{\alpha_2}*\cir<8pt>{}="11",
(-200,25) *+{\alpha_1}*\cir<8pt>{}="10",
"10", {\ar@{-}"11"},
"11", {\ar@{-}"12"},
"12", {\ar@{-}"13"},
"14", {\ar@{-}"13"},
"15", {\ar@{-}"12"},
(100,0) *+{c_2} ="2",
(100,50) *+{b_2} ="3",
(50,0) *+{c_1} ="4",
(50,50) *+{b_1} ="5",
(150,50) *+{b_3}="7",
(150,0) *+{c_3}="6",
(200,50) *+{b_4}="9",
(200,0) *+{c_4}="8",
(250,50) *+{b_5}="17",
(250,0) *+{c_5}="16",
(180,-10) *+{b_6}="19",
(180,-60) *+{c_6}="18",
"3", {\ar"2" <1.5pt>},
"3", {\ar"2" <-1.5pt>},
"2", {\ar"5"},
"4", {\ar"3"},
"5", {\ar"4"<1.5pt>},
"5", {\ar"4"<-1.5pt>},
"9", {\ar"8" <1.5pt>},
"9", {\ar"8" <-1.5pt>},
"7", {\ar"6" <1.5pt>},
"7", {\ar"6" <-1.5pt>},
"2", {\ar"7"},
"6", {\ar"3"},
"6", {\ar"9"},
"8", {\ar"7"},
"8", {\ar"17"},
"16", {\ar"9"},
"17", {\ar"16" <1.5pt>},
"17", {\ar"16" <-1.5pt>},
"19", {\ar"18" <1.5pt>},
"19", {\ar"18" <-1.5pt>},
"6", {\ar"19"},
"18", {\ar"7"},
\end{xy}\end{equation}


\subsection{BPS Spectra of Pure $SU(N)$ SYM}
In the following we will compute the BPS spectra of $SU(N)$ theories using the mutation method. We find a spectrum consisting of $N(N-1)$ BPS particles and their antiparticles at strong coupling in agreement with the identification of the spectrum in this region with CFT states of \cite{Lerche00}. 

For $N\ge 3$ these theories are not complete in the sense of ref. \cite{CV11} since their charge lattice has rank $2(N-1)$ while there are only $N$ physical moduli that can be varied corresponding to half of the charges and the coupling of the theory. We will therefore not have the freedom to adjust all the central charges as we wish since some of them will be fixed by special geometry. To apply the mutation method we therefore need to compute the central charges in a chamber in moduli space and find a basis which has central charges lying in a half plane.

\subsubsection{$SU(3)$}
We begin with an analysis of the $SU(3)$ theory starting from the quiver discussed in section \ref{sunweak}, which was obtained from a weak coupling analysis and which is verified by the $2d/4d$ correspondence \cite{CNV}. We identify the nodes of the quiver with cycles in the SW geometry and compute their central charges to determine the ordering of the mutations. Furthermore, we track these cycles to the strong coupling region where we produce the full BPS spectrum consisting of 6 particles.

The central charge function is part of the IR data of the theory, and is thus specified by the SW solution. The $SU(N)$ SW curve can be written as \cite{KLYT,AF,KLT}
\begin{equation}
y^2=(P_{A_{N-1}}(x,u_i))^2-\Lambda^{2N}\, , \quad P_{A_{N-1}}(x,u_i)=x^N-\sum_{i=2}^{N} u_{i} x^{N-i}\, ,
\end{equation}
where the $u_i$ are the Casimirs parametrizing the Coulomb branch and $\Lambda$ is the strong coupling scale. The SW differential is then given by  \cite{KLYT,AF,KLT}
\begin{equation}
\lambda(u_i)= \frac{1}{2  \pi i} \frac{\partial P_{A_{N-1}}(x,u_i)}{\partial x} \frac{x \,dx}{y} \, ,
\end{equation}
and a BPS particle which is represented by a cycle $\gamma$ on the SW curve has charge
\begin{equation}
Z_{u_i}(\gamma) = \int_\gamma \lambda (u_i) .
\end{equation}
Finally, the electric-magnetic inner product of two particles is computed by the intersection product of the associated cycles. We will use $\gamma$ to refer to both the particle and associated cycle, and $\circ$ to indicate both the electric-magnetic inner product and the intersection product.

We will calculate the central charge configuration for a weakly coupled point of the $SU(3)$ theory. For $SU(3)$ we set $u_2 = u$ and $u_3 = v$. The Casimirs $u_i$ determine the vevs of the Cartan elements of $SU(N)$ semi-classically, and it can be checked that $u \rightarrow -\infty$ and $v=0$ indeed corresponds to a weakly coupled point in $SU(3)$. 

\begin{figure}
\centering
\subfloat[Choice of cycles at weak coupling]{\label{swsu3weak}\includegraphics[height=2in]{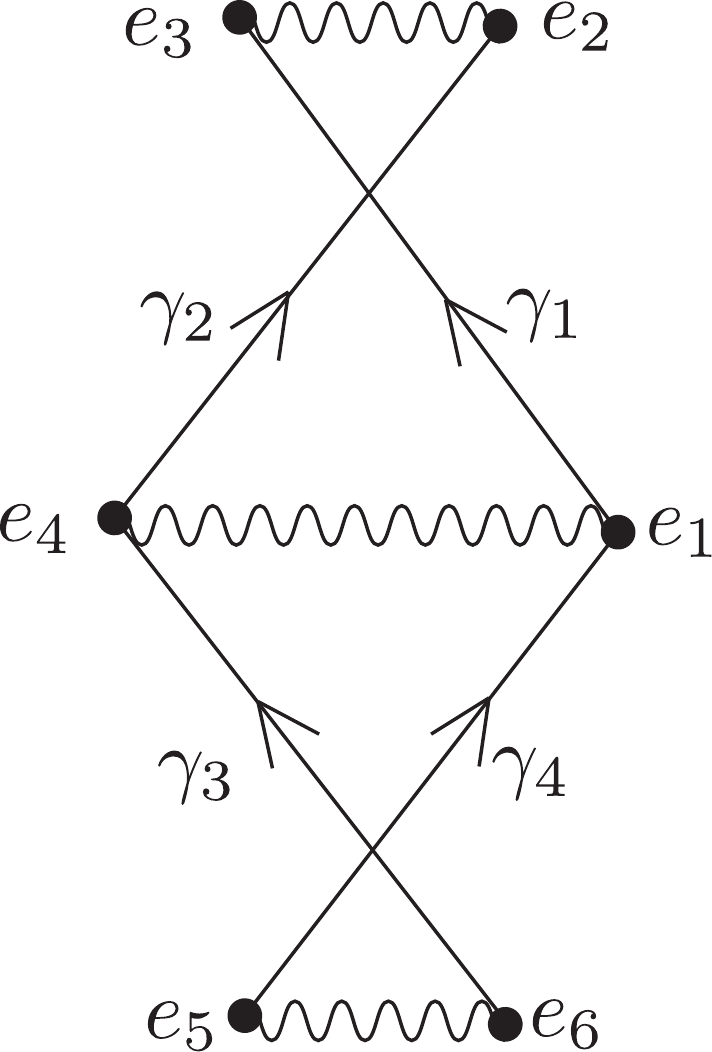}}\hspace{0.1\textwidth}
\subfloat[Cycles at strong coupling]{\label{swsu3strong}\includegraphics[height=2in]{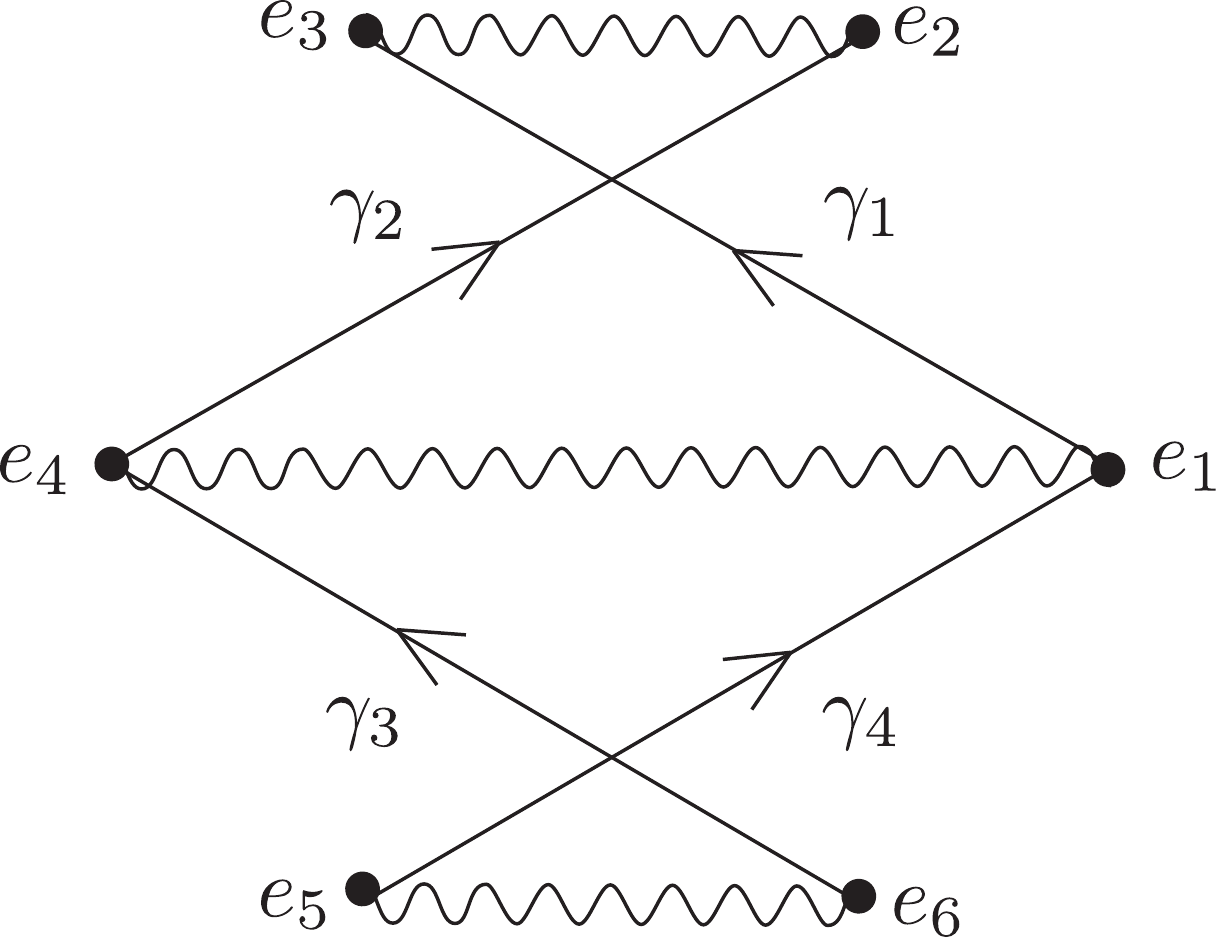}}
\caption{The choice of cycles in the $x$-plane at weak and strong coupling is shown in Figs. \ref{swsu3weak} ,\ref{swsu3strong} respectively. $e_i\,,i=1,\dots,6$ denote the roots of $(x^3-u x-v)^2-\Lambda^6$ and become the sixth roots of unity as we tune the moduli to strong coupling and set $\Lambda=1$.}
\end{figure}

The $SU(N)$ theory has an $Sp(2N-2,\mathbb{Z})$ duality which is manifest in the different possible choices of symplectic homology basis that could be identified with electric and magnetic charges. We postpone the charge labeling and identify the nodes of the quiver directly with a choice of cycles in the geometry as shown in Fig. \ref{swsu3weak}. 

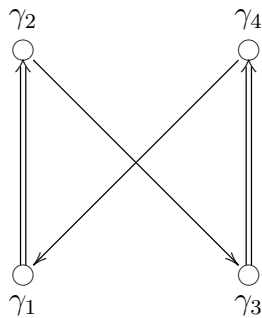
\begin{figure}
\centering
{\label{swsu3weakquiver}}
    \raisebox{10mm}{
    \begin{xy}
    (0,-4)*{\gamma_1}; (0,0)*{\Circle}="b1";
    (30,-4)*{\gamma_3}; (30,0)*{\Circle}="b2";
    (30,34)*{\gamma_4}; (30,30)*{\Circle}="c2";
    (0,34)*{\gamma_2}; (0,30)*{\Circle}="c1";
    {\ar@2{->} "b1";"c1"};{\ar "c1";"b2"};{\ar@2{->} "b2";"c2"};{\ar "c2";"b1"};
    \end{xy}    
}
\caption{Quiver obtained from the intersections of the cycles in Figs. \ref{swsu3weak},\ref{swsu3strong}.}
\end{figure}

The quiver obtained in this way at weak coupling should have a number of properties:
\begin{itemize}
\item The intersections of cycles must agree with the electric-magnetic inner product as defined by the quiver
\item The central charges of all the nodes must lie in a common half-plane
\item The apparent $SU(2)$ subquivers should be weakly coupled
\item The central charges of the $W$ bosons of the $SU(2)$ gauge groups should be vanishingly small compared to the central charges of the nodes in the $u\rightarrow -\infty$ limit
\end{itemize}

The last condition follows from the fact that the electrically charged objects should be parametrically light compared to the dyonic states of the theory at weak coupling, since here the electric particles are the fundamental degrees of freedom.

\begin{figure}
\centering
\subfloat[Weak coupling]{\label{su3weak}\frame{\includegraphics[height=2.25in]{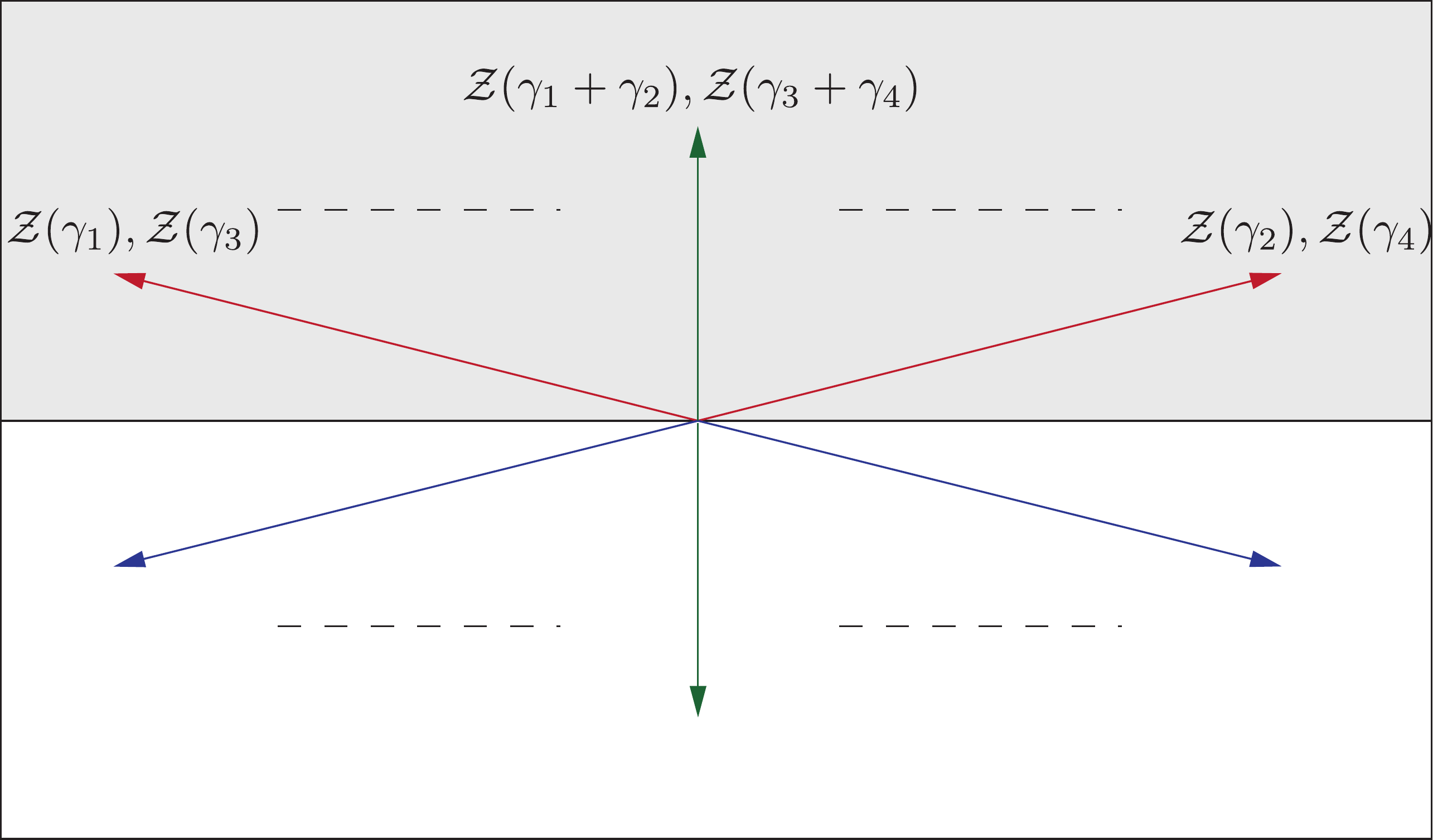}}}\hspace{0.02\textwidth}
\subfloat[Strong coupling]{\label{su3strong}\frame{\includegraphics[height=2.25in]{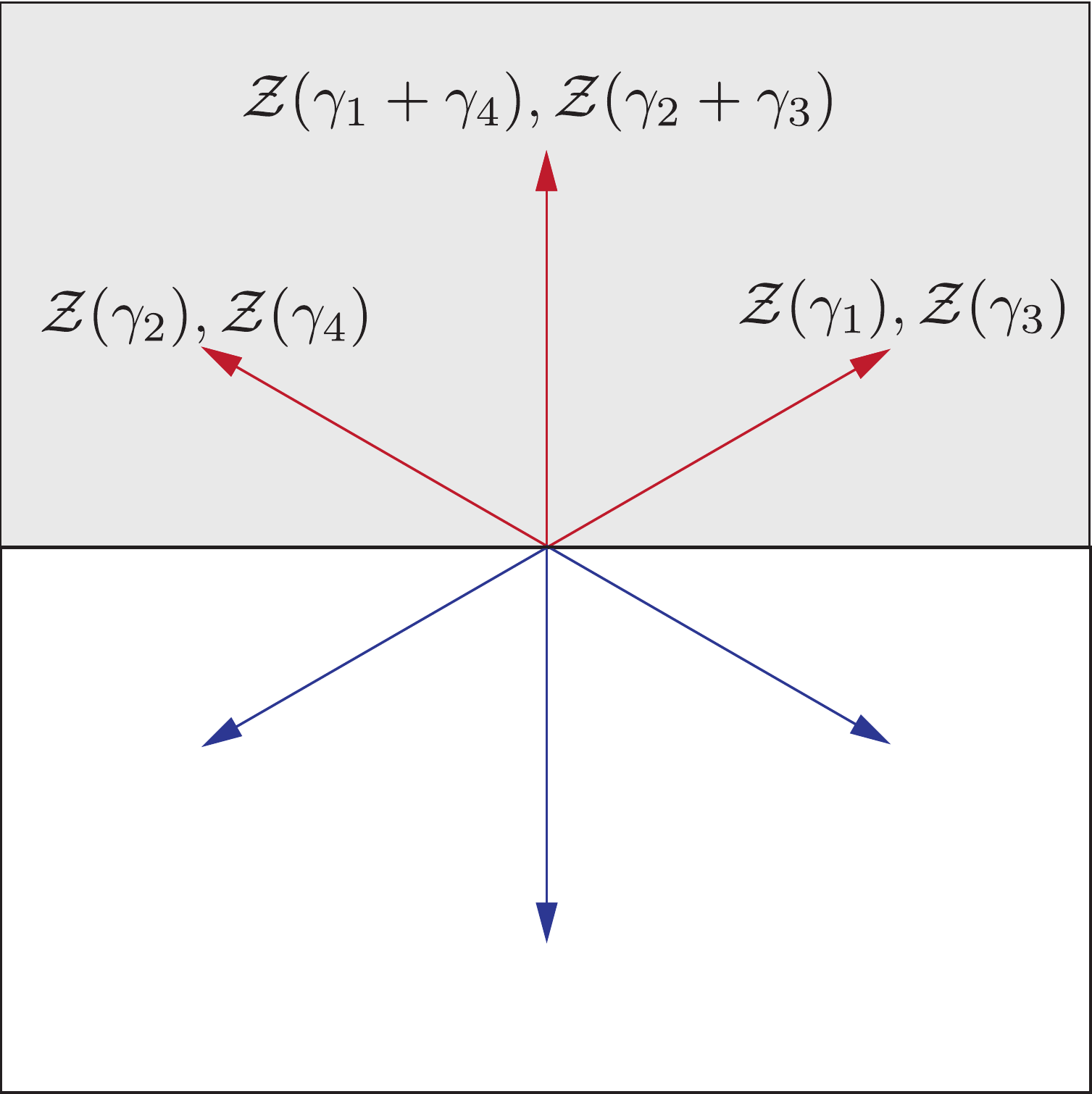}}}
\caption{The central charges of BPS states of $SU(3)$ are depicted at weak (a) and strong (b) coupling respectively. At weak coupling, the left- and right-most nodes, along with the weak coupling $W$ bosons are shown explicitly. The full spectrum at weak coupling is not known, but at least includes two infinite towers of dyons, which are not shown. In the limit of zero coupling, the left- and right-most nodes approach $\pi$ separation and infinite length. As we tune towards strong coupling, the states $\gamma_1,\gamma_3$ and $\gamma_2,\gamma_4$ approach and cross each other. At strong coupling, the full finite spectrum of BPS states is depicted; the $\mathbb{Z}_6$ symmetry is manifest.}
\end{figure}

The choice of cycles in Fig. \ref{swsu3weak} meets these conditions. That the first is met is obvious, and the latter three can be explicitly checked by numerically computing the associated integrals of the SW differential along the given curves. This has been done, and the values of the central charges for large but finite $u<0$ are as depicted in Fig. \ref{su3weak}. Since the $SU(2)$ subquivers are weakly coupled, we are in an infinite chamber, as expected at weak coupling. To apply the mutation method most efficiently we will tune the moduli to arrive in a chamber with a finite spectrum.

We can track the behavior of the quiver explicitly as we tune moduli. At walls of marginal stability nothing happens at the level of the quiver, while at walls of the second kind we must mutate to find a valid description on the other side. A generic path in the $SU(3)$ moduli space may pass through arbitrarily many - even infinitely many - walls of the second kind, thereby alluding an analysis. For $SU(3)$ there exists a path which takes us from weak coupling to the strongly coupled $u=0$ point and passes through no walls of the second kind, thereby allowing a quite seamless transition between the understood weak coupling chamber and the strongly coupled chamber containing the $u=0$ point.

We follow the straight line path with $v = \operatorname{Im} u = 0$ from $u=-\infty$ to $u=0$. The pairs of aligned central charges stay aligned along the entire path, and cross in tandem at a finite value of $u<0$. All the while, all central charges remain in the upper half-plane. At $u=0$, both $SU(2)$'s are strongly coupled, and the central charge configuration is as given in Fig. \ref{su3strong}. Now we simply apply the mutation algorithm with the central charges associated to this point in moduli space. What we find is a $N(N-1) = 6$ state chamber with states

\begin{equation}
\gamma_2,\gamma_4,\gamma_2+\gamma_3,\gamma_1+\gamma_4,\gamma_1,\gamma_3\,.
\end{equation}

Let us note some features of the strong coupling spectrum we have found. First of all, all states in the chamber correspond to vanishing cycles in the Seiberg-Witten geometry. That is, they all correspond to cycles which vanish somewhere on moduli space. This agrees with earlier intuition about the relation between the strong coupling $SU(N)$ spectrum and vanishing cycles of the SW geometry \cite{KLYT,KLT,Lerche96,Fiol}.

The second feature, which will become quite important in our $SU(N)$ analysis below, is that the chamber we have found respects the $\mathbb{Z}_{2N} = \mathbb{Z}_6$ symmetry of the IR solution.

In principle one would hope that the same story carried over for the $SU(N)$ case. We would ideally start from weak coupling and tune moduli until we arrived at the strongly coupled $u_i = 0$ point, and then see that this point lied in a finite chamber with $N(N-1)$ states. Unfortunately the situation becomes technically complicated, in a way we will briefly explain. Above, we chose a very particular path between the $u_i = 0$ point and weak coupling, along which the quiver passed through no walls of the second kind, where quiver mutation is necessary. This was a path which deformed the order 1 term in the defining polynomial of the Seiberg-Witten curve.

In the $SU(N)$ case it is always the $x^{N-2}$ deformation which has this nice property. That is, if we deform the coefficient of the $x^{N-2}$ term alone from the $u_i=0$ point along certain directions in $\mathbb{C}$, the quiver will be extremely well behaved, just as above. The issue is that it is only in the $N=3$ case that this deformation alone is sufficient to arrive at weak coupling. In all other cases there will be some unbroken subgroup which remains. Thus to get to weak coupling, we must deform lower order terms, but these are not nice in terms of the quiver description. In particular, no simple choice seems to get from strong to weak coupling while only passing through a small number of walls of the second kind. Potentially such a path remains to be found, and the same method can then be generalized to the $SU(N)$ case. At present, we will proceed with a discussion of the $SU(N)$ case at $u=0$ based on what we've learned in $SU(3)$.

 
\subsubsection{$SU(N)$ at Strong Coupling}\label{sunstrong}
We now consider the general case of $SU(N)$ at strong coupling. Our objective is to determine the quiver, charge labels of nodes, and ordering of central charges at some point of strong coupling, and then compute the resulting spectrum via the mutation method. Of course, to honestly produce the quiver we would need to somehow find a basis of BPS states. However, the quiver has already been derived from other considerations, and motivated from a purely $4d$ perspective in \ref{sunweak}. Here we will infer quiver along with charge labels at strong coupling by generalizing the results above for $SU(3)$. 

\begin{figure}\centering
\includegraphics[height=4in]{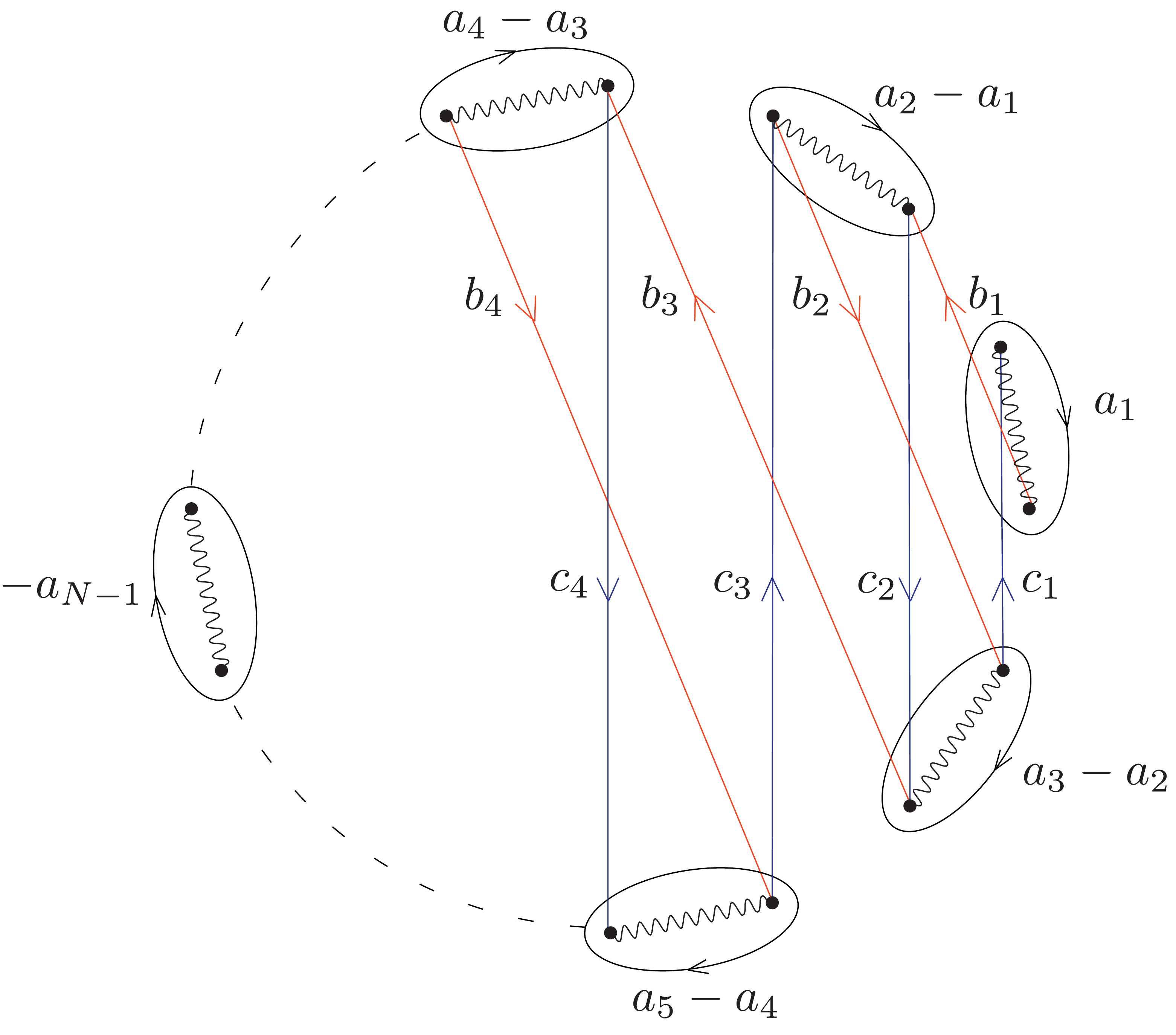}
\caption{The Seiberg-Witten curve described by (\ref{swcorigin}), shown as a double cover of the $x$-plane, with branch cuts as indicated. The labelled $a_i, b_i$ cycles give a symplectic homology basis. The action of the $\mathbb{Z}_{2N}$ symmetry rotates the plane by $e^{-i\pi/N},$ and thus rotates $b_i$ into $c_i.$ The $b_i,c_i$ cycles constitute the positive integral basis of states that appear as nodes of the quiver. Note that we have taken a different convention for branch cuts than the one used in Fig. \ref{swsu3weak}. This choice is more convenient for the strong coupling analysis, and agrees with the conventions used in \cite{KLT}.}
\label{swstrong}
\end{figure}

Fix the moduli $u_i=0$, so that the Seiberg-Witten curve is given as 
\begin{equation} y^2=x^{2N}-\Lambda^{2N},
\label{swcorigin}
\end{equation}
with Seiberg-Witten differential
\begin{equation}\lambda=\frac{1}{2\pi i}\frac{N x^N dx}{y}.
\end{equation}
We take a symplectic homology basis, $a_i, b_i$ for $i=1,\dots,N-1,$ with $a_i\circ a_j=b_i\circ b_j=0$ and $a_i\circ b_j=\delta_{ij}.$ The appropriate choice of cycles is shown in Figure \ref{swstrong}. We have chosen the $a_i$'s to be the cycles that collapse as $u_{N}\rightarrow \infty,$ since these are pure electric charges. There is still some ambiguity in choosing $b$ cycles, which are pure magnetic monopoles with charges given by simple roots of $SU(N).$ We fix the ambiguity by choosing the $b$ cycles to be ones that vanish somewhere in moduli space. This is a natural choice, since each of the simple roots has a full $SU(2)$ moduli space associated with it contained in the $SU(N)$ moduli space; by the original Seiberg-Witten $SU(2)$ analysis, the monopole associated to each simple root becomes massless at some locus of the $SU(N)$ moduli space.

At the origin of moduli space, the curve has a $\mathbb{Z}_{2N}$ discrete symmetry. If we denote $\xi$ the generator of the symmetry, we have
\begin{equation}
\xi(x)= e^{-i\pi/N} x.
\end{equation}
The action on the $x$-plane is simply a $-\pi/N$ rotation; on the central charge function $\mathcal{Z}$, this gives
\begin{align}
\xi(\lambda)&= -e^{-i\pi/N}\lambda \\
\xi\left(\mathcal{Z}(\gamma)\right)&= -e^{-i\pi/N}\mathcal{Z}(\gamma).
\end{align}
This induces an exact symmetry of the quantum theory that will be quite useful.
It indicates that BPS states will come in $\mathbb{Z}_{2N}$ orbits; the magnitude of their central charges of cycles in an orbit are all identical, and their phases are distributed $\mathbb{Z}_{2N}$ symmetrically in the complex plane. Again, by $SU(2)$ reasoning, each magnetic monopole with simple root charge will be a BPS state at the origin of moduli space. From Figure \ref{swstrong}, it is clear that all the $b_i$'s are in distinct orbits. Thus we have obtained $(N-1)$ distinct orbits, one for each simple root monopole with electric-magnetic charge $(0,\alpha_i);$ each orbit consists of $2N$ BPS states, $N$ of which are particles, and $N$ antiparticles.

\begin{figure}
\centering
\includegraphics{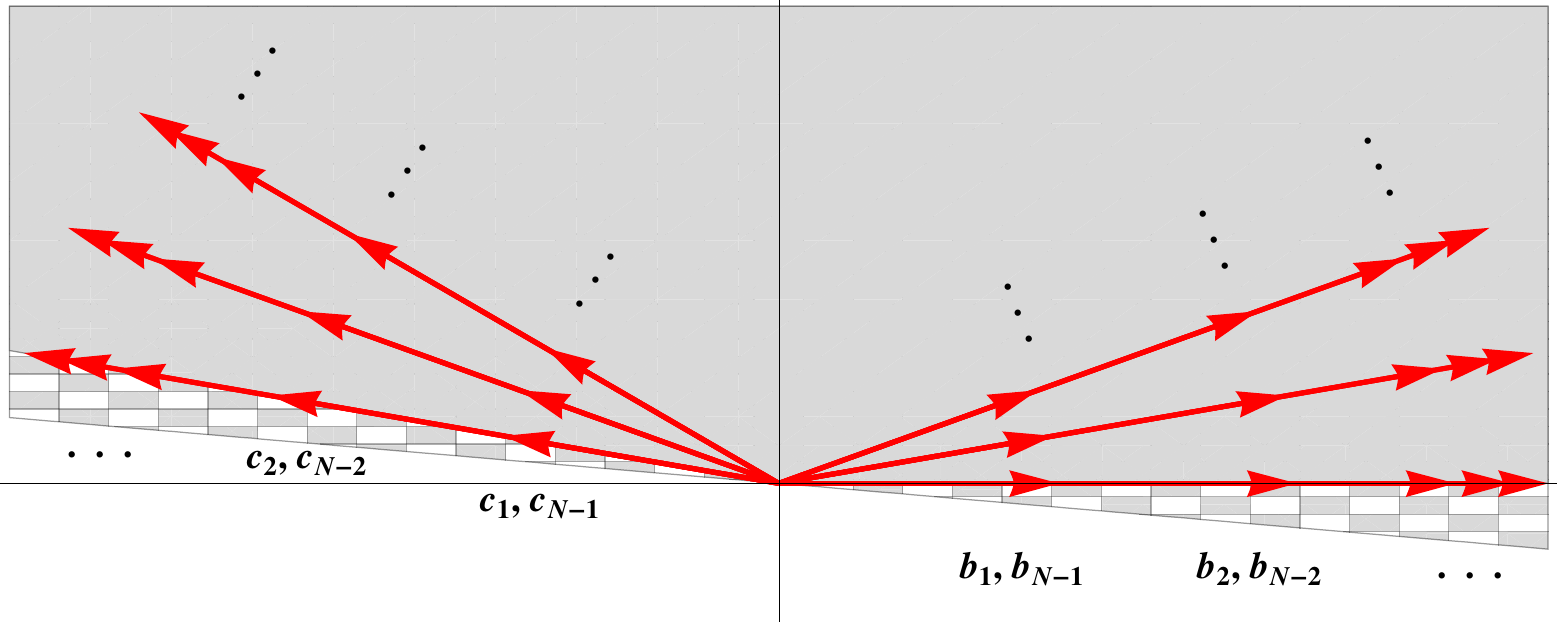}
\caption{Central charges of vanishing cycles plotted in the $\mathcal{Z}$-plane (where we have rotated by some overall phase $\arg(i\kappa(N)).$ The half-plane we use to construct the quiver is shown as the gray region. The $b_j$ cycles have $\mathcal{Z}(b_j)\sim \sin\frac{j\pi}{N}$; note that $\mathcal{Z}(b_j)=\mathcal{Z}(b_{N-j})$. The $b_j$ are therefore $N-1$ distinct collinear states shown on the positive real axis. Each ray of collinear red arrows is a $\mathbb{Z}_{2N}$ rotation of the $b_j$'s. There are $N$ such rays in the half-plane, situated at $2N$-roots of unity. In total we have $N(N-1)$ states depicted in the diagram. The antiparticles in the opposite half-plane are not shown. The half-plane is chosen so that $b_j$ are right-most BPS states, which forces $c_j$ to be left-most BPS states. As explained in the analysis, for such a half-plane to exist, the region checkered in white and gray must be free of BPS states.}
\label{sunzplane}
\end{figure}

To compute the periods, we integrate the Seiberg-Witten differential, to obtain
\begin{equation}
\int \lambda = \frac{1}{2 \pi}\frac{N}{N+1}x^{N+1}\,_2F_1\left(\frac{1}{2},\frac{N+1}{2N},\frac{1}{2N}+\frac{3}{2},1\right)=\kappa(N)x^{N+1},
\end{equation}
where $\kappa$ is some proportionality constant that depends on $N$ but is independent of $x.$
Evaluating the definite integral for the $b_i$'s shown in Figure \ref{swstrong}, we find
\begin{equation}
\mathcal{Z}(b_j)=2\kappa(N) i e^{i\pi/N} \sin\frac{j\pi}{N}
\end{equation}
From the action of the $\xi$, we see that the full $\mathbb{Z}_{2N}$ orbits of vanishing cycles will fill out all $2N$-roots of unity (up to some overall phase $\arg (ie^{i\pi/N}\kappa(N))$) in the $\mathcal{Z}$-plane. This configuration of central charges is depicted in Figure \ref{sunzplane}

To continue, we now generalize from the $ SU(2)$ and $SU(3)$ results. In those cases, the BPS spectra were precisely equivalent to the set of vanishing cycles of the Seiberg-Witten geometry. It is natural to imagine that for general $N$ it is at least possible to choose a positive integral basis for BPS states that consists of vanishing cycles. The vanishing cycles do in fact span the homology lattice, so this is sensible assumption. As we will see, this allows us to obtain a quiver that agrees with (\ref{sunweakquiv}), which was also proposed from other perspectives \cite{Fiol,CNV}. 
Thus, we seek a positive integral basis of vanishing cycles; to do so, we must first choose a half-plane. Since the $N-1$ $b_i$'s have the same phase, we may tune the half-plane to make them right-most vanishing cycles; then the $b_i$'s are forced to appear as $N-1$ nodes of the quiver.\footnote{In principle, a bound state of multiple $b_i$'s would also have the same phase, and one might worry that some of these $N-1$ states were actually bound states of the others. However, this is in fact impossible. The $b_i$ are linearly independent cycles, so none can occur as a linear combination of the others; furthermore  $b_i\circ b_j=0$, so there exist no bound states of the form $b_i+b_j.$ So \emph{all} of the $b_i$ cycles must appear as nodes of the quiver.} Having fixed this choice of half-plane, it is clear from Figure \ref{sunzplane} that $c_i\equiv\xi(b_i)$ form $N-1$ right-most vanishing cycles in the half-plane, and therefore must also appear in the quiver. These states are given as
\begin{equation}
c_i\equiv\xi(b_i)=\begin{cases} -a_{i-1}+2a_i-a_{i+1}+b_i=(\alpha_i,\alpha_i)& \text{if $i$ is even}\\
-a_{i-1}+2a_{i}-a_{i+1}-b_{i-1}-b_i-b_{i+1} =(\alpha_i,-\alpha_{i-1}-\alpha_i-\alpha_{i+1})& \text{if $i$ is odd}
\end{cases}
\end{equation}
We now have specified $2(N-1)$ nodes of the quiver; in fact, this is exactly the number of nodes in the quiver, by the counting $2r+f=2(N-1).$ At this point we have fully determined the quiver as follows:

\begin{equation}\label{sunstrongquiv}\begin{split}\begin{xy} 0;<1pt,0pt>:<0pt,-1pt>:: 
(240,3)*+{\;\;} ="12",
(240,78)*+{\;\;}="13",
(320,4)*+{\;\;} ="14",
(320,78)*+{\;\;}="15",
(280,40) *+{\dots},
(400,0) *+{c_{N-1}} ="0",
(400,80) *+{b_{N-1}} ="1",
(160,0) *+{c_{2}} ="2",
(160,80) *+{b_2} ="3",
(80,0) *+{c_1} ="4",
(80,80) *+{b_1} ="5",
"1", {\ar"0" <1.5pt>},
"1", {\ar"0" <-1.5pt>},
"3", {\ar"2" <1.5pt>},
"3", {\ar"2" <-1.5pt>},
"2", {\ar"5"},
"4", {\ar"3"},
"2",{\ar"13"},
"12",{\ar"3"},
"14",{\ar"1"},
"0",{\ar"15"},
"5", {\ar"4"<1.5pt>},
"5", {\ar"4"<-1.5pt>},
\end{xy}\end{split}\end{equation}
It is encouraging to note that mutation equivalences will allow us to make contact with the weak coupling discussion of section \ref{sunweak}. The quiver we have obtained (\ref{sunstrongquiv}) is already of the same form as (\ref{sunweakquiv}), but with different charge assignments. Mutating to the right on all $b_{2i}$ and to the left on all $b_{2i-1}$ will produce leave the quiver form unchanged, but transform the charges to $b_i=(0,-\alpha_i), c_i=(\alpha_i,\alpha_i).$ These are precisely the weak coupling charges proposed in section \ref{sunweak}, with some alternative choice of dyon pairs, $n_i=-1.$ Note, however, that in order to realize these mutations, we must go through a large number of wall crossings, since we took left-mutations of some $b_i,$ which, in our strong coupling calculation, are not left-most, but instead right-most.

We can use the quiver to compute the full BPS spectrum at this strong coupling chamber of moduli space. We begin by mutating on the left-most states, $c_i.$ This produces a new set of charges, $c_i\rightarrow -c_i,  b_i\rightarrow b_i+c_{i-1}+c_{i+1}.$ The new states that replace the $b_i$ are now left-most, again all at the same phase in the central charge plane. Focusing on the central charges of the nodes, we see that the charges of the new quiver are related to those of the original quiver by a rotation of $e^{-i \pi/N}$ (see Fig. \ref{sunzplane}). So as we continue mutating in phase order, this process of $N$ coincident mutations simply repeats itself.
Continuing in this way, a finite spectrum is exhibited by the mutation method with a mutation sequence of length $N(N-1)$,
\begin{equation}
c_1,c_2,\dots,c_{N-1},
b_1,b_2,\dots,b_{N-1}, 
c_1,c_2,\dots,c_{N-1},
b_1,b_2,\dots,b_{N-1},\dots
\end{equation} 

The states produced in this way are,
\begin{equation}
\begin{array}{ccccc}
c_1,&c_2,&c_3,&\dots,&c_{N-1},\\
b_{1}+c_2,&c_1+b_2+c_3,&c_{2}+b_3+c_{4},&\dots,&c_{N-2}+b_{N-1},\\
b_2+c_3,&b_1+c_2+b_3+c_4,&c_{1}+b_{2}+c_3+b_{4}+c_{5},&\dots,&c_{N-3}+b_{N-2}\\
b_3+c_4, & b_2+c_3+b_4+c_5, &b_{1}+c_{2}+b_{3}+c_{4}+b_{5}+c_{6},&\dots,&c_{N-4}+b_{N-2}\\
\vdots&\vdots&\vdots&\vdots&\vdots\\
b_{N-1},&b_{N-2},&b_{N-1}&,\dots,&b_{1}
\end{array}
\end{equation}
This array of states can be filled out iteratively after the first two rows are computed. The state $\mu_{ij}$ in position $(i,j)$ with $i\ge2$ is given by 
\begin{equation}\mu_{i-1,j-1}+\mu_{i-1,j+1}-\mu_{i-2,j},\end{equation} 
where we set $\mu_{ij}=0$ for $j<1$ and $j>N-1.$ It is slightly more economical to take as the base cases $i=0,1$ where we add $\mu_{0,j}=-b_j,$ along with $\mu_{1,j}=c_j$ as already given. The resulting states precisely fill out the full set of $N(N-1)$ vanishing cycles,
\begin{equation}
\boxed{|\mathcal{B}_{SU(N)}|=N(N-1).}
\end{equation}

This result agrees with the computation of strong coupling BPS states via CFT methods \cite{Lerche00} and is a strong confirmation of the techniques studied here.

\subsection{Adding Matter}\label{sunmatter}
Adding arbitrary hypermultiplet matter to pure SYM with ADE-type gauge group is quite analogous to the procedure described in \ref{su2matter} for $SU(2)$. Consider adding hypermultiplet matter charged under the gauge group $G$ in a representation $R.$ Again, we tune the mass of the matter to infinity. Here, by similar decoupling reasoning we would expect to add as a node a an electrically charged lowest weight state of the matter representation $R;$ ie we should have electric-magnetic charge $(-d,0)$ where $-d$ is the lowest weight of $R$. From this, positive linear combinations may generate the full representation $R$ by adding various $W$ bosons with charge $(\alpha_i,0)$ to the new state $(-d,0).$ 

Having determined the charge of the new node $f=(-d,0)$, it is straightforward to compute electric-magnetic inner products to fix the quiver. Explicitly, we may decompose the lowest weight $-d=-\sum_i d_i\omega_i$ where $d_i$ are positive integers. Then $f\circ b_j=(-d,0)\circ(0,\alpha_j)=-d_i (\omega_i\cdot\alpha_j )=-d_i$ and $f\circ c_j= (-d,0)\circ(\alpha_j,-\alpha_j)=d_i.$ Thus the new node has $d_i$ arrows connected to each node of the $i$th $SU(2)$ subquiver, forming an oriented three-cycle. Again we run into the subtlety seen in section \ref{n=2*}: this quiver can certainly generate the matter rep $R,$ but may in fact generate some additional matter representations. In fact, by adding such a node, we actually add the full tensor reducible representation $\otimes_i \mathbf{r_i}^{d_i},$ (where $\mathbf{r_i}$ are the fundamental reps of the gauge group) instead of adding only the irreducible rep, $R.$

We can propose one very clear consistency check on this procedure. Due to the structure of $\mathcal{N}=2$ hypermultiplets, adding a hypermultiplet in rep $R$ adds a multiplet of states in $R\oplus \bar{R}$. Thus, in principle, adding matter in rep $R$ is equivalent to adding matter in rep $\bar{R}.$ For the fundamental $\mathbf{N}$ of $SU(N),$ the lowest weight of $\mathbf{N}$ is $-\omega_{N-1},$ while the lowest weight of $\mathbf{\overline{N}}$ is $-\omega_{1}.$ This creates some ambiguity in defining the quiver of $SU(N)\;N_f>1.$

\begin{equation}\begin{split}\begin{xy} 0;<1pt,0pt>:<0pt,-1pt>:: 
(150,3)*+{\;\;} ="12",
(150,48)*+{\;\;}="13",
(200,4)*+{\;\;} ="14",
(200,48)*+{\;\;}="15",
(175,25) *+{\dots},
(250,0) *+{c_{N-1}} ="0",
(250,50) *+{b_{N-1}} ="1",
(100,0) *+{c_2} ="2",
(100,50) *+{b_2} ="3",
(50,0) *+{c_1} ="4",
(50,50) *+{b_1} ="5",
(300,-20) *+{f_{k+1}} ="6",
(300,10) *+{f_{k+2}} ="10",
(300,70) *+{f_{N_f}} ="11",
(300,40) *+{\vdots},
(0,-20) *+{f_1} ="7",
(0,10) *+{f_2} ="8",
(0,40) *+{\vdots},
(0,70) *+{f_k} ="9",
"1", {\ar"0" <1.5pt>},
"1", {\ar"0" <-1.5pt>},
"0", {\ar"6"},
"6", {\ar"1"},
"0", {\ar"10"},
"10", {\ar"1"},
"0", {\ar"11"},
"11", {\ar"1"},
"3", {\ar"2" <1.5pt>},
"3", {\ar"2" <-1.5pt>},
"2", {\ar"5"},
"4", {\ar"3"},
"2",{\ar"13"},
"12",{\ar"3"},
"14",{\ar"1"},
"0",{\ar"15"},
"5", {\ar"4"<1.5pt>},
"5", {\ar"4"<-1.5pt>},
"4", {\ar"7"},
"7", {\ar"5"},
"4", {\ar"8"},
"8", {\ar"5"},
"4", {\ar"9"},
"9", {\ar"5"},
\end{xy}\end{split}\end{equation}
By the above discussion, any choice of $0\le k\le N_f$ seems to give a possible quiver for this theory. 
For consistency, the representation theory of all of these quivers must be equivalent. One can easily check that the quivers are in fact mutation equivalent. To move node $f_i$ from the left to the right, apply the following sequence of mutations: $f_i,b_1,c_1,b_2,c_2,\dots,b_{N-1},c_{N-1};$ a similar reversed sequence $f_j,b_{N-1},c_{N-1},b_{N-2},c_{N-2},\dots,b_{1},c_{1}$ moves node $f_j$ from right to left. We can move the $f_i$ one by one across the quiver, and any two choices of $k$ will be connected via these mutation sequences.  Thus by the general reasoning of section \ref{MUT}, these quivers do in fact correspond to identical physical theories.


\subsection{BPS States of SQCD}\label{SQCD}
We now wish to extend our analysis of strong-coupling SYM to include arbitrary fundamental quark hypermultiplets coupled to the gauge group. Recall that our rule for coupling matter was valid with all masses tuned parametrically large. With a suitable definition of charges, only the $N_f$ flavor nodes will carry flavor charge,\footnote{Recall that in our analysis of $SU(2)$ with flavor, the natural assignment of charges gave flavor charge to the nodes of the $SU(2)$ subquiver, along with the additional flavor node. This was simply a familiar choice of convention; by redefining electric and magnetic charges, we can arrange a configuration in which only the additional matter node carries flavor charge.} and decouple from the pure gauge theory when masses are scaled up. We again study the origin of the Coulomb branch, and expect the light pure gauge degrees of freedom to reproduce the finite spectrum given above. Finally, we must fix the central charge phases of the flavor nodes; we choose all of them to be to the left of the $c_i;$ for definiteness, let $\arg\mathcal{Z}(f_1)>\arg\mathcal{Z}(f_2)>\dots>\arg\mathcal{Z}(f_{N_f}).$ Having fixed all parameters of the theory, we may use the mutation method to compute a finite spectrum. For each flavor $f_k$, we find, in phase order
\begin{equation}
f_k,f_k+b_1,f_k+b_1+c_1,f_k+b_1+c_1+b_2,\dots, f_k\sum_{i=1}^{N-1}b_i+c_i,
\end{equation}
given by mutation sequence
\begin{equation}
f_k,b_1,c_1,b_2\dots c_{N-1}.
\end{equation}
As discussed in section \ref{massivespec}, the charges assigned to nodes are dependent on some choice of `duality frame.' If we take the charge assignments found at weak coupling, $b_i=(0,\alpha_i), c_i=(\alpha_i,-\alpha_i),$ we can see a nice consistency check on this result.  With these charges, the flavor states found above contain $N$ pure electric (ie, zero magnetic charge) states with charges forming a fundamental $\mathbf{N}$ of the $SU(N),$ given by $f_k+\sum_{i=1}^k b_i+c_i,\, 0\le k \le N-1.$ The remaining states are then some additional $N-1$ additional flavor dyon states.

Since the flavor nodes are to the left with parametrically large masses, any state with flavor occurs before any of the light pure gauge degrees of freedom; by our choice of central charges, the flavor states occur in order. All states with flavor charge $f_1$ occur first, and then all states with charge $f_2$ and so on. Continuing with the mutation method, the set of $N(N-1)$ gauge dyons will be found after all the flavor states described above. The full spectrum is given by $N_f(2N-1)+N(N-1)$ BPS hypermultiplets, consisting of $2N-1$ flavor states for each fundamental, and $N(N-1)$ pure gauge strong coupling dyons,
\begin{equation}
\boxed{|\mathcal{B}_{\text{SQCD}}|=N_f(2N-1)+N(N-1).}
\end{equation}


\subsection{Further ADE examples}
Here, we briefly review some additional finite chambers of ADE-type gauge theories that may be obtained by the mutation method. For these examples, the period computation done in section \ref{sunstrong} becomes much more complicated. We will skip that calculation, and instead simply identify a finite mutation sequence that generalizes the one found there for $SU(N).$

For pure SYM with DE-type gauge group, the quiver was given in section \ref{ADEtype}. There exists a finite mutation sequence for any of the ADE-type quivers whose number of states is exactly the total number of roots of $G$,
\begin{equation}
\boxed{|\mathcal{B}_{ADE}|=\text{dim}( \text{adjoint}) -\text{rank} (G).}
\end{equation}
This spectrum can be interpreted as a monopole-dyon pair for every positive root. The mutation sequence is given as before 
\begin{equation}
c_1,c_2,\dots,c_n,b_1,b_2,\dots,b_n,c_1,c_2,\dots,c_n,\dots
\end{equation}

We can also study ADE-type groups with additional matter representations, by following the same strategy as \ref{SQCD}. We fix the pure gauge degrees of freedom at the strong coupling, finite chamber point discussed above, and take large mass limit for the matter. By choosing the phase of the matter nodes to be left-most, we force all states with flavor charge to be further left than the pure gauge states. For an A-type group (ie $SU(N)$), in addition to quarks, we may couple antisymmetric tensor representations, and find a finite chamber. Generalizing from the SQCD result, there is some duality frame for which the flavor states organize into $\frac{1}{2} N(N-1)$ pure electric states whose charges fill out the antisymmetric tensor of $SU(N)$, along with some number of additional dyon states.
Note that by contrast, an $SU(N)$ theory with matter in the symmetric tensor rep can never have a finite chamber. The symmetric tensor is given as a the highest weight representation of the tensor $\mathbf{N}\otimes\mathbf{N}.$ By the prescription of section \ref{sunmatter}, the resulting quiver would contain a subquiver of the form studied for the $SU(2),\,\mathcal{N}=2^*$ theory. In section \ref{n=2*}, we showed that this any chamber of this quiver contains at least two vector particles, and thus cannot have finitely many states. Furthermore, the presences of at least two accumulation rays obstructs the mutation method. The larger quiver for $SU(N)$ with a symmetric tensor will produce at least all the states obtained from its subquiver, and thus it will suffer from the same complications.

For a D-type group, $SO(2n)$ with matter in vector representation of $SO(2n)$, we find a finite chamber of $4(n+1)$ flavor states, along with the $2n(n-1)$ gauge states. Here the flavor states contain $2n$ pure electric states whose charges fill out a $\mathbf{2n}$-vector of $SO(2n),$ along with $2n+1$ additional flavor dyon states. With $N_v$ vector representations, we find
\begin{equation}
\boxed{|\mathcal{B}_{SO(2n)}|=N_v(4n+1)+2n(n-1).}
\end{equation}

We also find a finite chamber for $E_6$ with matter in the smallest fundamental representation, $\mathbf{27}$; the flavor states contain pure electric charges filling out the fundamental representation, along with 46 additional flavor dyon states; a theory with $N_f$ $\mathbf{27}$'s yields
\begin{equation}
\boxed{|\mathcal{B}_{E_6}|=73 N_f+72.}
\end{equation}
For $E_8$, one may not expect any finite chamber, since the smallest fundamental is the adjoint, and the resulting theory is $\mathcal{N}=2^*,$ that is, a massive deformation of a conformal $\mathcal{N}=4$ theory.


\section{Quivers for Theories of the Gaiotto Type}\label{tn}

A large class of $4d, \mathcal{N}=2$ theories emerging as the low energy limit of $n$ M5-branes wrapped on a punctured Riemann surface were studied in \cite{Gaiotto}. This class of theories contains, if one includes decoupling limits, all of the examples discussed herein, save the case of exceptional gauge groups. Here we begin a program aimed at systematically constructing the quivers for a general theory of the Gaiotto type.

In \cite{CV11,ACCERV1} the quivers for all rank two theories (i.e. theories with two M5-branes) with at least one puncture were found. The quivers were deduced directly from combinatorial data extracted from the Riemann surface \cite{KLMVW,SV,GMN09}. Here, we start by rediscovering the rank-2 quivers based on the methods in section \ref{su2gauge}. There we saw that glueing two rank 2 Gaiotto theories by a pair of punctures corresponds to diagonally gauging a pair of $SU(2)$ flavor symmetries, one per theory. Further, we saw that the corresponding gauging operation can be explicitly carried out on the quiver. Thus if we have the quivers for two rank two theories, we can glue them to produce the quiver of the combined theory.

We recall from \cite{Gaiotto} that there was in fact a basic building block from which all rank 2 theories can be constructed by glueing in this fashion. This building block, the so called $\mathcal{T}_2$ theory, corresponds to the thrice punctured sphere, and in 4d is the theory of a free half-hypermultiplet trifundamental of $SU(2)^3$. So, all we need to construct the quiver for any rank 2 Gaiotto theory is

\begin{itemize}
\item The quiver for the $\mathcal{T}_2$ theory, the theory of two M5-branes on a thrice punctured Riemann surface
\item A diagonal gauging rule for connecting two quivers associated with arbitrary punctured surfaces
\end{itemize}

The gauging rule is just as described in section \ref{su2gauge}, except if one of the two theories we are glueing happens to be $\mathcal{T}_2$ itself. Obviously since we plan on constructing everything by starting with $\mathcal{T}_2$ quivers, this difficulty must be overcome, and indeed we discuss the resolution in section \ref{t2quiv}.

For a theory with $n$ M5-branes there is no longer a single type of puncture; there is a classification of punctures by Young tableaux \cite{Gaiotto,Gaiotto:2009gz,BBT,CD1}. However, there is still a distinguished type of puncture: the kind that appears as the degeneration limit of surfaces, and correspondingly the ones which we can use to glue surfaces together. This is the puncture which corresponds to an $SU(n)$ flavor symmetry. In turn there is a special building block, namely the three punctured sphere, corresponding to the theory with (at least) $SU(N)^3$ flavor symmetry, called $\mathcal{T}_n$. 

Thus we begin here by proposing a quiver for the $\mathcal{T}_n$ theory. We consider checks on this proposal in the rank 3 case related to Argyres-Seiberg duality \cite{AS}. We then go on to describe how one can build a general quiver for a rank 3 surface, by considering how two $U(1)$ type punctures collide to give an $SU(3)$ puncture, and comment on the generalization to the rank $n$ case. 

\subsection{Quivers for $\mathcal{T}_n$}
We will first describe our proposal for the BPS quiver of the $\mathcal{T}_n$ theory. One reason why this theory is simple to study is that it is known to be related to M-theory on the Calabi-Yau singularity $ \mathbb{C}^{3}/\mathbb{Z}_{n}\times \mathbb{Z}_{n}$ \cite{BBT}.  Specifically, upon compactification along a circle we recover the $\mathcal{T}_{n}$ theory.  The quiver for the Calabi-Yau singularity can be computed by standard methods \cite{DM}, and indeed the result is known:
\begin{itemize}
\item The quiver has $n^{2}$ nodes.  We arrange them in a grid and index them accordingly as $A_{ij}$ where the indices $i,j$ run from $1$ to $n+1$ and are cylclicly identified so that $1$ is equal to $n+1$.
\item There are the following arrows
\begin{itemize}
\item Horizontal arrows: $A_{ij}\rightarrow A_{i+1,j}$
\item Vertical arrow: $A_{ij}\rightarrow A_{i,j+1}$
\item Diagonal arrows: $A_{ij}\rightarrow A_{i-1,j-1}$
\end{itemize}
\end{itemize}
An example of this structure for the case of $n=4$ is shown in below.
\begin{equation}\begin{split}\begin{xy}0;<1pt,0pt>:<0pt,-1pt>:: 
(0,0) *+{d}="11",
(30,0) *+{e}="12",
(60,0) *+{f}="13",
(90,0) *+{g}="14",
(120,0) *+{*_d}="15",
(0,30) *+{c}="21",
(30,30) *+{\Circle}="22",
(60,30) *+{\Circle}="23",
(90,30) *+{\Circle}="24",
(120,30) *+{*_c}="25",
(0,60) *+{b}="31",
(30,60) *+{\Circle}="32",
(60,60) *+{\Circle}="33",
(90,60) *+{\Circle}="34",
(120,60) *+{*_b}="35",
(0,90) *+{a}="41",
(30,90) *+{\Circle}="42",
(60,90) *+{\Circle}="43",
(90,90) *+{\Circle}="44",
(120,90) *+{*_a}="45",
(0,120) *+{*_d}="51",
(30,120) *+{*_e}="52",
(60,120) *+{*_f}="53",
(90,120) *+{*_g}="54",
"11", {\ar"12"},
"12", {\ar"13"},
"13", {\ar"14"},
"14", {\ar"15"},
"21", {\ar"22"},
"22", {\ar"23"},
"23", {\ar"24"},
"24", {\ar"25"},
"31", {\ar"32"},
"32", {\ar"33"},
"33", {\ar"34"},
"34", {\ar"35"},
"41", {\ar"42"},
"42", {\ar"43"},
"43", {\ar"44"},
"44", {\ar"45"},
"21", {\ar"11"},
"31", {\ar"21"},
"41", {\ar"31"},
"51", {\ar"41"},
"22", {\ar"12"},
"32", {\ar"22"},
"42", {\ar"32"},
"52", {\ar"42"},
"23", {\ar"13"},
"33", {\ar"23"},
"43", {\ar"33"},
"53", {\ar"43"},
"24", {\ar"14"},
"34", {\ar"24"},
"44", {\ar"34"},
"54", {\ar"44"},
"12",{\ar"21"},
"13",{\ar"22"},
"14",{\ar"23"},
"15",{\ar"24"},
"22",{\ar"31"},
"23",{\ar"32"},
"24",{\ar"33"},
"25",{\ar"34"},
"32",{\ar"41"},
"33",{\ar"42"},
"34",{\ar"43"},
"35",{\ar"44"},
"42",{\ar"51"},
"43",{\ar"52"},
"44",{\ar"53"},
"45",{\ar"54"},
\end{xy}\end{split}
\end{equation}

When we further compactify on a circle, the states carrying momentum along the circle become parametrically heavy.  As a result, the charge lattice of the theory has its rank reduced by one.  A natural way to achieve this is simply to delete a node of the above quiver, in which all nodes are identical. This is our proposal for the general $\mathcal{T}_{n}$ quiver.  In the following we will provide two explicit checks on this procedure by studying $\mathcal{T}_{2}$ and $\mathcal{T}_{3}$ cases in more detail.


\subsection{Rank 2 Quivers}
\label{t2quiv}
In the case of $\mathcal{T}_{2}$ the  quiver takes the form:
\begin{equation}\begin{split}\begin{xy}0;<1pt,0pt>:<0pt,-1pt>:: 
(0,0) *+{\Circle}="1",
(50,-10) *{X_1},
(50,10) *{Y_1},
(35.5,-27.5) *{X_2},
(14.5,-42.5) *{Y_2},
(64.5,-27.5) *{X_3},
(85.5,-42.5) *{Y_3},
(100,0) *+{\Circle}="2",
(50,-70) *+{\Circle}="3",
"1", {\ar "2" <-1.5pt>},
"2", {\ar "3" <-1.5pt>},
"3", {\ar "1" <-1.5pt>},
"2", {\ar "1" <-1.5pt>},
"3", {\ar "2" <-1.5pt>},
"1", {\ar "3" <-1.5pt>},
\end{xy}\end{split}
\end{equation}

Further, the general methods of \cite{LF1,LF2,ACCERV1} determine the superpotential
\begin{equation}
W=X_{1}Y_{1}+X_{2}Y_{2}+X_{3}Y_{3}+X_{1}X_{2}X_{3}+Y_{1}Y_{2}Y_{3}.
\end{equation}
This quiver is novel in that it provides our first example of a quiver with canceling arrows where the potential is not strong enough to integrate out the corresponding fields.

As described above, this theory has a BPS spectrum given by a half-hypermultiplet trifundamental of the flavor group $SU(2)^3$.  For generic values of the central charges of the nodes, this flavor symmetry is broken. However, the number of states, namely eight, is the same.  Of these eight, only half are particles, and of these four particles, three are manifest as nodes of the quiver.  Thus, consistency demands that our $\mathcal{T}_{2}$ quiver, together with its given superpotential, supports exactly one hypermultiplet bound state.  

The existence of this single state can be checked explicitly using quiver representation theory.  The unique representation with the required charges is a bound state of one of each of the three node particles.  The F-term equations of motion in this case are 
\begin{eqnarray}
X_{1}+Y_{2}Y_{3} & = & 0, \\
X_{2}+Y_{3}Y_{1} & = & 0, \\
X_{3}+Y_{1}Y_{2} & = & 0, \\
Y_{1}+X_{2}X_{3} & = & 0, \\
Y_{2}+X_{3}X_{1} & = & 0, \\
Y_{3}+X_{1}X_{2} & = & 0.
\end{eqnarray}
The solution of interest to us has all fields non-vanishing with
\begin{equation}
X_{1}X_{2}X_{3}=Y_{1}Y_{2}Y_{3}=-1.
\end{equation}
The moduli space can easily be determined by noting that, since $X_{1}$ and $X_{2}$ are non-zero, we can eliminate all gauge redundancy by setting $X_{1}=X_{2}=1.$  Then all remaining field values are fixed by the F-term equations and hence the moduli space is a point.  Thus this representation results in a single hypermultiplet.  Noting that this representation admits no non-trivial subrepresentations, we further conclude that this hypermultiplet is always stable and provides the required state in the spectrum.

Now that we have examined the $\mathcal{T}_2$ quiver itself, the next step is to begin glueing copies of it together by diagonally gauging $SU(2)$ flavor symmetries, as described above. We found a way to gauge generic quivers in section \ref{su2gauge}. However, that analysis in fact does not apply to gauging a factor of the $\mathcal{T}_2$ quiver itself, as we will see. It does apply to quivers associated with any other pair of rank 2 surfaces, so once we find how to gauge $\mathcal{T}_2$, the procedure for constructing a general rank 2 surface will be clear, and indeed agree with the results found by alternative methods in \cite{ACCERV1}.

The obvious obstruction to gauging $\mathcal{T}_2$ in the naive way can be seen by considering the charges of the quiver's nodes under the $SU(2)^3$ flavor symmetry. Recall that the content of the theory is a half-hypermultiplet transforming in the $2\otimes 2 \otimes 2$ representation of $SU(2)^3$. We will see that no $SU(2)$ action can be made manifest on nodes of the quiver, something which is required to use the gauging procedure of section \ref{su2gauge}. Suppose it could. Then, without loss of generality, we can take two of the nodes to have charges $(\frac 1 2, \frac 1 2, \frac 1 2)$ and $(- \frac 1 2, \frac 1 2, \frac 1 2)$. Further, we can without loss of generality assign the third node charge $(\frac 1 2, - \frac 1 2, \frac 1 2)$. Then clearly neither the state $(\frac 1 2, \frac 1 2, -\frac 1 2)$ nor the state $(-\frac 1 2, - \frac 1 2, \frac 1 2)$ is a positive integral combination of the nodes. Thus indeed no $SU(2)$ symmetry can be made manifest in the quiver, and so we can't simply apply the rules of section \ref{su2gauge}.

Why is this example at odds with the general framework? We note that for a generic quiver, to make the states associated with some symmetry appear as nodes, we could simply go to the symmetric point in the theory and then rotate the half-plane so that they were all left-most. Usually we have such freedom because of mass parameters which accompany flavor symmetries. However, in this case the charges of the states associated with the symmetry are not independent directions in the charge lattice, and can't be independently tuned. This is related to the fact that the $SU(2)$ symmetries mix particles and anti-particles, as the symmetry acts on half-hypermultiplets.

While our general analysis does not apply here, we can still gauge an $SU(2)$ ``by hand," since we know very clearly the content of this theory. After a single $SU(2)$ is gauged, the difficultly above disappears and all the subsequent quivers arrived at can be gauged in the naive way. Let us start with the $\mathcal{T}_2$ quiver with charges $(-\frac 1 2, -\frac 1 2, -\frac 1 2)$, $(\frac 1 2, \frac 1 2, -\frac 1 2)$ and $(-\frac 1 2, \frac 1 2, \frac 1 2)$. The bound states of this quiver fill out, with anti-particles, the trifundamental of $SU(2)^3$.

Say we gauge the first $SU(2)$ factor. Then our quiver should be the quiver for an $SU(2)$ gauge group coupled to a basis for those states with first $SU(2)$ charge $-\frac 1 2$ (since the $SU(2)$ will produce the $\frac 1 2$ states as bound states in the usual way).\footnote{Of course we may find additional bound states now that we have an interacting theory.} Thus we can simply take $(-\frac 1 2, -\frac 1 2, -\frac 1 2)$ in addition to $(-\frac 1 2, \frac 1 2, -\frac 1 2)$. Since the $SU(2)_1$ electromagnetic charges of these states will be identical, they couple only to the $SU(2)$ factor in the usual way, and we are left with the quiver

  \[
  \begin{xy}
  (0,0)*{\Circle}="a";
(0,30)*{\Circle}="c";
(15,15)*{\Circle}="b";
(25,15)*{\Circle}="d";
  {\ar@2{->} "a";"c"};
    {\ar "c";"b"};{\ar "b";"a"};
    {\ar "c";"d"};{\ar "d";"a"}
  \end{xy}
  \]

Now we have a procedure for building up the quiver for any rank 2 Gaiotto theory from its pair of pants decomposition. Quivers obtained via this method indeed agree with the results found in \cite{ACCERV1}.

\subsection{Rank 3 Quivers}
In the case of $\mathcal{T}_{3}$ the  quiver takes the form:
\begin{equation}\begin{split}\begin{xy}0;<1pt,0pt>:<0pt,-1pt>:: 
(-50,0) *+{a_1}*\cir<8pt>{}="a3",
(0,0) *+{a_2}*\cir<8pt>{}="a2",
(50,0) *+{a_3}*\cir<8pt>{}="a1",
(-50,150) *+{b_1}*\cir<8pt>{}="b1",
(50,150) *+{b_2}*\cir<8pt>{}="b2",
(0,150) *+{b_3}*\cir<8pt>{}="b3",
(-170,100) *+{c_1}*\cir<8pt>{}="c1",
(-170,50) *+{c_2}*\cir<8pt>{}="c2",
"a1",{\ar"b1"},
"a1",{\ar"b2"},
"a1",{\ar"b3"},
"a2",{\ar"b1"},
"a2",{\ar"b2"},
"a2",{\ar"b3"},
"a3",{\ar"b1"},
"a3",{\ar"b2"},
"a3",{\ar"b3"},
"b1",{\ar"c1"},
"b1",{\ar"c2"},
"b2",{\ar"c2"},
"b2",{\ar"c1"},
"b3",{\ar"c2"},
"b3",{\ar"c1"},
"c1",{\ar"a3"},
"c1",{\ar"a2"},
"c1",{\ar"a1"},
"c2",{\ar"a3"},
"c2",{\ar"a2"},
"c2",{\ar"a1"},
\end{xy}\end{split}
\end{equation}
The quiver's structure can be better understood by grouping the nodes into three sets labeled above as $\{a_{1}, a_{2}, a_{3}\}$, $\{b_{1}, b_{2}, b_{3}\}$, $\{c_{1}, c_{2}\}$.  In terms of the quiver there is a permutation symmetry on the $a$-type nodes, and similarly on the $b$ and $c$-type nodes.  Thus for the purposes of illustrating the general structure we can simply draw one member of each group, in which case the quiver appears as:
\begin{equation}\begin{split}\begin{xy}0;<1pt,0pt>:<0pt,-1pt>:: 
(0,0) *+{a_i}*\cir<8pt>{}="a",
(60,0) *+{b_j}*\cir<8pt>{}="b",
(30,-40) *+{c_k}*\cir<8pt>{}="c",
"a", {\ar "b"},
"b", {\ar "c"},
"c", {\ar "a"}
\end{xy}\end{split}
\end{equation}

We first find a finite chamber of this quiver using the mutation method. In decreasing phase order, we find the 24 state chamber:
\begin{equation}\begin{array}{l}
c_1,c_1+a_1,c_1+a_2,c_1+a_3,b_1,b_2,b_3,c_2+b_1+b_2+b_3,2c_1+a_1+a_2+a_3,\\c_1+a_1+a_2,c_1+a_2+a_3,c_1+a_1+a_3,c_2+b_1+b_2,c_2+b_2+b_3,c_2+b_1+b_3, \\c_1+a_1+a_2+a_3, 2c_2+b_1+b_2+b_3,a_1,a_2,a_3,c_2+b_1,c_2+b_2,c_2+b_3,c_2.
\end{array}\end{equation}

We can provide a strong consistency check on our proposal for this quiver by recalling that the $\mathcal{T}_3$ theory coincides with the $E_6$ Minahan-Nemeschansky theory \cite{MN}. In particular, the flavor symmetry group $SU(3)^3$ sits inside a full $E_6$ flavor group. The $E_6$ theory enjoys an Argyres-Seiberg type duality \cite{AS}. In particular, there is an equivalence between the $E_6$ theory with an $SU(2)$ subgroup of its flavor symmetry gauged and coupled to an additional fundamental, and $SU(3)$ SYM coupled to 6 fundamentals.

This duality has a strict implication for our $T_3$ quiver. On one side, we can gauge an $SU(2)$ global symmetry in the $\mathcal{T}_{3}$ quiver following the considerations of section \ref{su2gauge}, and couple to it and additional fundamental in the obvious way; on the other side, we have proposed and studied quivers of arbitrary $SU(N)$ SQCD theories in section \ref{sunmatter}. Since these theories are to be connected by a single moduli space, there must exist some mutation equivalence between their quivers.

Indeed, in the process of checking this duality, we find an additional check that we can perform. The $\mathcal{T}_{3}$ quiver exhibits $S_2\times S_3\times S_3$ discrete symmetries, acting on the $c$, $b,$ and $a$-type nodes which by the reasoning of section \ref{su2gauge} indicates a global $SU(2)\times SU(3)\times SU(3)$ of the resulting physics. The actual theory admits a full $E_6$ symmetry, which contains three identical $SU(3)$s; in the quiver, however, we only see the $SU(2)$ subgroup of one of these $SU(3)$'s. The physics, on the other hand, does not distinguish between the three $SU(3)$'s, and thus applying the quiver gauging rules to any of the three $SU(2)$'s available should give mutation equivalent results. It is a nontrivial fact that the quivers obtained from these three gauging procedures, 
\begin{equation}\begin{xy} 0;<1pt,0pt>:<0pt,-1pt>:: 
(0,0) *+{3} ="1",
(35,20) *+{1} ="2",
(35,-20) *+{2}="3",
(70,0) *+{c_1}="4",
(105,20) *+{a_i} ="a2",
(105,-20) *+{b_i} ="b2",
"2", {\ar"3" <1.5pt>},
"2", {\ar"3" <-1.5pt>},
"3", {\ar"4"},
"4", {\ar"2"},
"3", {\ar"1"},
"1", {\ar"2"},
"4", {\ar"a2"},
"b2", {\ar"4"},
"a2", {\ar"b2"},
(150,0) *+{3} ="5",
(185,20) *+{1} ="6",
(185,-20) *+{2}="7",
(220,-10) *+{a_2}="8",
(220,10) *+{a_1}="9",
(255,20) *+{b_i} ="c2",
(255,-20) *+{c_i} ="d2",
"6", {\ar"7" <1.5pt>},
"6", {\ar"7" <-1.5pt>},
"8", {\ar"6"},
"7", {\ar"5"},
"7", {\ar"8"},
"5", {\ar"6"},
"8", {\ar"c2"},
"d2", {\ar"8"},
"9", {\ar"c2"},
"d2", {\ar"9"},
"c2", {\ar "d2"},
(300,0) *+{3} ="10",
(335,20) *+{1} ="11",
(335,-20) *+{2}="12",
(370,-10) *+{b_2}="13",
(370,10) *+{b_1}="14",
(405,-20) *+{a_i} ="e2",
(405,20) *+{c_i} ="f2",
"11", {\ar"12" <1.5pt>},
"11", {\ar"12" <-1.5pt>},
"13", {\ar"11"},
"12", {\ar"10"},
"12", {\ar"13"},
"10", {\ar"11"},
"13", {\ar"f2"},
"e2", {\ar"13"},
"14", {\ar"f2"},
"e2", {\ar"14"},
"f2", {\ar "e2"},
\end{xy}\end{equation}
are all mutation equivalent, and hence describe the same 4d field theory.\footnote{In view of what's to come, we have already coupled additional fundamentals to our $SU(2)$s.} In these quivers, we have represented the triplets of identical nodes as $a_i,b_i,\, i=1,2,3$, and the pair of identical nodes $c_i,\, i=1,2.$ The arrows incident on these duplicated nodes of course indicate sets of arrows, one incident on each of the duplicated nodes. Beginning with the quiver on the left, we can obtain the middle quiver by the following mutation sequence:
$$3,c,b_1,a_1,c,3,2,1,b_1,b_2,c,$$
and we obtain the quiver on the right via:
$$3,c,b_1,a_1,c,3,2,1,b_1,b_2,1.$$

Now we return to our check of Argryes-Seiberg duality. Since the three quivers obtained by gauging an $SU(2)$ subgroup are mutation equivalent, we now focus on the left-most quiver. Argyres-Seiberg duality indicates that this quiver should be mutation equivalent to:
\begin{equation}\begin{split}\begin{xy} 0;<1pt,0pt>:<0pt,-1pt>:: 
(100,0) *+{c_2} ="2",
(100,50) *+{b_2} ="3",
(50,50) *+{c_1} ="4",
(50,0) *+{b_1} ="5",
(150,-10) *+{f_4} ="6",
(150,25) *+{f_5} ="10",
(150,60) *+{f_6} ="11",
(0,-10) *+{f_1} ="7",
(0,25) *+{f_2} ="8",
(0,60) *+{f_3} ="9",
"2", {\ar"6"},
"6", {\ar"3"},
"2", {\ar"10"},
"10", {\ar"3"},
"2", {\ar"11"},
"11", {\ar"3"},
"3", {\ar"2" <1.5pt>},
"3", {\ar"2" <-1.5pt>},
"2", {\ar"5"},
"4", {\ar"3"},
"5", {\ar"4"<1.5pt>},
"5", {\ar"4"<-1.5pt>},
"4", {\ar"7"},
"7", {\ar"5"},
"4", {\ar"8"},
"8", {\ar"5"},
"4", {\ar"9"},
"9", {\ar"5"},
\end{xy}\end{split}\end{equation}
Beginning again with the left-most quiver above, we find a mutation equivalence given by
$$3,c,b_1,a_1,2,a_1,a_2,a_3,b_2,b_3.$$
This is a very robust check. Argyres-Seiberg duality is manifest at the level of quivers by a non-trivial sequence of mutation dualities.

Now we consider how one would write the quiver for a general rank 3 Gaiotto theory. First we recall the new punctures on Riemann surfaces that arise in the rank 3 case. For rank 2, there was a single type of puncture, which indicated an $SU(2)$ flavor symmetry of the theory. In rank 3, we have two types of punctures; punctures of the first kind indicate $U(1)$ flavor symmetries, and punctures of the second kind indicate $SU(3)$ flavor symmetries.

We require two quiver gauging rules to build up new theories. The first is the analog of the glueing/gauging rule developed in section \ref{su2gauge}. We consider glueing two Riemann surfaces along punctures of the second kind. This again corresponds to gauging the diagonal subgroup $SU(3)_d\subset SU(3)_1\times SU(3)_2$ of two flavor symmetries associated with the two punctures. At the level of the quiver, we have a straightforward rule - identify the two $SU(3)$ flavor symmetries as $S_3$ discrete symmetries of the respective quivers, delete two of the three symmetric nodes from each quiver, add a $SU(3)$ SYM subquiver, and couple the remaining nodes from each triplet as fundamental $\mathbf{3}$'s of the new $SU(3)$ subquiver. The other operation on Riemann surfaces we must understand is that of splitting punctures. The prototypical example of this operation is exactly the Argyres-Seiberg duality just discussed. Argyres-Seiberg duality related the Gaiotto theory on a sphere with two punctures of each kind (which gives $SU(3),\, N_f=6$) to the theory on a sphere with three punctures of the second kind (the $\mathcal{T}_{3}$ theory), with a gauged $SU(2)$ flavor symmetry. We may interpret this as splitting a puncture of the second kind into two punctures of the first kind. The effect on the resulting physics is to gauge an $SU(2)$ subgroup of the $SU(3)$ flavor symmetry corresponding to the split puncture, and couple a fundamental to the gauged $SU(2).$ Again we have a straightforward gauging procedure for the quiver.

Although we have described quite explicit rules, we must now face some limitations in implementing them. Recall from section \ref{su2gauge} that in the rank 2 case, for a surface with $p$ punctures, there exists a mutation form with $p-1$ global $SU(2)$ symmetries visible in the quiver. Because of this fact, we can gauge all but one $SU(2)$, and thus build up any surface with at least one puncture. In the rank 3 case, all of the quivers have infinite mutation classes \cite{CV11}. As a result, there is no easy way to systematically search the mutation classes and identify a quiver that makes the maximal number of symmetries visible. As we glue quivers, we actually are losing visible symmetries. For example, consider glueing two $\mathcal{T}_{3}$ theories to form the sphere with four punctures of the second kind. We start with two visible $SU(3)$ symmetries in each quiver; after the glueing, the resulting theory has only two remaining visible $SU(3)$'s. In order to find the maximal three $SU(3)$'s, we would need to go through some mutation sequence, which could involve arbitrarily many mutations. Finding such a mutation sequence is quite a difficult computational problem that scales as $\text{(number of nodes)}^\text{(length of sequence)}$. Unfortunately, this is an obstruction to implementing this procedure in practice if one wants to obtain surfaces of genus two or higher.

Nonetheless, these techniques allow us to propose quivers for spheres and tori with sufficient punctures. Let $(p_1,p_2)$ denote the number of punctures of the first and second kind respectively. Glueing a $\mathcal{T}_{3}$ surface to an existing theory takes $(p_1,p_2)\rightarrow (p_1,p_2+2),$ and the splitting rule takes $(p_1,p_2)\rightarrow (p_1+2,p_2-1).$ We take as our base cases the $\mathcal{T}_{3}$ theory (a sphere with $(0,3)$ punctures). The latter theory has a weak coupling description as an $SU(3)^2$ gauge theory with a bifundamental and three fundamental quarks for each $SU(3).$ This leads to a proposal for the associated quiver, 
\begin{equation}\begin{split}\begin{xy} 0;<1pt,0pt>:<0pt,-1pt>:: 
(100,0) *{\Circle} ="2",
(100,50) *{\Circle} ="3",
(50,50) *{\Circle}="4",
(50,0) *{\Circle} ="5",
(150,25) *{\Circle} ="10",
(0,-10) *{\Circle} ="7",
(0,25) *{\Circle} ="8",
(0,60) *{\Circle} ="9",
(250,0) *{\Circle} ="12",
(250,50) *{\Circle} ="13",
(200,50) *{\Circle} ="14",
(200,0) *{\Circle} ="15",
(300,-10) *{\Circle} ="17",
(300,25) *{\Circle} ="18",
(300,60) *{\Circle} ="19",
"2", {\ar"10"},
"10", {\ar"3"},
"3", {\ar"2" <1.5pt>},
"3", {\ar"2" <-1.5pt>},
"2", {\ar"5"},
"4", {\ar"3"},
"5", {\ar"4"<1.5pt>},
"5", {\ar"4"<-1.5pt>},
"4", {\ar"7"},
"7", {\ar"5"},
"4", {\ar"8"},
"8", {\ar"5"},
"4", {\ar"9"},
"9", {\ar"5"},
"14", {\ar"10"},
"10", {\ar"15"},
"13", {\ar"12" <1.5pt>},
"13", {\ar"12" <-1.5pt>},
"12", {\ar"15"},
"14", {\ar"13"},
"15", {\ar"14"<1.5pt>},
"15", {\ar"14"<-1.5pt>},
"12", {\ar"17"},
"17", {\ar"13"},
"12", {\ar"18"},
"18", {\ar"13"},
"12", {\ar"19"},
"19", {\ar"13"},
\end{xy}\end{split}\end{equation}
From these two base cases it is relatively easy to see that the glueing and splitting rules will allow us to construct spheres satisfying $p_1+2p_2\ge 6$ and tori satisfying $p_1+2p_2 \ge 2.$


\section*{Acknowledgements}

We would like to thank Andy Neitzke and Greg Moore for discussing their upcoming work  with us \cite{GMN12}. We thank Dan Xie for interesting discussions of his work. We thank the 2011 Simons workshop in Mathematics and Physics and the Simons
Center for Geometry and Physics for hospitality during the completion of this
work. The work of MA is supported by DFG fellowship AL 1407/1-1. The work of CV
is supported by NSF grant PHY-0244821

\end{document}